\documentclass[a4paper,10pt,abstracton,openbib,final]{scrartcl}
\makeatletter
\DeclareOldFontCommand{\rm}{\normalfont\rmfamily}{\mathrm}
\DeclareOldFontCommand{\sf}{\normalfont\sffamily}{\mathsf}
\DeclareOldFontCommand{\tt}{\normalfont\ttfamily}{\mathtt}
\DeclareOldFontCommand{\bf}{\normalfont\bfseries}{\mathbf}
\DeclareOldFontCommand{\it}{\normalfont\itshape}{\mathit}
\DeclareOldFontCommand{\sl}{\normalfont\slshape}{\@nomath\sl}
\DeclareOldFontCommand{\sc}{\normalfont\scshape}{\@nomath\sc}
\makeatother
\usepackage[a4paper,includehead,top=20mm,bottom=20mm,right=20mm,left=20mm]{geometry}
\usepackage{amssymb}
\usepackage{amsthm}
\usepackage{amsmath, amsthm, amsxtra}
\usepackage{authblk}
\usepackage{graphicx}

\usepackage{bm, upgreek}
\usepackage{booktabs, multirow}
\usepackage{subfigure}
\usepackage{float}

\newcommand{\begeq}{\begin{equation}\begin{gathered}}
\newcommand{\eqend}{\end{gathered}\end{equation}}
\renewcommand{\t}[1]{\bm{#1}}
\renewcommand{\tt}[1]{\pmb{#1}}
\renewcommand{\d}{\, \mathrm d }

\newcommand{\p}{\partial}

\newcommand{\del}{\updelta}

\newcommand{\pd}[2]{\frac{\p #1}{\p #2}}

\newcommand{\degree}{^\circ}

\newcommand{\cen}{\overset{\text{c}}}
\newcommand{\ten}[1]{\bm{#1}}
\newcommand{\begal}{\begin{equation}\begin{aligned}}
\newcommand{\alend}{\end{aligned}\end{equation}}
\newcommand{\ma}{^\text{M}}
\graphicspath{{Figures/}}

\usepackage{setspace}

\title{\huge Verification of asymptotic homogenization method developed for periodic architected materials in strain gradient continuum}
\author[a]{Hua Yang} 
\author[b]{B. Emek Abali \thanks{Corresponding author: bilenemek@abali.org}} 
\author[a]{Wolfgang H. M\"uller } 
\author[c]{\\ Salma Barboura}
\author[c]{Jia Li}

\affil[a]
{Technische Universit\"at Berlin, 
	Berlin Germany} 

\affil[b]
{Uppsala University, 
	 Uppsala, Sweden}

\affil[c]
{Sorbonne Paris North University, 
	Paris, France}

\date{} 
\date{}

\begin{document}
	
\maketitle
\begin{abstract}
Strain gradient theory is an accurate model for capturing the size effect and localization phenomena. However, the challenge in identification of corresponding constitutive parameters limits the practical application of the theory.
We present and utilize asymptotic homogenization herein. All parameters in rank four, five, and six tensors are determined with the demonstrated computational approach. Examples for epoxy carbon fiber composite, metal matrix composite, and aluminum foam illustrate the effectiveness and versatility of the proposed method. The influences of volume fraction of matrix, the stack of RVEs, and the varying unit cell lengths on the identified parameters are investigated. The homogenization computational tool is applicable to a wide class materials and makes use of open-source codes in FEniCS. We make all of the codes publicly available in order to encourage a transparent scientific exchange.

\end{abstract}
\section{Introduction}
\label{Introduction}

Composite materials have been widely used in engineering practice.  Due to the heterogeneous nature of composites, the mechanical properties of such materials are dependent on their substructures, for example, the material properties of matrix and reinforcements, the shape of inclusions, or the volume fraction of matrix, etc. 
An accurate determination of effective properties of these heterogeneous media plays an important role in the design and analysis of composites. Experiments could be conceived to evaluate their effective properties, but it is also possible to compute effective material parameters by means of homogenization methods \cite{ boutin1996microstructural, dirrenberger2019computational}, which reduce demands for experiments and enable to comprehend microstructure influence on the macroscale in any complex geometries.

Homogenization techniques \cite{arabnejad2013mechanical, chen2020extended, yvonnet2020computational, jakabvcin2020periodic} allow to represent a heterogeneous elastic material, at the microscale, as an equivalent homogeneous  elastic material at the macroscale. Although of primary importance, the conventional homogenization fails to describe the mechanical response when the heterogeneity of the material is of the same order of the macroscale. This inaccuracy is due to the fact that the conventional homogenization methods are based on a separation of scales, given by $\epsilon = l/L $, $l \ll L$. Here, $l$ represents the typical length scale
characteristic of the microstructural heterogeneity and $L$ stands for the macroscopic length scale. If the microstructure consists of relatively small heterogeneity, or the macroscopic length is infinitely large, classical homogenization gives an adequate estimate of the average macroscopic properties \cite{hollister1992comparison}. However, if the size of heterogeneity is of the same order of magnitude as that of the macroscopic problem, conventional homogenization technique fails.  For example, the size effect occurs when the length scale of the macroscopic heterogeneous materials ($L$) approaches the length scale of the underlying heterogeneity ($l$).  An up-scaling of Cauchy theory indicates that additional terms are necessary in the constitutive equations in order to predict the size effect observed in experiments \cite{muller2020experimental}.

In order to incorporate these additional terms, different homogenization techniques are proposed in the literature, for example in the framework of generalized mechanics \cite{mindlin1968first, eringen1999theory, altenbach2016generalized} such as micropolar theory \cite{kumar2004generalized, dos2012construction}, couple stress \cite{skrzat2020effective}, strain gradient theory \cite{yvonnet2020computational, kouznetsova2002multi, goda2016construction, abdoul2019homogenization, weeger2021numerical, forest2011generalized}, and micromorphic continuum \cite{rokovs2019micromorphic}. The task of obtaining homogenized constitutive equations for generalized continua is challenging and a number of debates are active in the literature \cite{kumar2004generalized, dos2012construction, liu2009effective, eremeyev2016effective, ganghoffer2021variational}. Many methods have been proposed to construct strain gradient continua by means of asymptotic homogenization approaches \cite{bacigalupo2018identification, boutin2017linear}, multi-scale computational approaches \cite{kouznetsova2004multi}, dynamic methods \cite{rosi2018validity,rosi2020waves}, and several other identification techniques \cite{dell2019advances, dell2019pantographic, misra2015identification, alibert2019homogenization, rahali2015homogenization}. Asymptotic homogenization method improves descriptions by exploiting higher order terms and considering their role in macroscopic behaviors. In \cite{li2011micromechanics, li2013numerical, barboura2018establishment, yang2019determination, abali2020additive}, an asymptotic homogenization based solution has been utilized to determine parameters of composite materials.  Two issues were addressed therein. One	is that the identified strain gradient parameters are all zero when structures are homogeneous. The other one is that these parameters are independent of stack of RVEs. 

In this paper, we briefly recall the homogenization method described in \cite{yang2019determination}, which is based on the formal analysis in \cite{barboura2018establishment}. A complete computational methodology determining all parameters in 2D and 3D has been achieved recently in \cite{abali2020additive}. We basically use the same procedure and verify the computational implementation by several numerical sanity checks. In this way, we reveal an important limitation that remains undetected during a formal analysis. Homogenization begins with two materials of different properties. In the case of one material with a nearly zero stiffness, the difference in properties may cause numerical problems in the implementation. For example, a structure with voids is a benchmark case for this issue. Indeed, we propose a change in the formulation in order to make the numerical implementation robust and the methodology more general. We apply the method to determine 2D and 3D composite materials effective parameters as well as verify the results by additional simulations. For determining parameters, we use epoxy carbon fiber composite material, SiC/Al metal matrix composite, and aluminum foam. As a benchmark, we choose aluminum foam.

The content of this paper is structured as follows:  In Section \ref{sect:homo}, the underlying method is explained in order to clearly present the addition proposed herein. In Section \ref{sect:Iden}, the details of a complete numerical implementation are demonstrated. In Section \ref{Example}, effective parameters for 2D and 3D composite materials including epoxy-carbon fiber composites, metal matrix composite, and aluminum foam are identified as well as the aforementioned two challenges have been exploited for checking the robustness of the implementation. In Section 5 we discuss the positive definiteness and in Section 6 we verify the parameters by using a strain gradient simulation.  The homogenization computational tool is developed based on open-source codes in FEniCS. It allows for all kinds of 2D or 3D composite materials constructed by periodic microstructures. The codes are made publicly available in \cite{compreal} in order to enable a transparent scientific exchange.

\section{Homogenization method} \label{sect:homo}

We start from an assertion that the deformation energy for the domain representing RVE, $\Omega_P$, at the microscale is equal to the energy for the RVE at the macroscale, namely
\begeq  \label{equivalence of energy1}
\int_{\Omega_P} { w^\text{m}} \d V = \int_{\Omega_P} { w^\text{M}} \d V \ .
\eqend 
The superscripts ``$\text{m}$'' and ``$\text{M}$'' are used to denote microscopic and macroscopic quantities, respectively. At the microscale, detailed microstructures are present in the RVE. At the macroscale, the same domain is modeled by a homogeneous ``metamaterial.''  We emphasize that an RVE may be different from a unit cell. A unit cell is the simplest repeating unit of heterogeneity. Spatial repetition of unit cells composes an RVE. At the microscale, the first order theory is used, as a consequence, we need to have a second order theory at the macroscale \cite{075}. Now by starting with a linear strain measure, in the case of a linear material model, we obtain a quadratic deformation energy,
\begeq  \label{equivalence of energy2}
\int_{\Omega_P} \frac{1}{2} C^\text{m}_{ijkl}  u^\text{m}_{i,j}  u^\text{m}_{k,l} \d V = \int_{\Omega_P}  \big(  \frac{1}{2}C^\text{M}_{ijkl}  u^\text{M}_{i,j}  u_{k,l}^\text{M} +  G_{ijklm}^\text{M}u^\text{M}_{i,j} u^\text{M}_{k,lm} + \frac{1}{2}D_{ijklmn}^\text{M}  u_{i,jk}^\text{M}  u_{l,mn}^\text{M} \big) \d V \ .
\eqend 
Displacement fields at micro- and macroscales are indicated by "m" and "M," respectively. $ C^\text{m}_{ijkl}  $ is given in each material point of the RVE. We begin with the known microscale and search for its corresponding homogenized effective parameters. The effective coefficients, $C^\text{M}_{ijkl}$, $G_{ijklm}^\text{M}$, and $D^\text{M}_{ijklmn}$ are the unknowns that we are searching for. We emphasize that the quadratic energy and symmetric strain measure lead to the minor symmetries $ C^\text{m}_{ijkl} = C^\text{m}_{jikl} = C^\text{m}_{ijlk}, C^\text{M}_{ijkl} = C^\text{M}_{jikl} = C^\text{M}_{ijlk} $, $G_{ijklm}^\text{M} = G_{jiklm}^\text{M} = G_{ijlkm}^\text{M} $, $D^\text{M}_{ijklmn} = D^\text{M}_{jiklmn} = D^\text{M}_{ijkmln} $ and major symmetries $ C^\text{m}_{ijkl} = C^\text{m}_{klij}$, $ C^\text{M}_{ijkl} = C^\text{M}_{klij}$, $D^\text{M}_{ijklmn} = D^\text{M}_{lmnijk} $ of the classical and strain gradient stiffness tensors. In what follows, the connections between the microscopic material parameters and macroscopic ones are established.

Let us investigate the macroscopic case for an RVE, $\Omega_P$. Firstly, the geometric center of the RVE is defined as $\cen{\ten X} = \frac1V \int_{\Omega_P} \t{X} \d V $.  A \textsc{Taylor} expansion of the macroscopic displacement around the center of the RVE is written as
\begal  \label{Taylor's series of displacements}
{u^\text{M} _i}(\t{X}) &= {u^\text{M} _i}\Big|_{\cen{\t X}} 
+  u^\text{M}_{i,j} \Big|_{\cen{\t X}} (X_j-\cen X_j) 
+ \frac{1}{2}  u^\text{M}_{i,jk} \Big|_{\cen{\t X}} (X_j-\cen X_j) (X_k-\cen X_k)  
\ , \\
u^\text{M}_{i,l}(\t{X}) &= u^\text{M}_{i,j} \Big|_{\cen{\t X}} \delta_{jl}
+ \frac{1}{2}  u^\text{M}_{i,jk} \Big|_{\cen{\t X}} ( \delta_{jl} (X_k-\cen X_k) + (X_j-\cen X_j) \delta_{kl} ) 
\ , \\
&= u^\text{M}_{i,l} \Big|_{\cen{\t X}} +  u^\text{M}_{i,lk} \Big|_{\cen{\t X}} (X_k-\cen X_k)
\ , \\
u^\text{M}_{i,lm}(\t{X}) &= u^\text{M}_{i,lk} \Big|_{\cen{\t X}} \delta_{km} = u^\text{M}_{i,lm} \Big|_{\cen{\t X}} \ .
\alend
Then by using the spatial averaging, we have
\begal \label{displacement average}
\langle {u}_{i,j}^\text{M} \rangle &=\frac{1}{V}\int_{\Omega^P} u^\text{M} _{i,j} \d V= u^\text{M}_{i,j}\Big|_{\cen{\t X}} 
\ , \\
\langle {u}_{i,jk}^\text{M} \rangle &=\frac{1}{V}\int_{\Omega^P} u^\text{M} _{i,jk}\d V= u^\text{M}_{i,jk} \Big|_{\cen{\t X}} \ .
\alend 
Therefore, the macroscopic energy of an RVE reads as follows (the detailed derivation can be found in \cite{yang2019determination}), as the macroscopic stiffness tensors are constant in space, 
\begeq \label{strain energy for homogenized material} 
\int_{\Omega_P } \big(  \frac{1}{2}C^\text{M}_{ijkl}  u^\text{M}_{i,j}  u_{k,l}^\text{M} +  G_{ijklm}^\text{M}u^\text{M}_{i,j} u^\text{M}_{k,lm} + \frac{1}{2}D_{ijklmn}^\text{M}  u_{i,jk}^\text{M}  u_{l,mn}^\text{M} \big) \d V
\ , \\ =  \frac{V}{2}   C^\text{M}_{ijlm} \langle{u}_{i,j}^\text{M}\rangle \langle{u}_{l,m}^\text{M}\rangle + V G_{ijklm}^\text{M}\langle u^\text{M}_{i,j} \rangle  \langle u^\text{M}_{k,lm} \rangle +\frac{V}{2} ( C^\text{M}_{ijlm} \bar{I}_{kn} +   D^\text{M}_{ijklmn} ) \langle{u}_{i,jk}^\text{M}\rangle \langle{u}_{l,mn}^\text{M}\rangle  \ , \\
\eqend
\begeq
\bar{I}_{kn} = \frac{1}{V} \int_{\Omega_P} (X_k-\cen X_k)(X_n-\cen X_n) \d V \ .
\eqend 

\begin{figure}[H]
	\centering
	\includegraphics[width=0.85\textwidth]{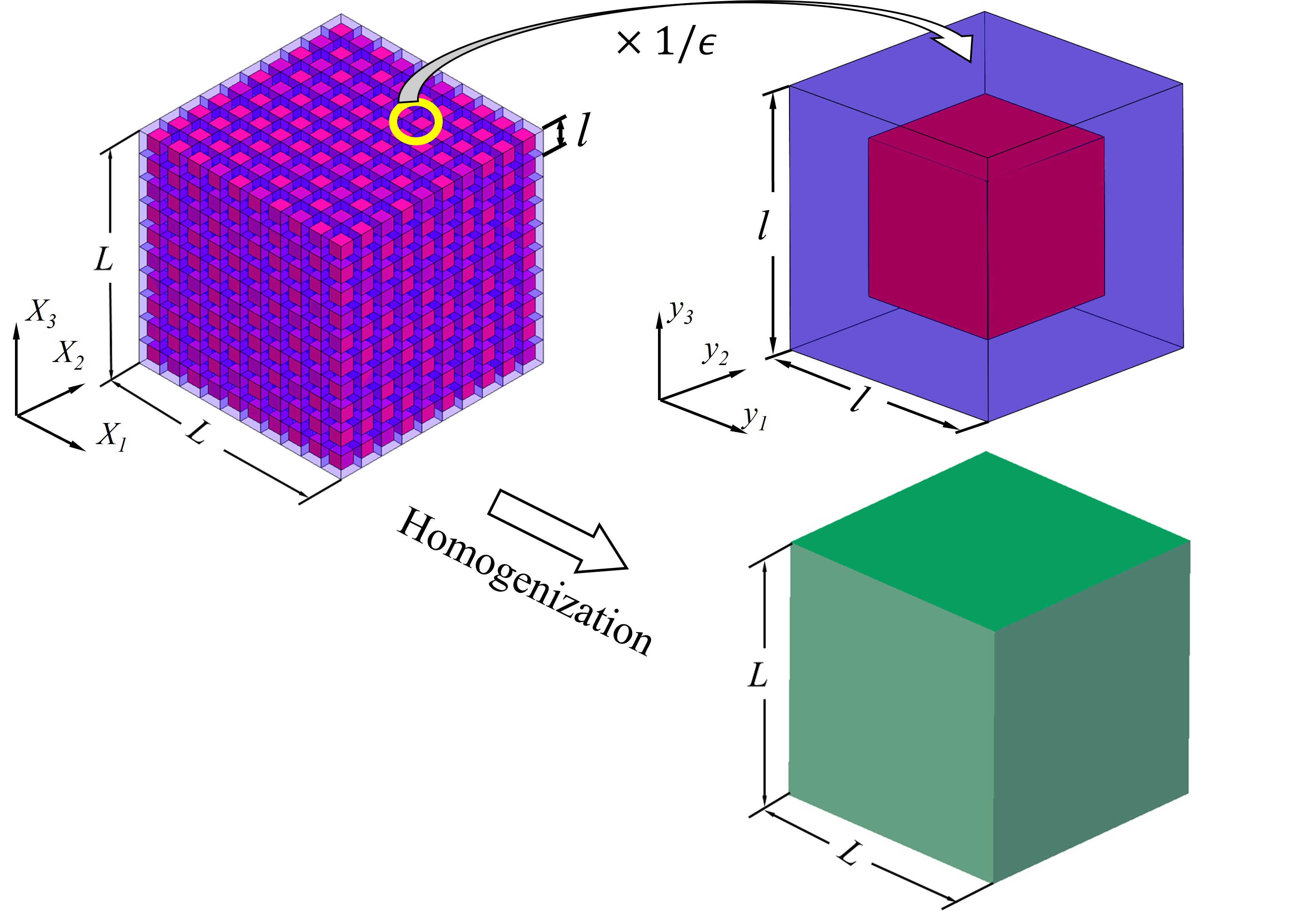}
	\caption{The heterogeneous continuum and its equivalent homogenized continuum.}
	\label{Fig1} 
\end{figure}

At the microscale, the asymptotic homogenization method is used to approximate the deformation energy for the RVE. We introduce a small parameter $\epsilon$, which is defined as $\epsilon = \frac{l}{L}$, where $l$ is the characteristic length of the microstructure, $L$ is the length of the macroscopic structure as shown in Fig. \ref{Fig1}. We remark that $\epsilon$ is the so-called homothetic ratio, which shows the scaling law for strain gradient moduli. This property will be illustrated later. A local coordinate is then introduced as
\begin{equation} \label{link_coord}
	y_j= \frac1\epsilon ( X_j - \cen X_j ) \ ,
\end{equation}
which is used to describe the local fluctuations caused by microscopic heterogeneity.  Variable  $\ten X$ is associated with the macroscopic scale. The displacement field for the RVE at the microscale is thus approximated with regard to $\epsilon$ as
\begin{equation} \label{displacement function}
	\t{u ^\text{m}}(\t X) = \overset{0} {\t u}( \t X, \t y) + \epsilon \overset{1}{\t u}(\t X,\t y) + \epsilon^2 \overset{2} {\t u}(\t X, \t y) + \dots \ .
\end{equation}
For a linear elastostatics problem, we propose to write the governing equations within the RVE, as follows:
\begeq \label{local govening equation}
\big( C_{ijkl}^\text{m}  u^\text{m}_{k,l} \big)_{,j} + \rho^{\text{m}} f_{i} =  0  \ ,
\eqend
where $ \rho^{\text{m}}  f_i$ are volume forces, $\rho^{\text{m}}$ is the mass density at the microscale, hence it is a function in $\t X$ effected by the heterogeneous structure. We stress that this interpretation of using mass density at the microscale has an important remedy to the generality of the computational implementation. This change in the formulation is for the first time in the literature and numerically beneficial in the case of voids. Since voids possess numerically zero mass density, their contribution to the homogenization is weakened by exploiting this amendment in the formulation. By substituting Eqn.~\eqref{displacement function} to Eqn.~\eqref{local govening equation} and gathering terms having the same order in $\epsilon$ leads to the following equations:
\begin{itemize}
	\item
	in the order of $\epsilon^{-2}$
	\begin{equation} \label{first terms}
		\frac{\partial }{\partial  y_j} \Big( C_{ijkl}^\text{m}   \frac{\partial \overset{0}u_k}{\partial y_l} \Big) = 0 \ ;
	\end{equation}
	\item
	in the order of $\epsilon^{-1}$
	\begin{equation} \label{second terms}
		\Big( C_{ijkl}^\text{m} \frac{\partial \overset{0}u_k}{\partial y_l} \Big)_{,j} 
		+  \frac{\partial }{\partial  y_j} \big( C_{ijkl}^\text{m}   \overset{0}u_{k,l} \big)  
		+ \frac{\partial }{\partial  y_j} \Big( C_{ijkl}^\text{m}   \frac{\partial \overset{1}u_k}{\partial y_l} \Big)   
		= 0 \ ;
	\end{equation}
	\item
	in the order of $\epsilon^{0}$
	\begeq \label{third terms}
	\big( C_{ijkl}^\text{m}   \overset{0}u_{k,l} \big)_{,j} 
	+  \Big( C_{ijkl}^\text{m}   \frac{\partial \overset{1}u_k}{\partial y_l} \Big)_{,j}
	+  \frac{\partial }{\partial  y_j} \big( C_{ijkl}^\text{m}  \overset{1}u_{k,l} \big)
	+ \frac{\partial }{\partial  y_j} \Big( C_{ijkl}^\text{m}  \pd{\overset{2}u_k}{y_l} \Big)   
	+ \rho^{\text{m}} f_i = 0 \ .
	\eqend
\end{itemize}

The only possible solution of Eqn.\,\eqref{first terms} is to restrict $\overset{0}u_i(\t X)$ as
\begin{equation} \label{first terms solution}
	\overset{0}u_i = \overset{0}u_i(\t X) \ .
\end{equation}
Because $\overset{0}u_i(\t X)$ is only dependent on the macroscopic coordinates, from Eqn.\,\eqref{displacement function},  by a coefficient comparison, we obtain that it may be chosen as the macroscopic displacement $\overset{0}{u}_i(\t X) =u^\text{M}_i(\t X)$. After substituting Eqn.\,\eqref{first terms solution} into Eqn.\,\eqref{second terms}, and introducing $\varphi_{abc}=\varphi_{abc}(\t y)$  which is $\ten y$-periodic with zero average value $\int_{\Omega^P} \varphi_{abi} \d V = 0$, we obtain
\begeq \label{diff.eq.phi}
\frac{\partial }{\partial  y_j} \bigg( C_{ijkl}^\text{m} \Big(  \frac{\p \varphi_{abk}}{\p y_l} + \delta_{ak} \delta_{bl} \Big) \bigg)
= 0 \ .
\eqend
Consequently, the solution of Eqn.\,\eqref{second terms} is given as
\begin{equation} \label{second terms solution}
	\overset{1}{u}_i = \varphi_{abi} {u}_{a,b}^\text{M}(\ten X)  + \overset{1}{\bar u}_i(\t X) \ ,
\end{equation}
where $\overset{1}{\bar u}_i =\overset{1}{\bar u}_i(\t X)$ are integration constants. 

By recalling the governing equation at the macroscale with an analogous suggestion to use a macroscale mass density, we have
\begeq \label{governing equation}
C^\text{M}_{ijkl}  u^\text{M}_{k,lj}
- D_{ijklmn}^\text{M}    u_{l,mnkj}^\text{M}   + \rho^{\text{M}} f_i= 0 \ ,
\eqend
with $   \rho^{\text{M}} = \frac1V \int_{\Omega_P} \rho^{\text{m}} \d V  $ and the usual axiom that body forces are scale independent such that $ f_i$ remains the same at the micro- and macroscales.  By neglecting the fourth order term in Eqn.\,\eqref{governing equation}, we obtain
\begeq \label{body force}
f_i =  -\frac{C^\text{M}_{ijkl}  u^\text{M}_{k,lj}}{\rho^{\text{M}}} \ .
\eqend
By plugging Eqn.\,\eqref{first terms solution}, Eqn.\,\eqref{second terms solution} (with $\overset{1}{\bar u}_i(\t X) = 0$), and Eqn.\,\eqref{body force} into Eqn.\,\eqref{third terms} and introducing $\psi_{abci}$ which is $\ten y$-periodic with zero average $\int_{\Omega^P} \psi_{abci} \d V = 0$, the solution of $ \overset{2}{u}_i $ may be given as:
\begeq \label{third terms solution}
\overset{2}{u}_i = \psi_{abci} {u}_{a,bc}^\text{M}(\ten X)  + \overset{2}{\bar u}_i(\t X) \ ,
\eqend
where $\overset{2}{\bar u}_i(\t X)$ are integration constants in $\t y$. The fourth order tensor $\psi_{abcd}$ must satisfy
\begeq  \label{diff.eq.psi}
\frac{\partial }{\partial  y_j} \bigg( C_{ijkl}^\text{m} \Big(  \frac{\p \psi_{abck}}{\p y_l} +  \varphi_{abk} \delta_{lc} \big) \bigg)
+  C_{ickl}^\text{m} \Big(  \frac{\p \varphi_{abk}}{\p y_l} + \delta_{ka} \delta_{lb}  \Big) - \frac {\rho^{\text{m}}}{\rho^{\text{M}}}{{C}}_{icab}^\text{M} = 0  \ .
\eqend
We emphasize that the last term is a source term simply applying the loading to the system at the microscale by considering mass densities. By neglecting the mass density ratio, one applies a source term even in the case of voids that may lead to numerically inconsistent results for $\t \psi$ parameters. The microscale displacement field is rewritten as
\begin{equation} \label{displacement function macro}
	u^\text{m}_i(\ten X,\ten y) 
	= {u}_i^\text{M} (\ten X)
	+ \epsilon \varphi_{abi}(\ten y) {u}_{a,b}^\text{M}(\ten X) 
	+ \epsilon^2 \psi_{abci}(\ten y) {u}_{a,bc}^\text{M}(\ten X) 
	+ \dots \ .
\end{equation}
By using Eqn.\,\eqref{displacement function macro} and the latter on the left-hand side of Eqn.\,\eqref{equivalence of energy2} the microscopic energy becomes 
\begeq \label{strain energy density2}
\int_{\Omega_P} \frac12 C^\text{m}_{ijkl} u^\text{m}_{i,j} u^\text{m}_{k,l} \d V
= \\
\frac{V}{2}\Big( \bar{C}_{abcd} \langle{u}_{a,b}^\text{M}\rangle \langle{u}_{c,d}^\text{M}\rangle 
+
\bar{G}_{abcde}\langle{u}_{a,b}^\text{M}\rangle \langle{u}_{c,de}^\text{M}\rangle
+ \bar{D}_{abcdef}\langle{u}_{a,bc}^\text{M}\rangle \langle{u}_{d,ef}^\text{M}\rangle 
\Big)  \ ,
\eqend
with
\begeq \label{relation_to_epsilon}
\bar{C}_{abcd} = \frac{1}{V} \int_{\Omega_P}  C_{ijkl}^\text{m} L_{abij} L_{cdkl} \d V \ , \\
\bar{G}_{abcde} = \frac{2\epsilon}{V} \int_{\Omega_P}  C_{ijkl}^\text{m} L_{abij} M_{cdekl} \d V \\
\bar{D}_{abcdef} = \frac{\epsilon^2}{V} \int_{\Omega_P} C_{ijkl}^\text{m} M_{abcij} M_{defkl} \d V \ , 
\eqend 
The appearance of $\epsilon^2$ is due to the fact that Eqn.\,\eqref{homogenized_SG} is expressed in the local coordinate $\t y$ (The fifth order tensor $\t M$ is only related to $\t y$).
\begal \label{homogenized_L_M}
L_{abij} &= \delta_{ia} \delta_{jb} + \frac{\p \varphi_{abi}}{\p y_j} 
\ , \\
M_{abcij} &= y_{c} \Big( \delta_{ia} \delta_{jb} + \frac{\p \varphi_{abi}}{\p y_j} \Big) 
+ \Big( \varphi_{abi} \delta_{jc} + \frac{\p \psi_{abci} }{\p y_j} \Big) \ .
\alend
Based on Eqn.~\eqref{equivalence of energy2} the effective parameters are calculated by
\begeq \label{homogenized_tensors}
C^\text{M}_{abcd} = \frac{1}{V} \int_{\Omega_P}  C_{ijkl}^\text{m} L_{abij} L_{cdkl} \d V \ , \\
\eqend 
\begeq \label{homogenized_G}
G^\text{M}_{abcde} = \frac{\epsilon}{V} \int_{\Omega_P}  C_{ijkl}^\text{m} L_{abij} M_{cdekl} \d V \ , \\
\eqend
\begeq \label{homogenized_SG}
D^\text{M}_{abcdef} =  \frac{\epsilon^2}{V} \Bigg(  \int_{\Omega_P} C_{ijkl}^\text{m} M_{abcij} M_{defkl} \d V  - 	C^\text{M}_{abde} \int_{\Omega_P}y_c y_f \d V  \Bigg)\ .
\eqend 

It should be remarked that the Eqn.\,\eqref{homogenized_tensors} coincides with the well known asymptotic homogenization method. The classical stiffness tensor is scale independent. However, as observed from the Eqn.\,\eqref{homogenized_SG}, strain gradient stiffness parameters depend on $\epsilon^2$. Indeed, these parameters emerge related to the substructure and vanish as $\epsilon = 0 $ meaning that the substructure diminishes. We stress that this distinction is of importance and comes out of the proposed methodology quite naturally. 
As obvious in Eqn.\,\eqref{link_coord}, the homothetic ratio, $\epsilon$, acts as a multiplier between the macroscopic length scale (in global coordinates, $\t X$)  and microscopic length scale (in local coordinates, $\t y$). In this way, we acquire different $\t G^{\text M}$ and $\t D^{\text M}$ coefficients for the same RVE in larger structures without repeating the calculations. The role of $\epsilon$ will be further illustrated by using numerical examples.

\section{Numerical implementation} \label{sect:Iden}

\begin{figure}[H]
	\centering
	\includegraphics[width=0.85\textwidth]{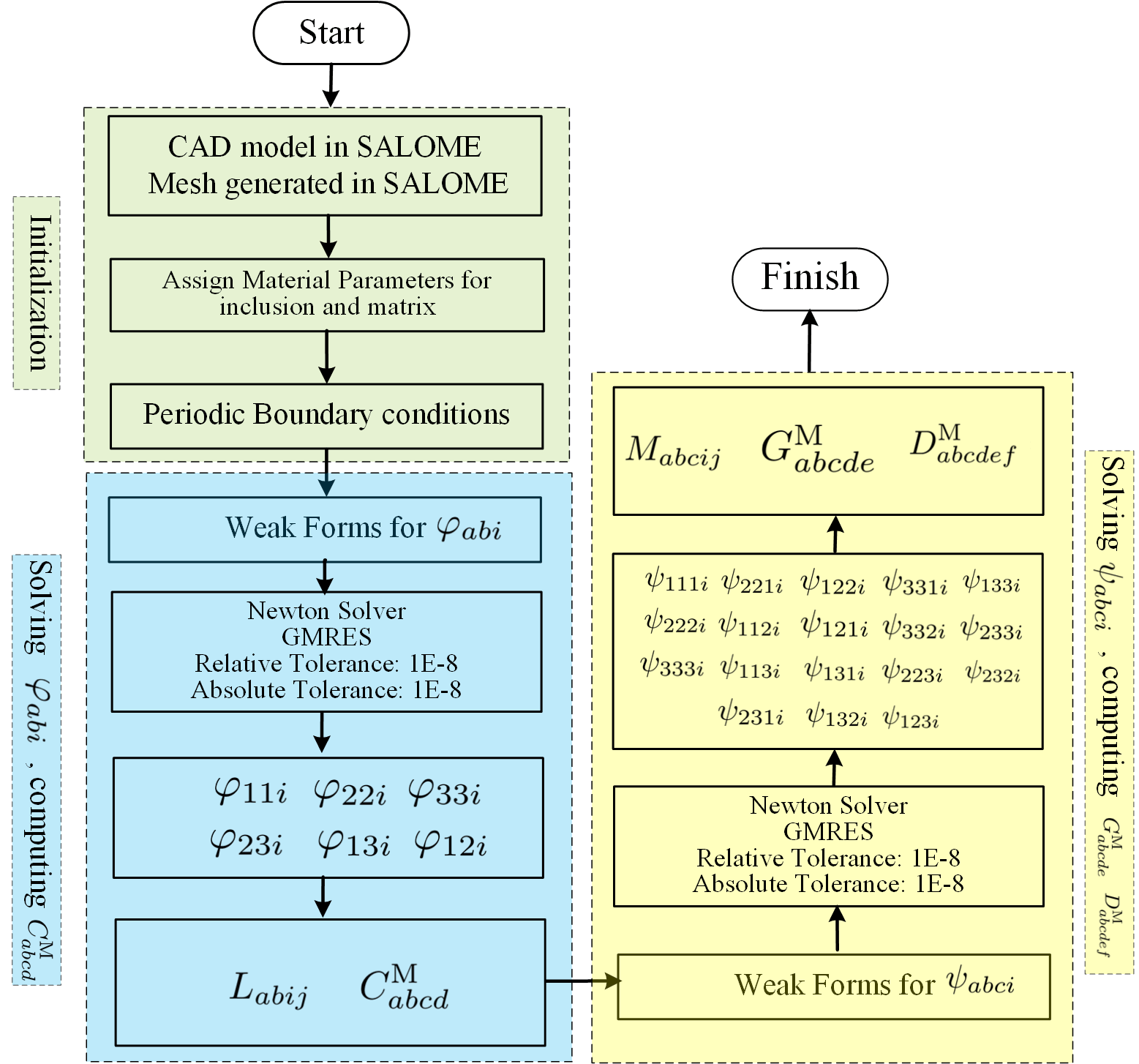}
	\caption{The flowchart of the numerical implementation.}
	\label{flowchart} 
\end{figure}

\begin{figure}[H]
	\centering
	\includegraphics[width=0.7\textwidth]{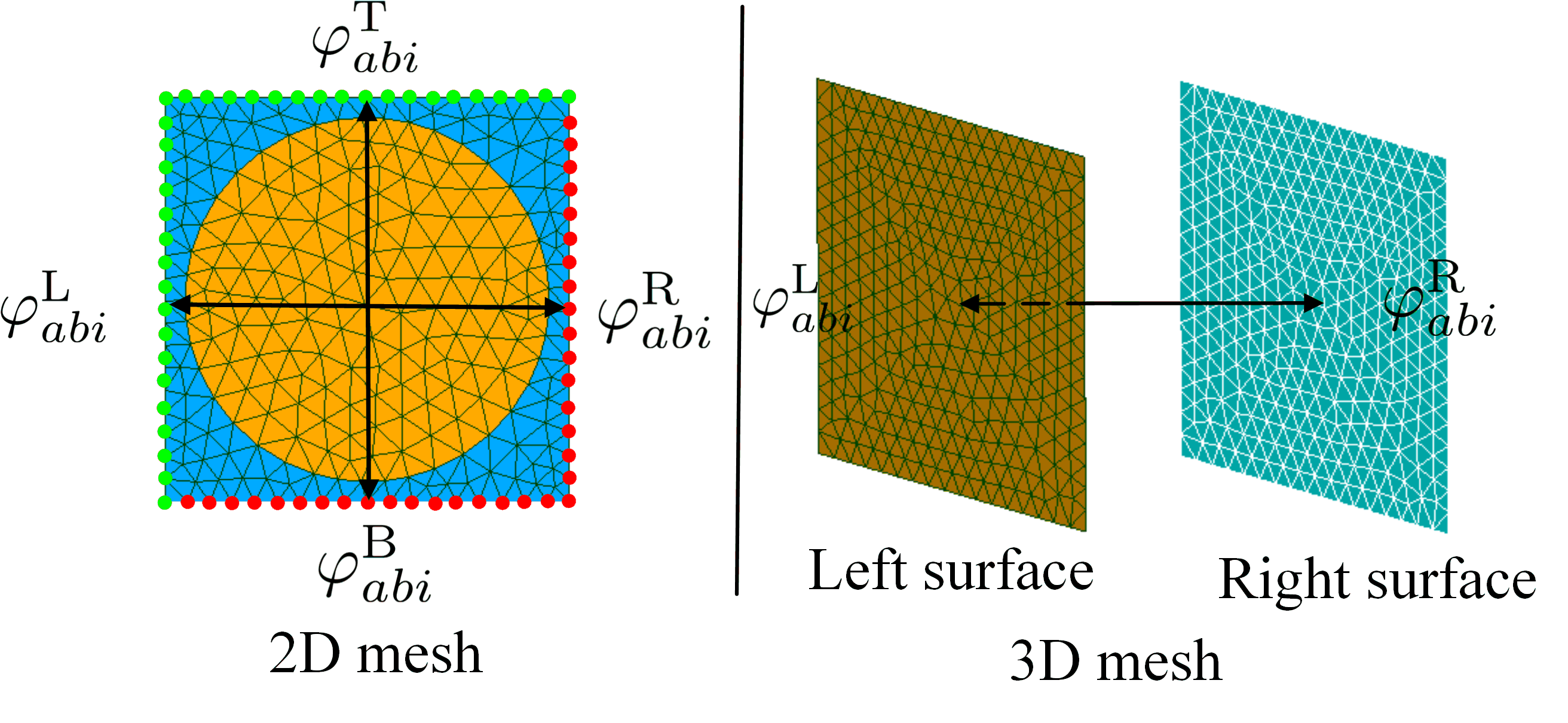}
	\caption{Periodic boundary conditions applied in FEM. Left: Right edge and green edge have the same mesh. Right: Only corresponding surfaces are shown, and so-called left surface and right surface have the same mesh. Same mesh is necessary for implementing periodic boundary conditions.}
	\label{PBC} 
\end{figure}

In order to identify effective parameters, Eqn.\,\eqref{homogenized_tensors} and Eqn.\,\eqref{homogenized_SG} need to be resolved, which requires $\ten\varphi$ and $\ten\psi$. The tensors $\ten\varphi$ and $\ten\psi$ are the solutions of Eqn.\,\eqref{diff.eq.phi} and Eqn.\,\eqref{diff.eq.psi}, which are solved numerically by the finite element method. As shown in Figure \ref{flowchart}, six cases $\varphi_{11i}, \varphi_{22i}, \varphi_{33i}, \varphi_{23i}, \varphi_{13i}, \varphi_{12i}$ in total in 3D need to be computed under periodic boundary conditions. After using integration by parts, considering the constraints of zero average for $\t \varphi$, the following weak form for $\varphi_{abk}$ is generated
\begeq
\int_{\Omega^P}  \bigg( C_{ijkl}^\text{m} \Big(  \frac{\p \varphi_{\underline{a}\underline{b}k}}{\p y_l} + \delta_{\underline{a}k} \delta_{\underline{b}l} \Big) \bigg) \frac{\partial \del \varphi_{\underline{a}\underline{b}i}}{\partial  y_j} \d V + \updelta \int \lambda_{abi} \varphi_{abi}  \d V = 0 \,
\eqend
and then immediately we have
\begeq
\int_{\Omega^P}  \bigg( C_{ijkl}^\text{m} \Big(  \frac{\p \varphi_{\underline{a}\underline{b}k}}{\p y_l} + \delta_{\underline{a}k} \delta_{\underline{b}l} \Big) \bigg) \frac{\partial \del \varphi_{\underline{a}\underline{b}i}}{\partial  y_j} \d V + \int_{\Omega^P} \lambda_{\underline{a}\underline{b}i} \del \varphi_{\underline{a}\underline{b}i} \d V + \int_{\Omega^P} \del\lambda_{\underline{a}\underline{b}i}  \varphi_{\underline{a}\underline{b}i} \d V= 0 \ ,
\eqend
where over underlined indices, no summation convention is applied. All fields with a variational delta, $\updelta$, denote a corresponding test function such that $\t \varphi$ and $\t \lambda$ are unknowns. For each case of $\varphi_{11i}, \varphi_{22i}, \varphi_{33i}, \varphi_{23i}, \varphi_{13i}, \varphi_{12i}$, a corresponding \textsc{Lagrange} multiplier $\lambda_{11i}, \lambda_{22i}, \lambda_{33i}, \lambda_{23i}, \lambda_{13i}, \lambda_{12i}$, is employed in order to enforce the zero average constrains of  $\t \varphi$ \cite{bleyer2018numericaltours}. Likewise, the weak form for calculating $\psi_{abci}$ reads
\begeq
\int_{\Omega^P} \Bigg( \bigg( C_{ijkl}^\text{m} \Big(  \frac{\p \psi_{\underline{abc}k}}{\p y_l} +  \varphi_{\underline{ab}k} \delta_{l\underline{c}} \Big) \bigg) \frac{\partial \del \psi_{\underline{abc}i}}{\partial  y_j} 
- \\
-  C_{i\underline{c}kl}^\text{m} \Big(  \frac{\p \varphi_{\underline{ab}k}}{\p y_l} + \delta_{k\underline{a}} \delta_{l\underline{b}}  \Big) \del \psi_{\underline{abc}i}
+ \frac {\rho^{\text{m}}}{\rho^{\text{M}}}{C}_{i\underline{cab}}^\text{M} \del \psi_{\underline{abc}i} \Bigg) \d V + 
\\ +
\int_{\Omega^P} \lambda_{\underline{abc}i} \del \psi_{\underline{abc}i} \d V  + \int_{\Omega^P} \del\lambda_{\underline{abc}i}  \psi_{\underline{abc}i} \d V= 0 \ .
\eqend
There are 18 weak forms in 3D to be solved for $\psi_{111i}$, $\psi_{112i}$, \dots $\psi_{123i}$..
The weak forms have been solved by the FEniCS platform. CAD model and mesh files are created by using an open-source software SALOME \cite{027}. Triangle  for surfaces elements and tetrahedron for volume elements are used to discretize the system by using the algorithms from NetGen Mesh Generator. An RVE needs to fulfill periodic boundary conditions such that the corresponding edges (in 2D) or surfaces (in 3D) are matching for nodes to be defined as the same degree of freedom in order to enforce the periodic boundary conditions as shown in Figure \ref{PBC}.

\section{Numerical examples} \label{Example}

The proposed homogenization method provides a unified analysis for general 2D and 3D composites. It may be used to homogenize fiber reinforced composites, particulate composites, and porous materials. In order to show the predictive capability of the proposed method, four examples are demonstrated in the following.

\subsection{2D epoxy-carbon fiber composite}
\begin{figure}[H]
	\centering
	\includegraphics[width=0.65\textwidth]{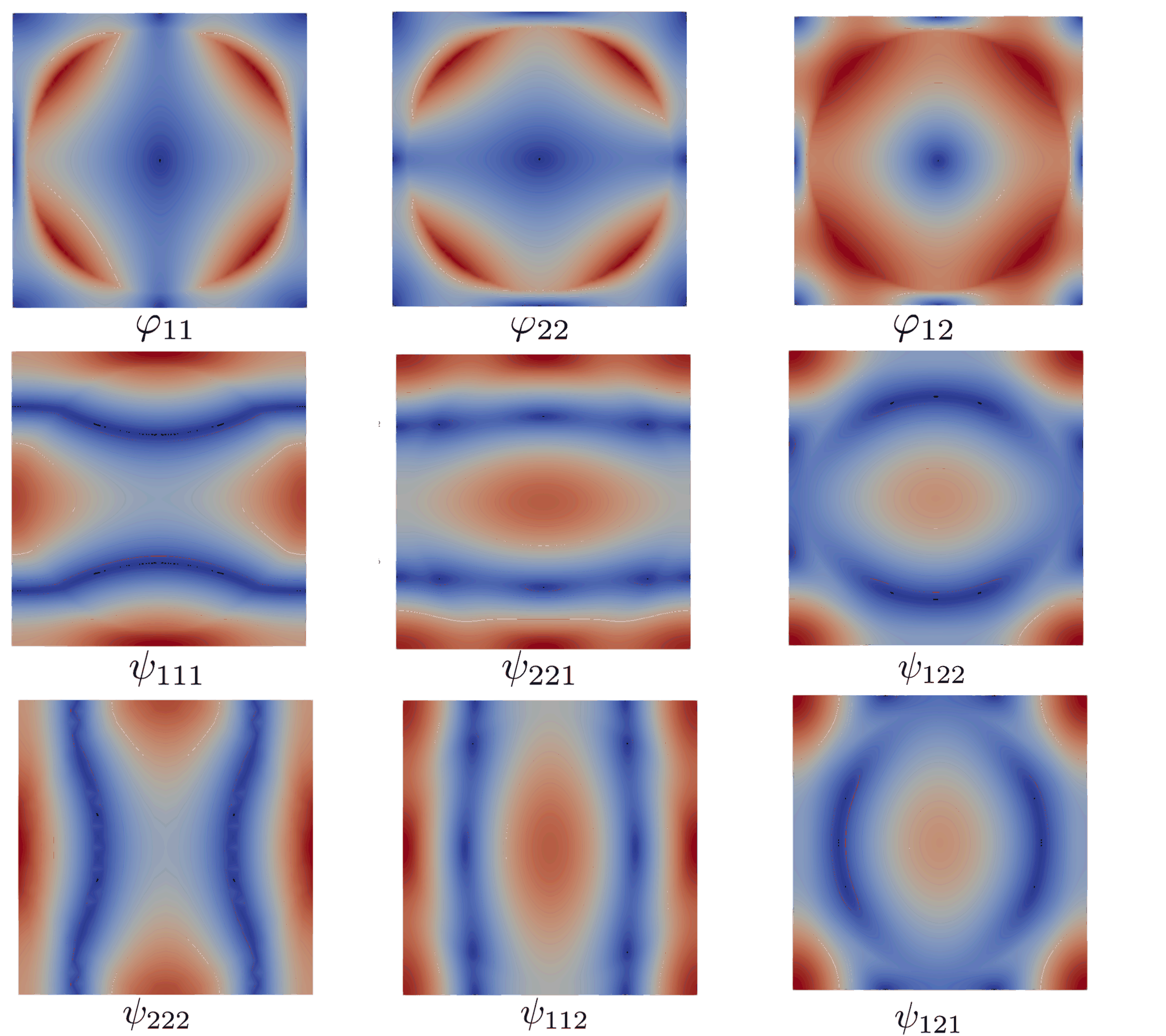}
	\caption{Solutions for $\t \varphi$ and $\t \psi$. Color distribution showing the cubic symmetry resulted local fluctuation in $\t \varphi$ and $\t \psi$ fields. Color bars are omitted since we analyze qualitatively. }
	\label{PhiPsi} 
\end{figure}

A 2 dimensional carbon fibers reinforced epoxy composite structure is investigated. The material properties  \footnote{Values of material properties are taken from matweb.com} for both constituents (matrix and inclusion) are shown in Table \ref{tab:2D_EC}. The size of the unit cell is 1 mm. The fiber is of circular shape, its radius is 0.45 mm, thus, the volume fraction of matrix is $36.4\%$.
\begin{table}[H]
	\centering
	\caption{Material properties used for 2D epoxy-carbon fiber composite. $E$ Young's modulus, $\nu$ Poisson's ratio, and $\rho$ mass density. }
	\begin{tabular}{lccc}
		\toprule
		Type & $E$ in GPa & $\nu$  & $\rho$ in kg/m$^3$\\
		\midrule
		Matrix (Epoxy)  & 17.3  & 0.35  &  1780 \\
		Inclusion (Carbon fiber) & 35.9 & 0.30 & 1650 \\
		\bottomrule
	\end{tabular}
	\label{tab:2D_EC}
\end{table}
Voigt notations as presented in Table \ref{tab:Voigt2D}, Table \ref{tab:2DSG1} are used to represent rank four, five, six tensors as matrices (analogous to \textsc{Voigt}'s notation).
\begin{table}[H]
	\centering
	\caption{\textsc{Voigt} notation used for 2D strain tensors.}
	\begin{tabular}{p{1cm}p{1cm}p{1cm}p{1cm}p{1cm}}
		\toprule
		& $A$ & 1 & 2 & 3  \\
		\midrule
		& $ij$  & 11 & 22& 12    \\
		\bottomrule
	\end{tabular}
	\label{tab:Voigt2D}
\end{table}
\begin{table}[H]
	\centering
	\caption{\textsc{Voigt} kind-notation used for 2D strain-gradient tensors.}
	\begin{tabular}{p{1cm}p{1cm}p{1cm}p{1cm}p{1cm}p{1cm}p{1cm}p{1cm}p{1cm}}
		\toprule
		& $\theta$ & 1 & 2 & 3 & 4 & 5 & 6 \\
		\midrule
		& $ijk$  & 111 & 112& 221 & 222 & 121 & 122  \\
		\bottomrule
	\end{tabular}
	\label{tab:2DSG1}
\end{table}
A bottleneck in the homogenization may be the missing convergence criteria. We propose a simple yet effective approach by using the material symmetry class of the analyzed microstructure. Owing to the cubic material symmetry, we know that $C_{1111} = C_{2222}$, $D_{111111} = D_{222222}$. The convergence analysis is conducted as shown in Table \ref{tab:convergence}, by comparing the ratios $C_{1111}/C_{2222}$, and $D_{111111} / D_{222222}$. When they tend to be 1, the computation is converged.
\begin{table}[H]
	\centering
	\caption{Convergence analysis. With the increasing of degrees of freedom, the ratios $C_{1111}/C_{2222}$, and $D_{111111} / D_{222222}$ reach 1.}
	\resizebox{1.15\textwidth}{!}{$\displaystyle
		\begin{tabular}{p{1cm}p{2cm}p{2cm}p{2cm}p{2cm}p{2cm}p{2cm}p{0.5cm}}
			\toprule
			DOFs & $C_{1111}$ GPa & $C_{2222}$ GPa & $C_{1111}/C_{2222}$ & $D_{111111}$ N & $D_{222222}$ N & $D_{111111}/ D_{222222}$ &\\
			\midrule
			1342 & 38.6 & 38.7 & 99.7 \% & 510.4 & 496.0 & 103.0 \% &\\
			22362 & 38.9 & 38.9 & 100.0 \% & 505.3 & 506.1 & 100.0 \% &\\
			90226  & 39.0 & 39.0& 100.0 \% & 506.4 & 505.8 & 100.0 \%  &\\
			\bottomrule
		\end{tabular} $ }
	\label{tab:convergence}
\end{table}
The solutions for $\t \varphi$ and $\t \psi$ are presented in Figure \ref{PhiPsi}. It is observed as expected that these fluctuations are all periodic. Furthermore, due to the fact that the material is cubic, rotating $\varphi_{22}$, $\psi_{111}$, $\psi_{221}$, $\psi_{122}$ by 90 $\degree$ gives the same shapes as $\varphi_{11}$, $\psi_{222}$, $\psi_{112}$, $\psi_{121}$. 
The identified effective classical and stain gradient stiffness tensors are shown as follows:
\begeq
C\ma_{AB} = \begin{pmatrix}
	39.0 & 18.0 & 0.0 \\ 
	18.0 & 39.0 & 0.0 \\ 
	0.0 & 0.0 & 10.0 \\ 
\end{pmatrix} \text{\,GPa}\ , \notag
\eqend
\begeq
D\ma_{\theta\gamma} =
\begin{pmatrix}
	506.4 & 181.9 & 0.0 & 
	-0.0 & 0.0 & -182.2 & 
	\\
	181.9 & -299.4 & 0.0 & 
	0.0 & 0.0 & -176.2 & 
	\\
	0.0 & 0.0 & 181.2 & 
	-175.4 & -183.0 & 0.0 & 
	\\
	0.0 & 0.0 & -175.4 & 
	-298.5 & 181.2 & 0.0 & 
	\\
	0.0 & 0.0 & -183.0 & 
	181.2 & 505.8 & 0.0 & 
	\\
	-182.2 & -176.2 & 0.0 & 
	0.0 & 0.0 & 181.0 &  
\end{pmatrix} \text{\,N}\ . \notag
\eqend
It is found that there are three independent parameters in the stiffness tensor and six independent parameters in the strain gradient stiffness tensor. This observation is consistent with \cite{auffray2015complete, auffray2009derivation} for cubic materials. Albeit we circument of showing, the implementation successfully computes all parameters of $\t G$ as (numerical) zeros, as expected from the cubic material symmetry as well. By using the \textsc{Voigt} notation similar to the approach as in \cite{auffray2015complete, auffray2009derivation, auffray2013matrix} in Table \ref{tab:2DSG2},  the strain gradient stiffness matrix is made to be block-diagonal; each diagonal block matrix includes only non-zero parameters, and each diagonal block matrix is invariant under every cyclic permutation of $\t X$ axis, $\t Y$ axis, and $\t Z$ axis \cite{auffray2013matrix}. Therefore, the \textsc{Voigt} notaion proposed in \cite{auffray2013matrix} will be used throughout the paper.
\begin{table}[H]
	\centering
	\caption{\textsc{Voigt} notation used for 2D strain-gradient tensors proposed in \cite{auffray2013matrix}.}
	\begin{tabular}{p{1cm}p{1cm}p{1cm}p{1cm}p{1cm}p{1cm}p{1cm}p{1cm}p{1cm}}
		\toprule
		& $\alpha$ & 1 & 2 & 3 & 4 & 5 & 6 \\
		\midrule
		& $ijk$  & 111 & 221& 122 & 222 & 112 & 121  \\
		\bottomrule
	\end{tabular}
	\label{tab:2DSG2}
\end{table}
\begeq
D\ma_{\alpha\beta} =
\begin{pmatrix}
	506.4 & 181.9 & -182.2 & 
	0.0 & 0.0 & 0.0 & 
	\\
	181.9 & -299.4 & -176.2 & 
	0.0 & 0.0 & 0.0 & 
	\\
	-182.2 & -176.2 & 181.0 & 
	0.0 & 0.0 & 0.0 & 
	\\
	0.0 & 0.0 & 0.0 & 
	505.8 & 181.2 & -183.0 & 
	\\
	0.0 & 0.0 & 0.0 & 
	181.2 & -298.5 & -175.4 & 
	\\
	0.0 & 0.0 & 0.0 & 
	-183.0 & -175.4 & 181.2 & 
\end{pmatrix} \text{\,N}\ . \notag
\eqend
\subsection{Interpretation of the homothetic ratio}
\begin{figure}[H]
	\centering
	\subfigure[The homothetic ratio $\epsilon = \frac{1}{4}$. ]{\includegraphics[width=0.75\textwidth]{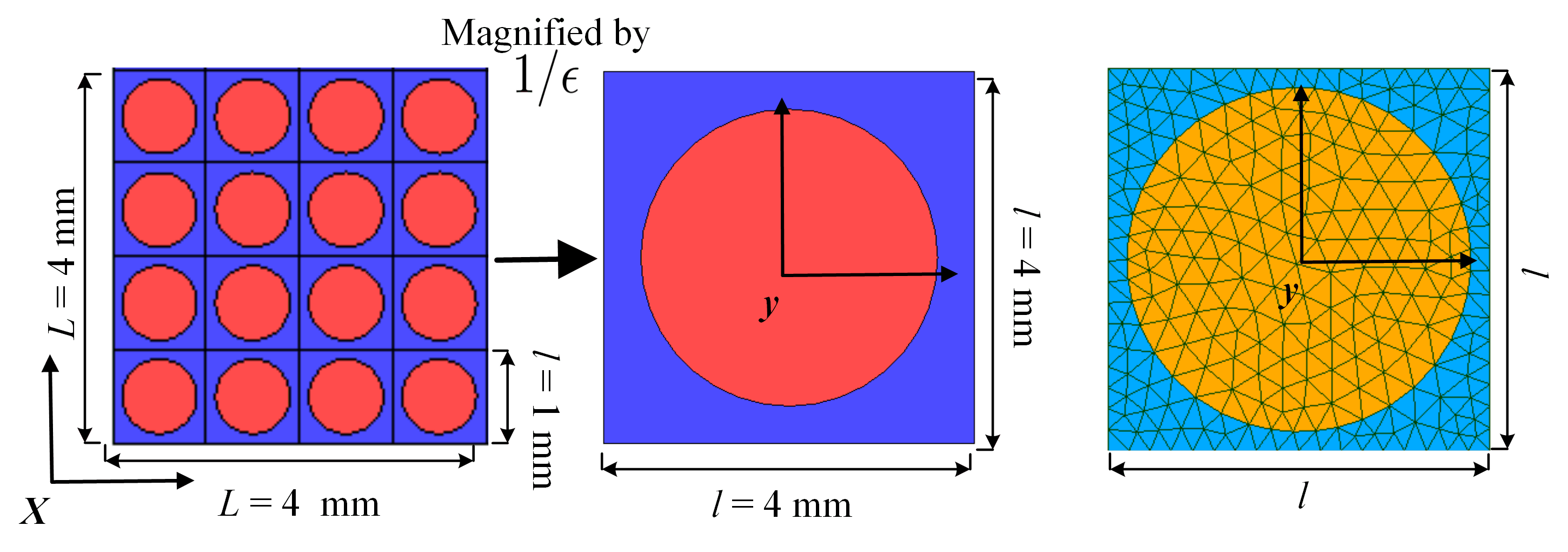}}
	\subfigure[The homothetic ratio $\epsilon = \frac{1}{8}$. ]{\includegraphics[width=0.75\textwidth]{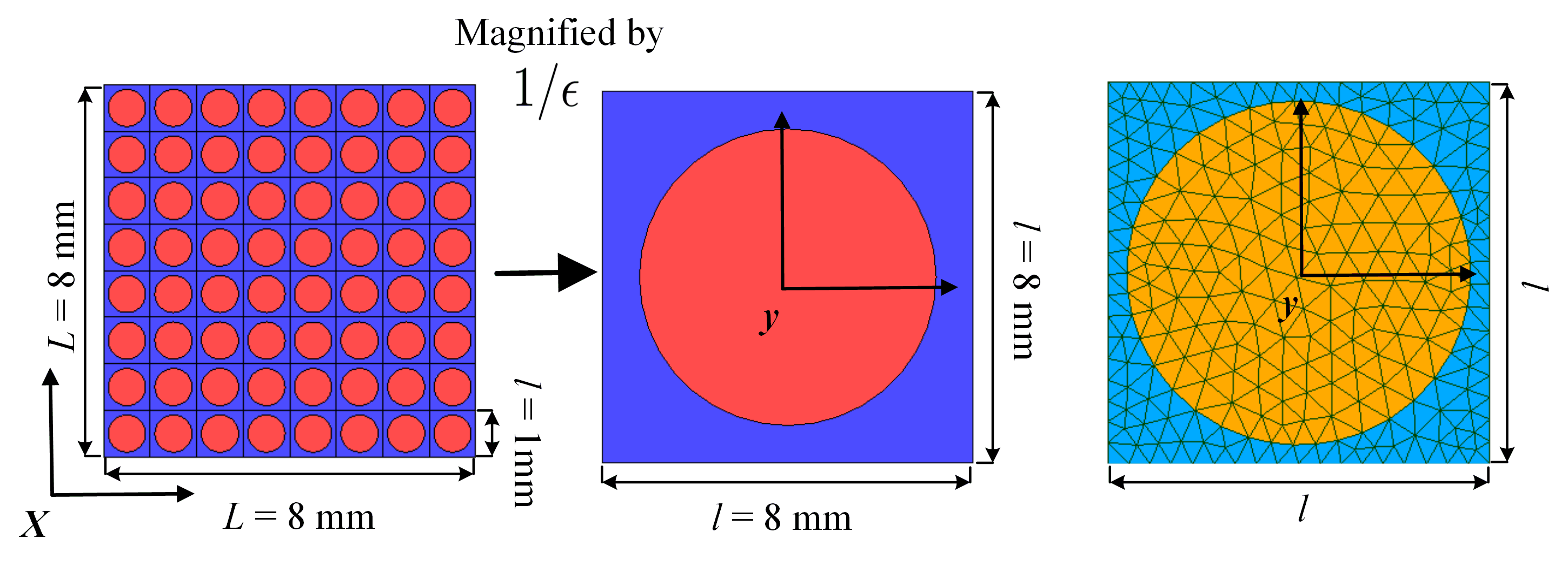}}
	\subfigure[The homothetic ratio $\epsilon = \frac{1}{8}$. ]{\includegraphics[width=0.75\textwidth]{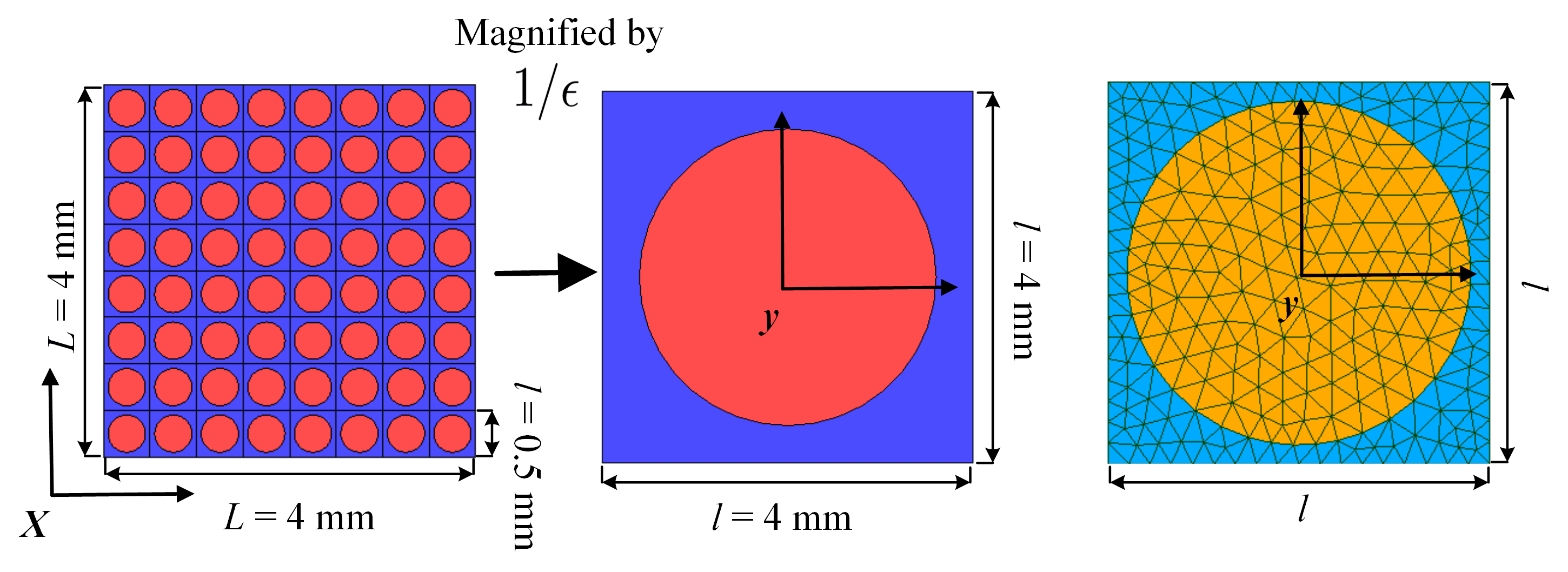}}
	\caption{ Visualization regarding the meaning of the homothetic ratio $\epsilon$.} 
	\label{epsilon}
\end{figure}

When determining the strain gradient moduli, physical relevance of the so-called homothetic ratio, $\epsilon$, is needed for determining the correct value. Two coordinate systems are scaled to each other by this homothetic ratio. Let us consider specific cases as shown in Figure \ref{epsilon}. In Figure \ref{epsilon} (a), the macroscopic length is $L = 4$ mm and the microscopic length is $l = 1$ mm with $\epsilon = \frac{l}{L} = \frac{1}{4}$.  RVE is of length $L$ = 1 mm in global coordinates $\t X$, but it is measured as $l$ = 4 mm in local coordinate $\t y$. Since Eqn.\,\eqref{homogenized_SG} are expressed in the local coordinate $\t y$, the parameters in $D^\text{M}_{abcdef} $ are calculated in the local coordinate. Thus, the length of the computational domain in Eqn.\,\eqref{homogenized_SG} is 4 mm. Likewise, in Figure \ref{epsilon} (b), the length of integration domain is 2 times larger than that in Figure \ref{epsilon} (a). However $\epsilon =\frac{1}{8}$  is half of the former one. This leads to the equal values for strain gradient moduli. Consequently, in the last section, $\epsilon$ can be chosen as, for example, $\frac{1}{4}$ or $\frac{1}{8}$, as long as the corresponding length of integration domain is chosen accordingly. 

Indeed, a scaling rule occurs for the strain gradient moduli. For example, in Figure \ref{epsilon} (c),  the length of RVE is half of that in Figure \ref{epsilon} (a). The same macroscopic length equals calculated integrals in Eqn.\,\eqref{homogenized_SG}. The differences of the obtained strain gradient parameters originate from the $\epsilon^2$ as presented in Eqn.\,\eqref{homogenized_SG}. The strain gradient parameters for Figure \ref{epsilon} (a) are 4 times larger than those for Figure \ref{epsilon} (c). This scaling factor is calculated as the ratio between $\epsilon^2$, also equal to the square of ratio of the unit cell lengths. Therefore, herein, we conclude that the strain gradient moduli are indeed not related to the macroscopic length but the microscopic length. This interpretation is indeed in coincidence with the well-known size effect in the literature. We emphasize that the substructure affects the values in $C_{ijkl}^{\text M}$, but not its ratio with respect to the macroscale. Therefore, for different substructures, $C_{ijkl}^{\text M}$ needs to be recalculated. For the same substructure but different homothetic ratios, they remain the same.
\subsection{3D cases}
In the followings, we consider 3D cases. The effective parameters in the classical stiffness tensor and strain gradient stiffness tensor for a carbon fibers reinforced epoxy composite, a (hard) spherical particles reinforced (soft) matrix, a metal matrix composite, and an aluminum foam will be investigated. The used \textsc{Voigt} notations for these 3 dimensional cases are displayed in Table \ref{tab:Voigt_strain} and Table \ref{tab:Voigt_strain_gradient}.
\begin{table}[H]
	\centering
	\caption{\textsc{Voigt} notation used for 3D strain tensors.}
	\begin{tabular}{p{1cm}p{1cm}p{1cm}p{1cm}p{1cm}p{1cm}p{1cm}p{1cm}}
		\toprule
		& $A$ & 1 & 2 & 3 & 4 & 5 & 6 \\
		\midrule
		& $ij$  & 11 & 22 & 33 & 23 & 13 & 12  \\
		\bottomrule
	\end{tabular}
	\label{tab:Voigt_strain}
\end{table}
\begin{table}[H]
	\centering
	\caption{\textsc{Voigt} notation used for 3D strain-gradient tensors.}
	\resizebox{1.15\textwidth}{!}{$\displaystyle
		\begin{tabular}{cccccccccccccccccccc}
			\toprule
			& $\alpha$ & 1 & 2 & 3 & 4 & 5 & 6 & 7 & 8 & 9 & 10 & 11 & 12 & 13 & 14 & 15 & 16 & 17 & 18\\
			\midrule
			& $ijk$  & 111 & 221& 122 & 331 & 133 & 222 & 112 & 121 & 332 & 233 & 333 & 113 & 131 & 223 & 232 & 231 & 132 & 123 \\
			\bottomrule
		\end{tabular} $}
	\label{tab:Voigt_strain_gradient}
\end{table}
\subsubsection{3D fiber reinforced composite}

\begin{figure}[H]
	\centering
	\includegraphics[width=1.1\textwidth]{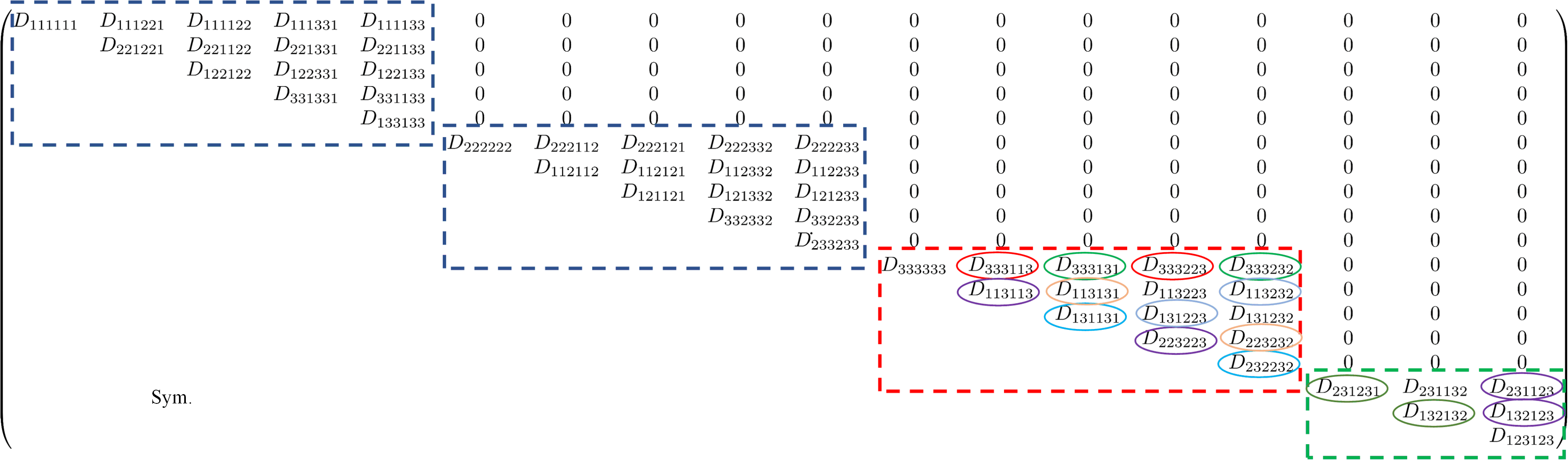}
	\caption{The structure of strain gradient stiffness tensor for transverse isotropic materials. It is found that the first two $5 \times 5$ matrices in the diagonal are equal, for example, $D_{111111} = D_{222222} $. In the third 5 $\times$ 5 matrix in the diagonal, it is also observed that $D_{333113} = D_{333223} $, $ D_{333131} = D_{333232}$, $D_{113113}=D_{223223}$, $D_{113131}=D_{223232} $, $ D_{113232} = D_{131223}$, $ D_{131131} = D_{232232 }$. In the 3 $ \times$ 3 matrix,  $D_{231231} = D_{132132} $, $ D_{231123} = D_{132123}$.}
	\label{TV_D} 
\end{figure}

\begin{figure}[H]
	\centering
	\includegraphics[width=0.75\textwidth]{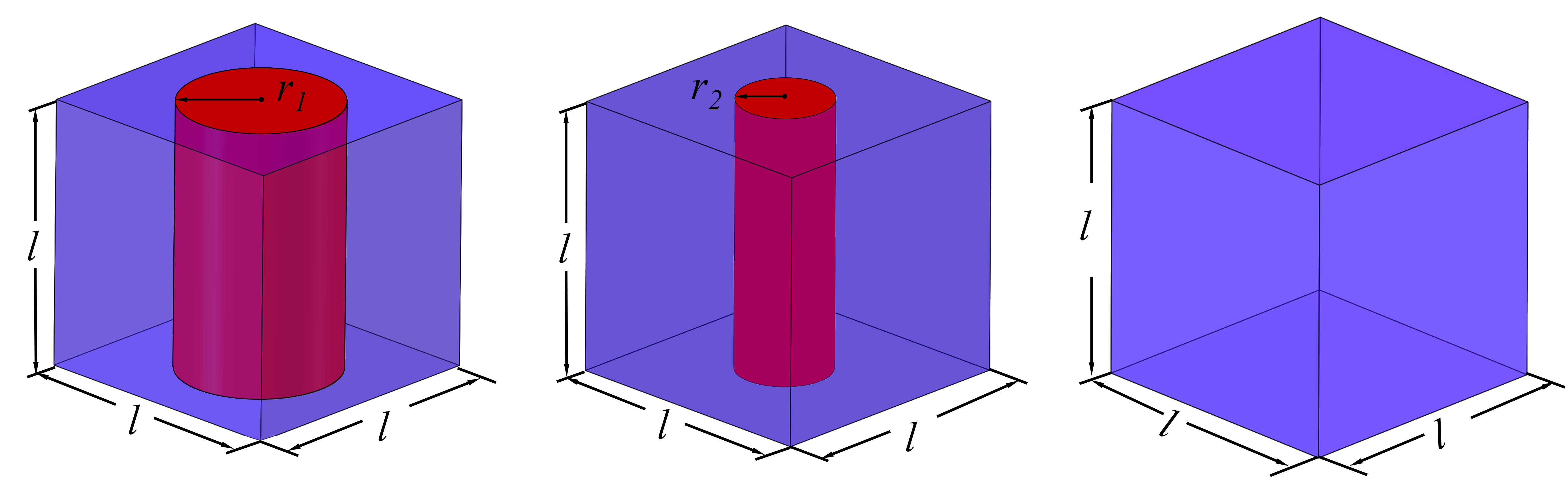}
	\caption{Different volume fraction of matrix. $l = 1$ mm, $r_1 = 0.45$ mm, $r_2 = 0.35$ mm. }
	\label{Vol_Cy} 
\end{figure}

\begin{figure}[H]
	\centering
	\subfigure[Effective classical stiffness parameters. ]{\includegraphics[width=0.47\textwidth]{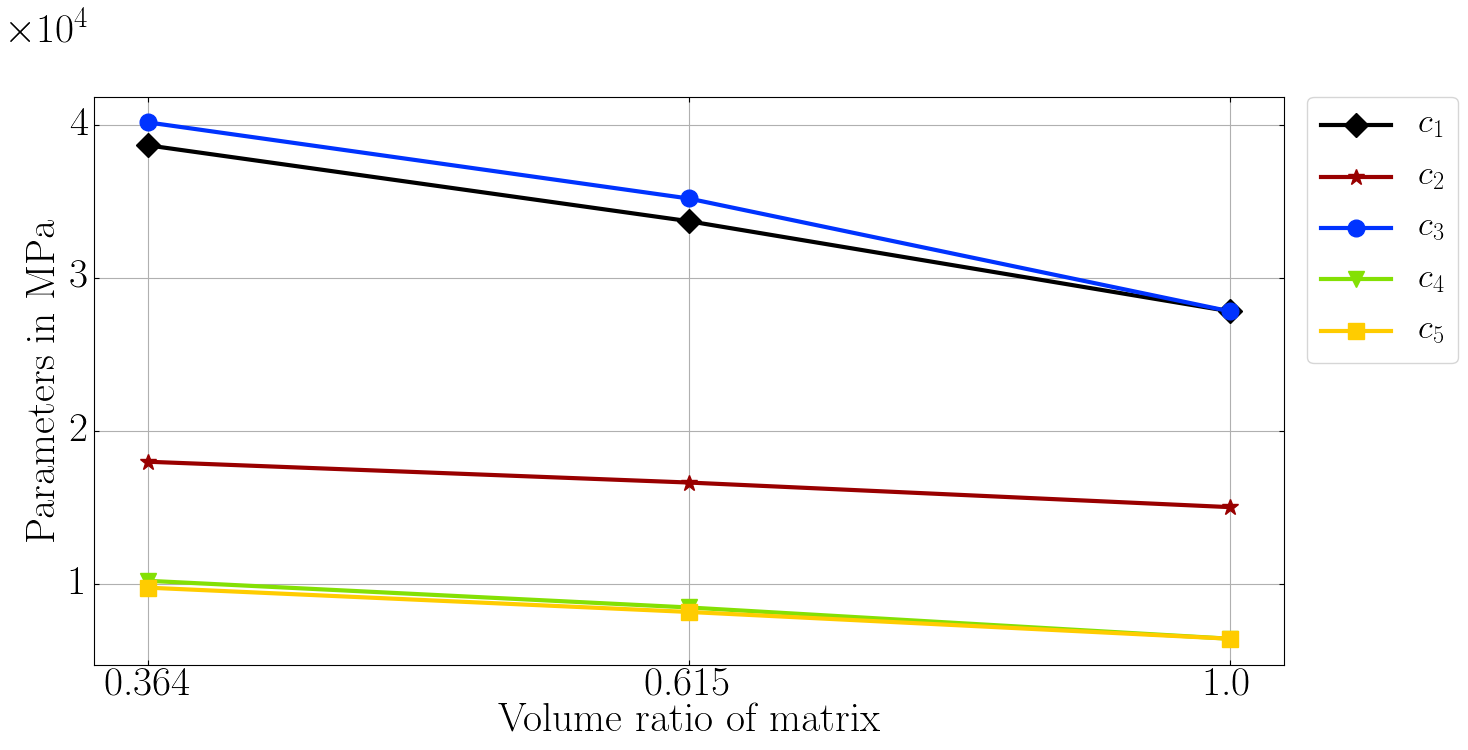}}
	\subfigure[Effective strain gradient stiffness parameters ($d_1$ - $d_7$). ]{\includegraphics[width=0.48\textwidth]{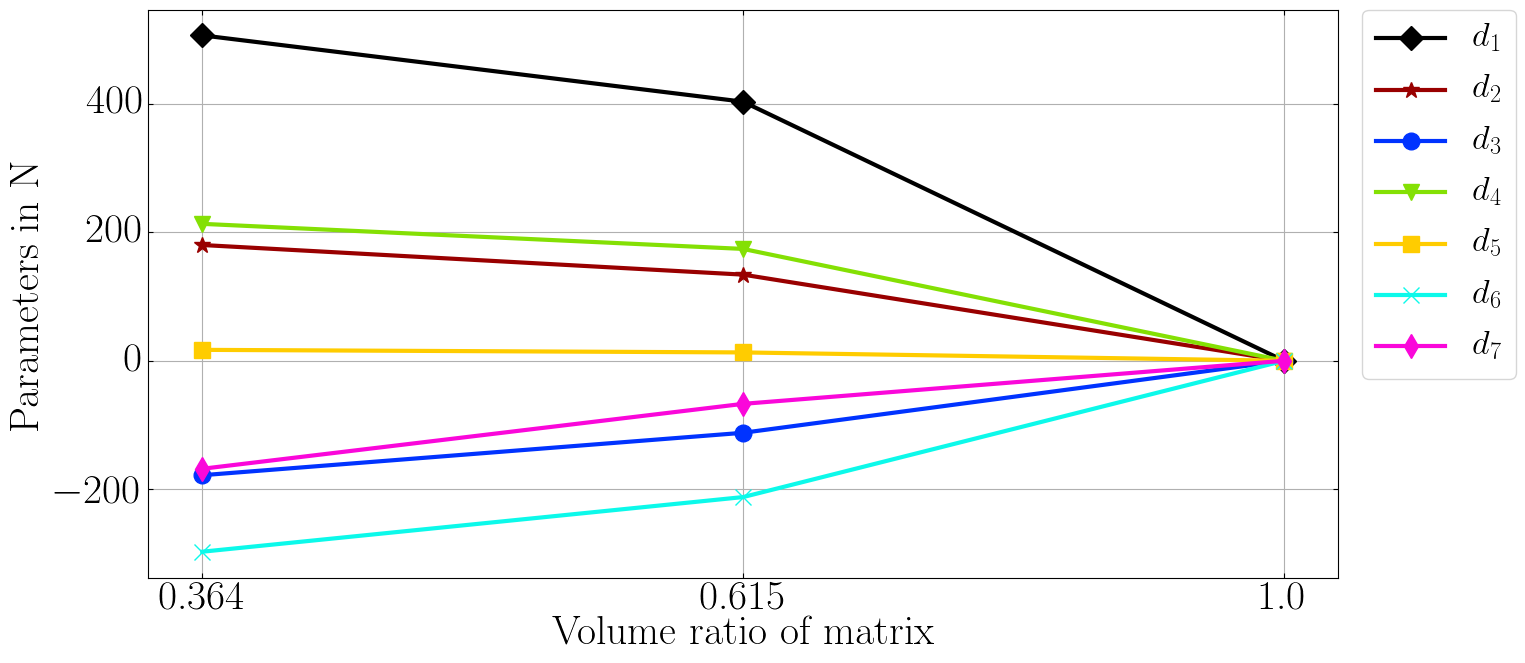}}
	\subfigure[Effective strain gradient stiffness parameters ($d_8$ - $d_{14}$). ]{\includegraphics[width=0.48\textwidth]{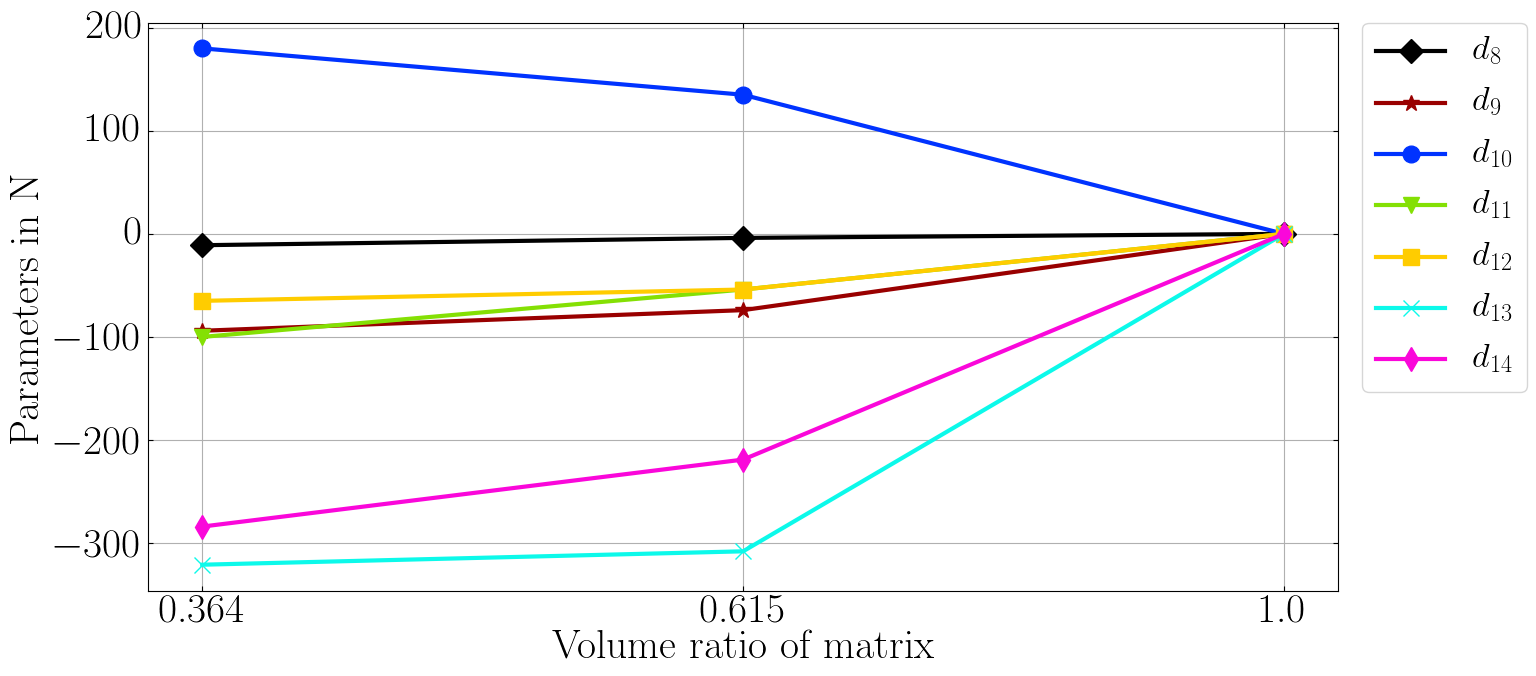}}
	\subfigure[Effective strain gradient stiffness parameters ($d_{15}$ - $d_{21}$). ]{\includegraphics[width=0.48\textwidth]{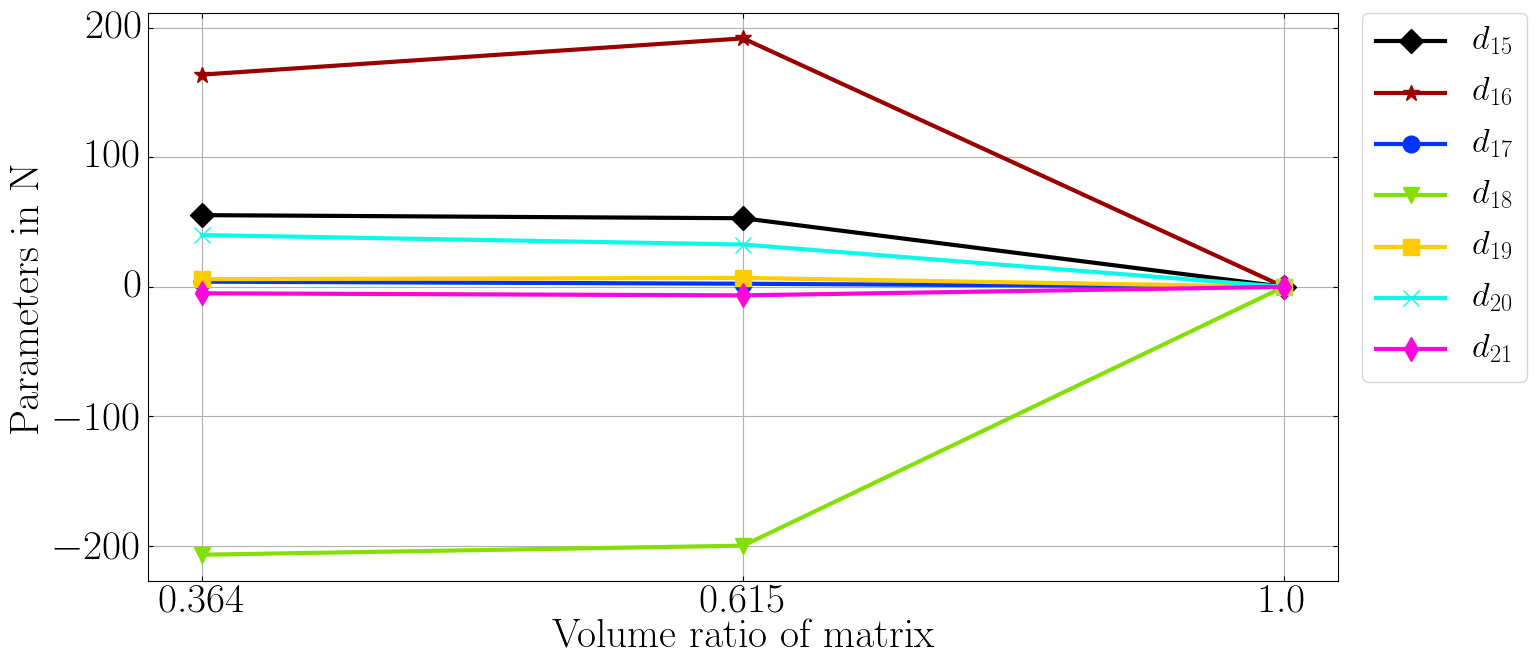}}
	\subfigure[Effective strain gradient stiffness parameters ($d_{22}$ - $d_{28}$). ]{\includegraphics[width=0.48\textwidth]{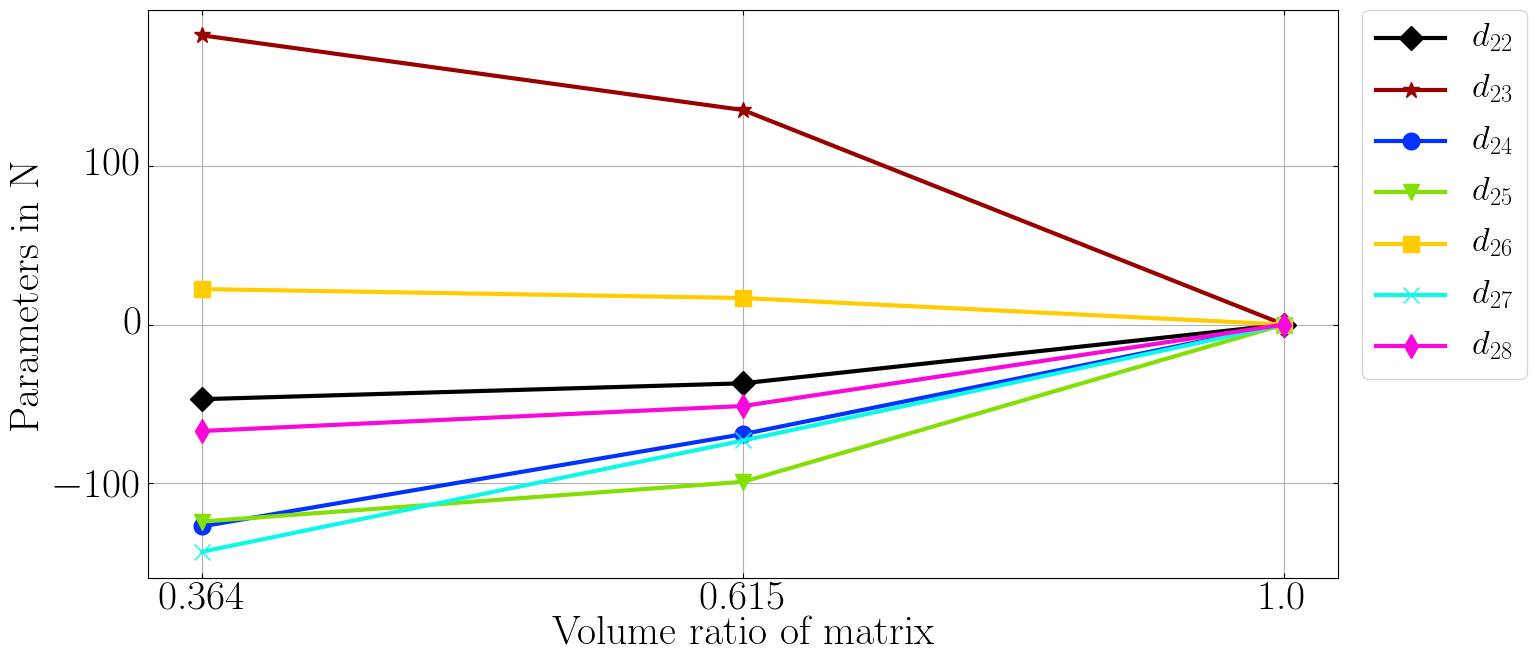}}
	\caption{ Effective material parameters with the changing of volume fraction of matrix. It should be noted that when the material is purely homogeneous (volume fraction of matrix is 1), all higher order parameters vanish.} 
	\label{vol_Cy2}
\end{figure}

\begin{figure}[h]
	\centering
	\includegraphics[width=0.75\textwidth]{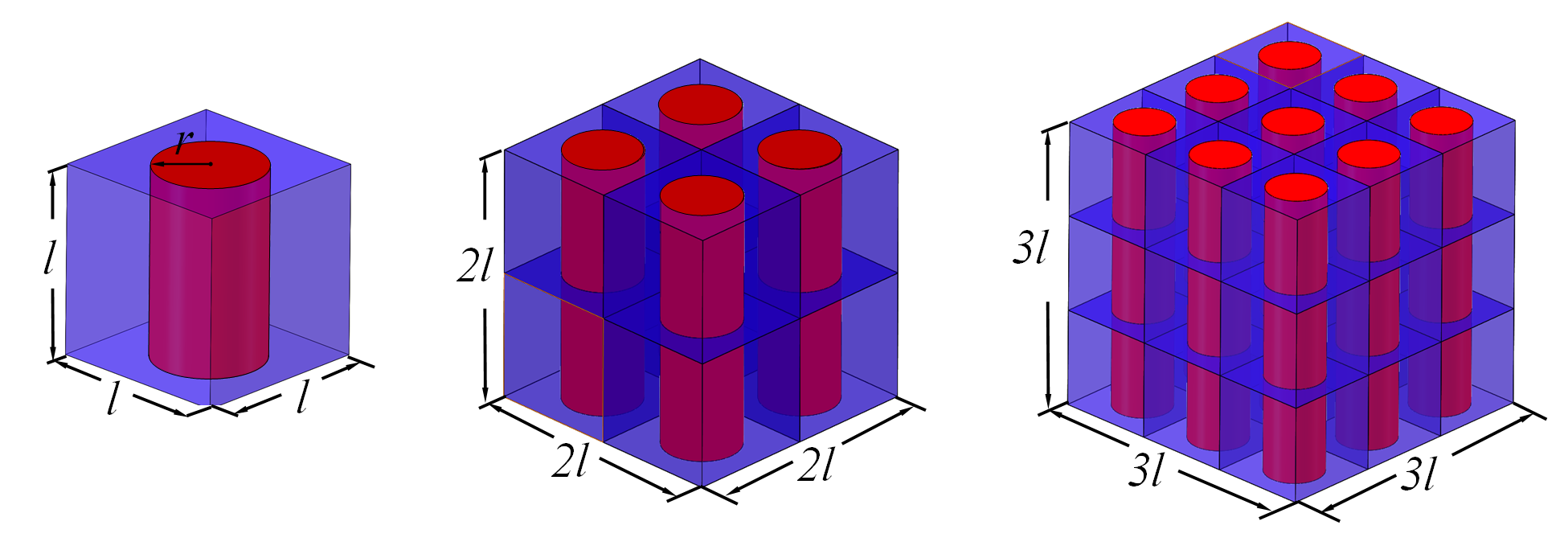}
	\caption{RVEs constructed by 1 unit cell, 8 unit cells, 27 unit cells. $l$ = 1 mm, the radius of fiber is 0.45 mm.}
	\label{Cylinder_RVE} 
\end{figure}

\begin{figure}[H]
	\centering
	\subfigure[Effective classical stiffness parameters. ]{\includegraphics[width=0.47\textwidth]{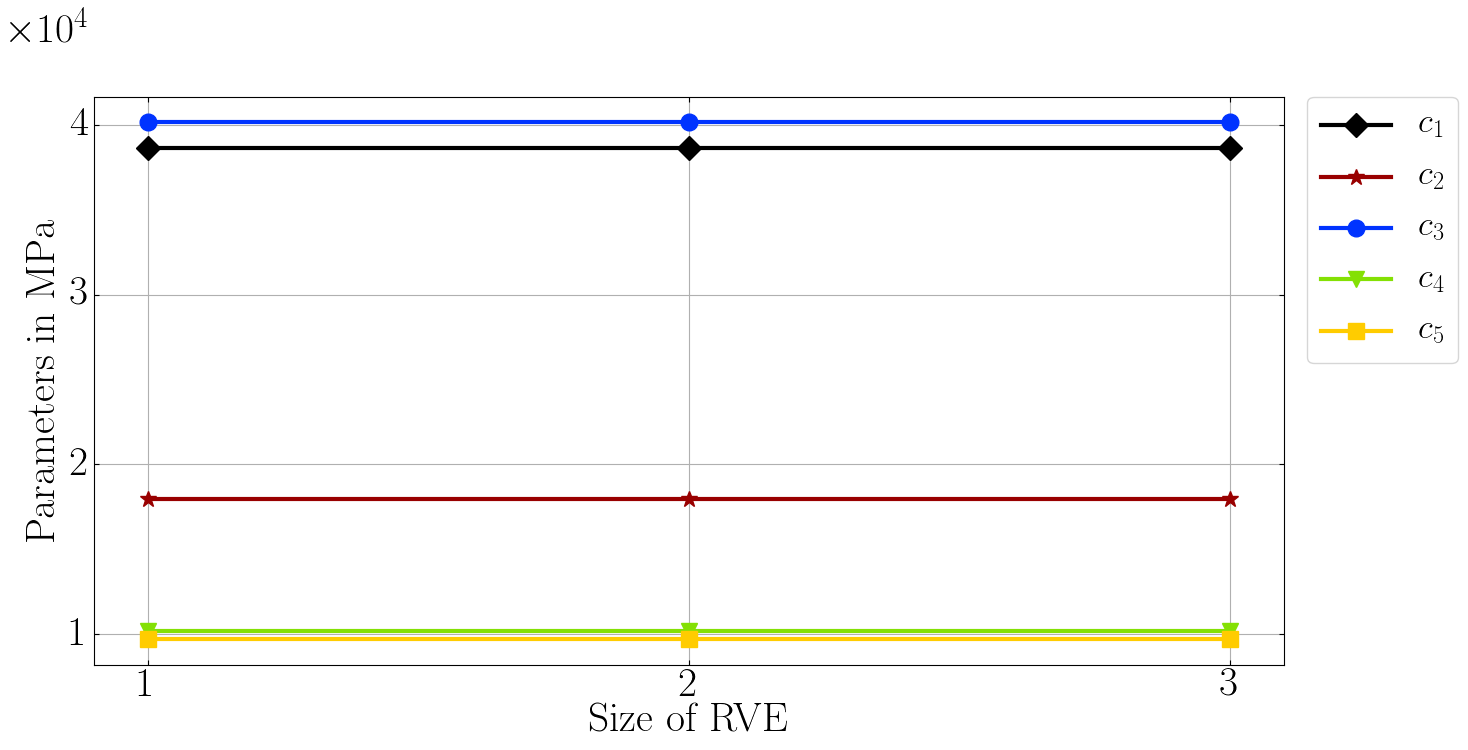}}
	\subfigure[Effective strain gradient stiffness parameters ($d_1$ - $d_7$). ]{\includegraphics[width=0.48\textwidth]{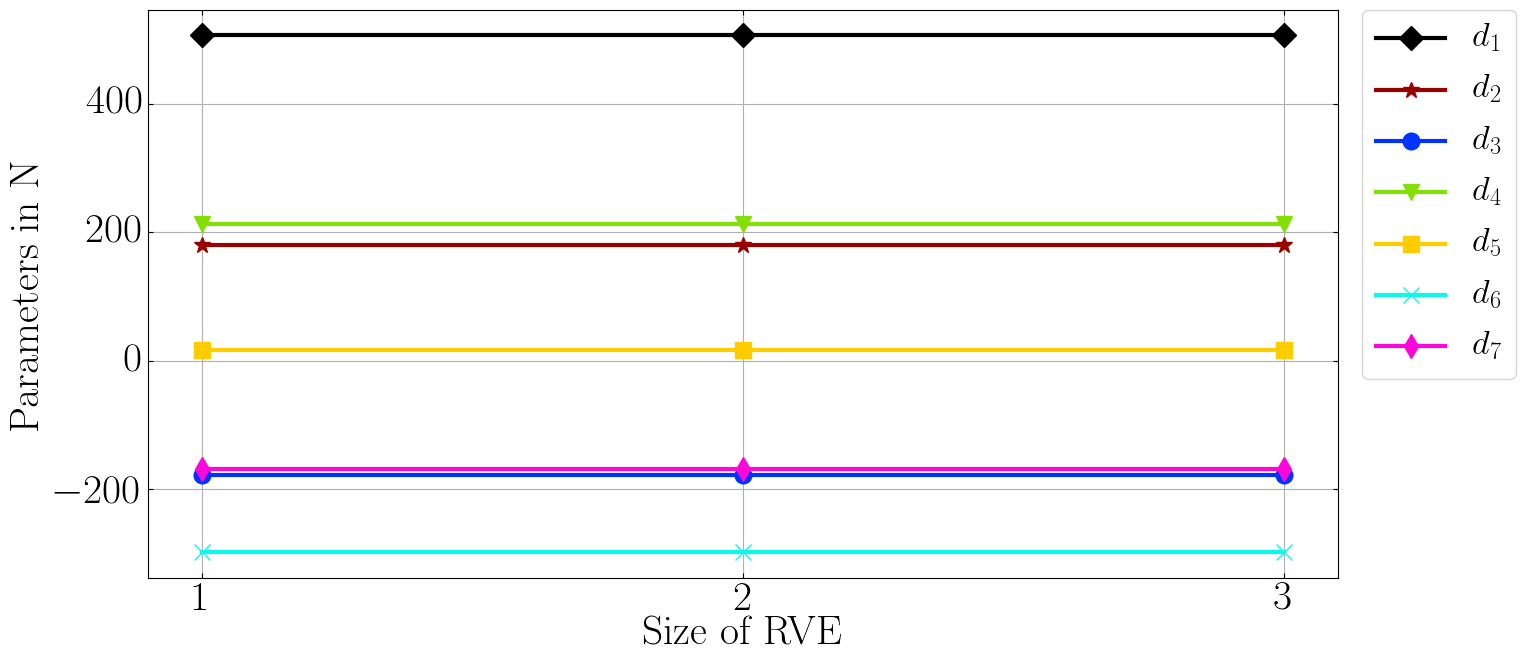}}
	\subfigure[Effective strain gradient stiffness parameters ($d_8$ - $d_{14}$). ]{\includegraphics[width=0.48\textwidth]{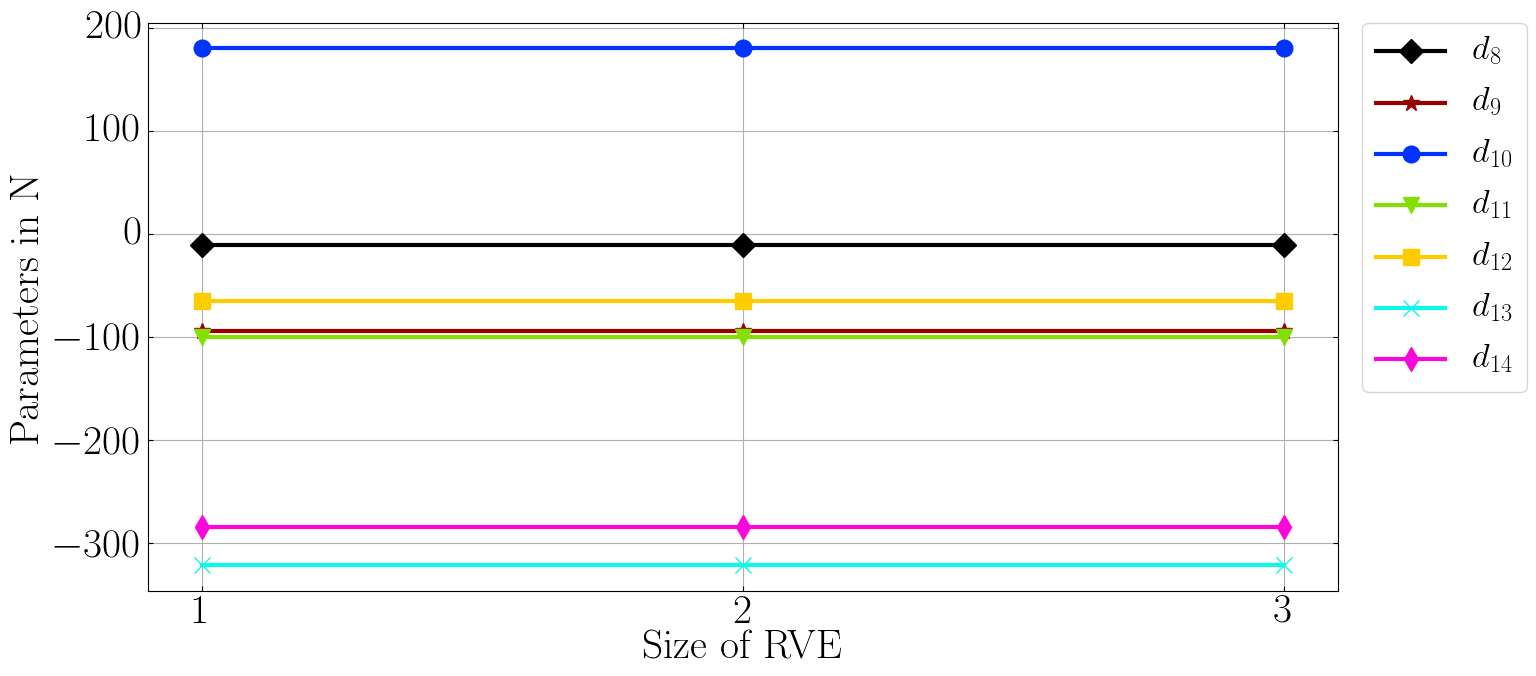}}
	\subfigure[Effective strain gradient stiffness parameters ($d_{15}$ - $d_{21}$). ]{\includegraphics[width=0.48\textwidth]{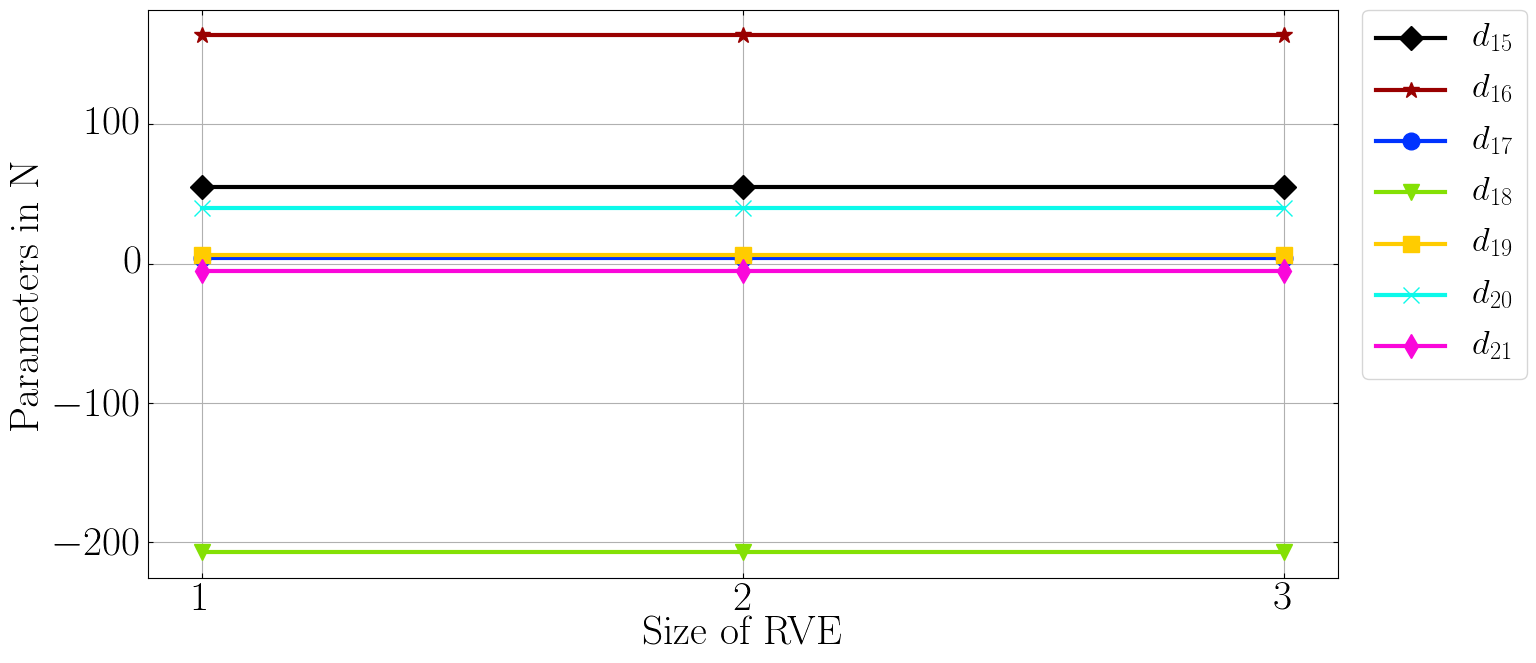}}
	\subfigure[Effective strain gradient stiffness parameters ($d_{22}$ - $d_{28}$). ]{\includegraphics[width=0.48\textwidth]{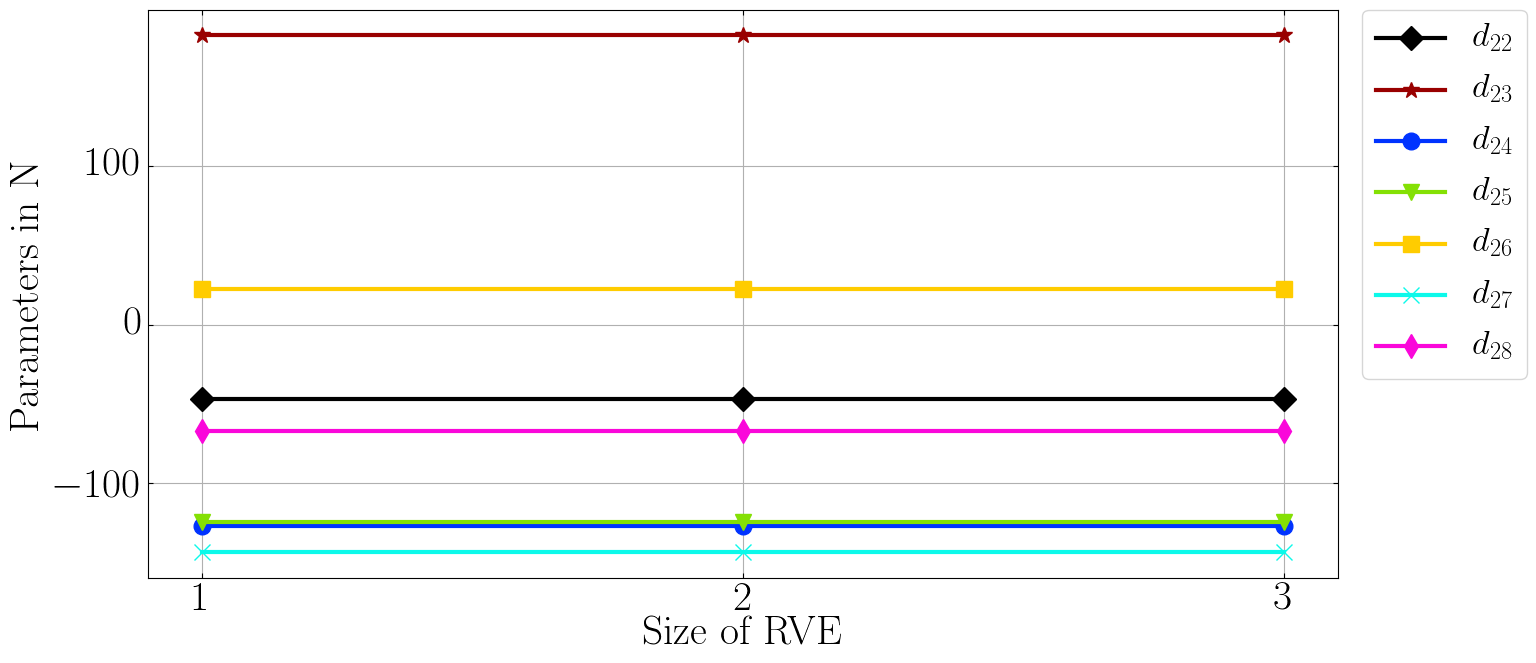}}
	\caption{ Effective material parameters with the repetition of RVEs (1\,mm$\times$1\,mm$\times$1\,mm, 2\,mm$\times$2\,mm$\times$2\,mm, 3\,mm$\times$3\,mm$\times$3\,mm ). } 
	\label{Cylinder_RVE2}
\end{figure}

\begin{figure}[h]
	\centering
	\includegraphics[width=0.68\textwidth]{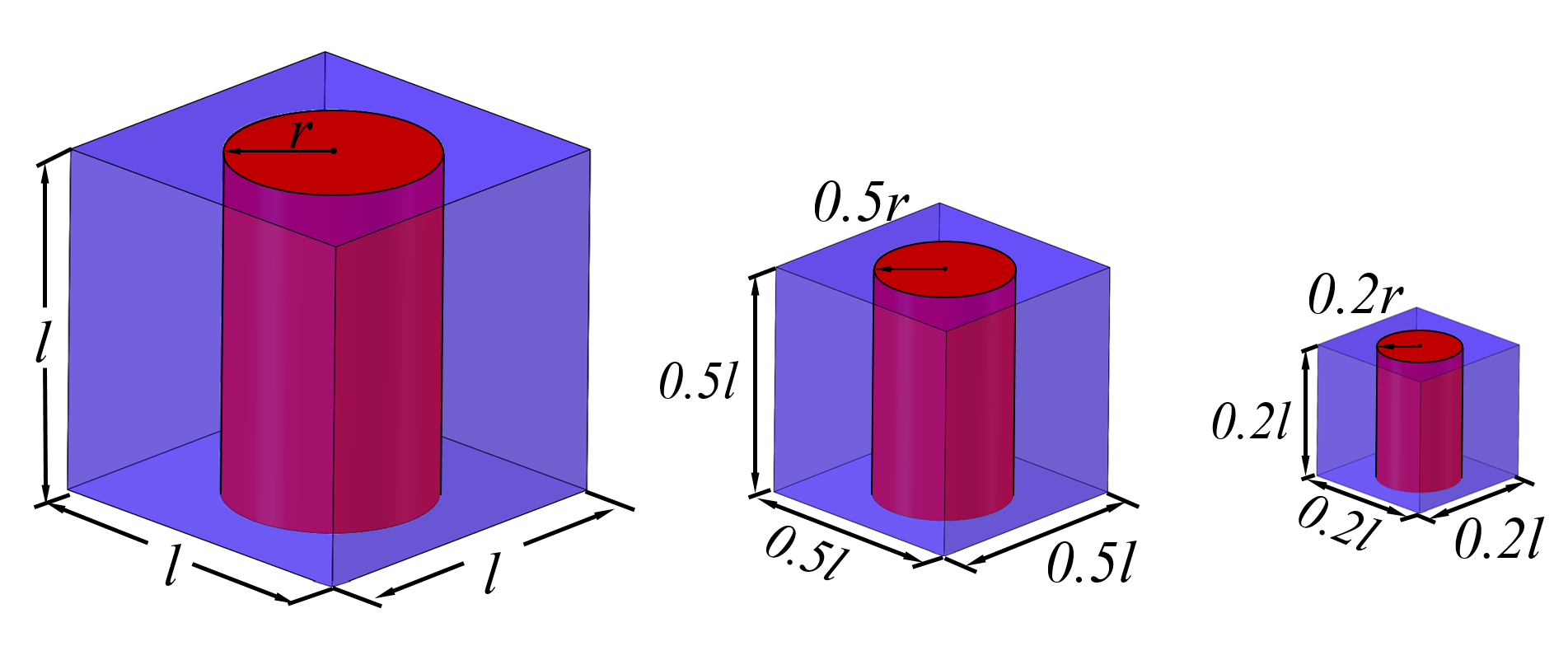}
	\caption{Unit cells with the changing sizes. $l$ = 1 mm The volume fraction of matrix are kept equal.}
	\label{Cylinder_UC} 
\end{figure}

\begin{figure}[H]
	\centering
	\subfigure[Effective classical stiffness parameters. ]{\includegraphics[width=0.47\textwidth]{CyRVE_c1c5}}
	\subfigure[Effective strain gradient stiffness parameters ($d_1$ - $d_7$). ]{\includegraphics[width=0.48\textwidth]{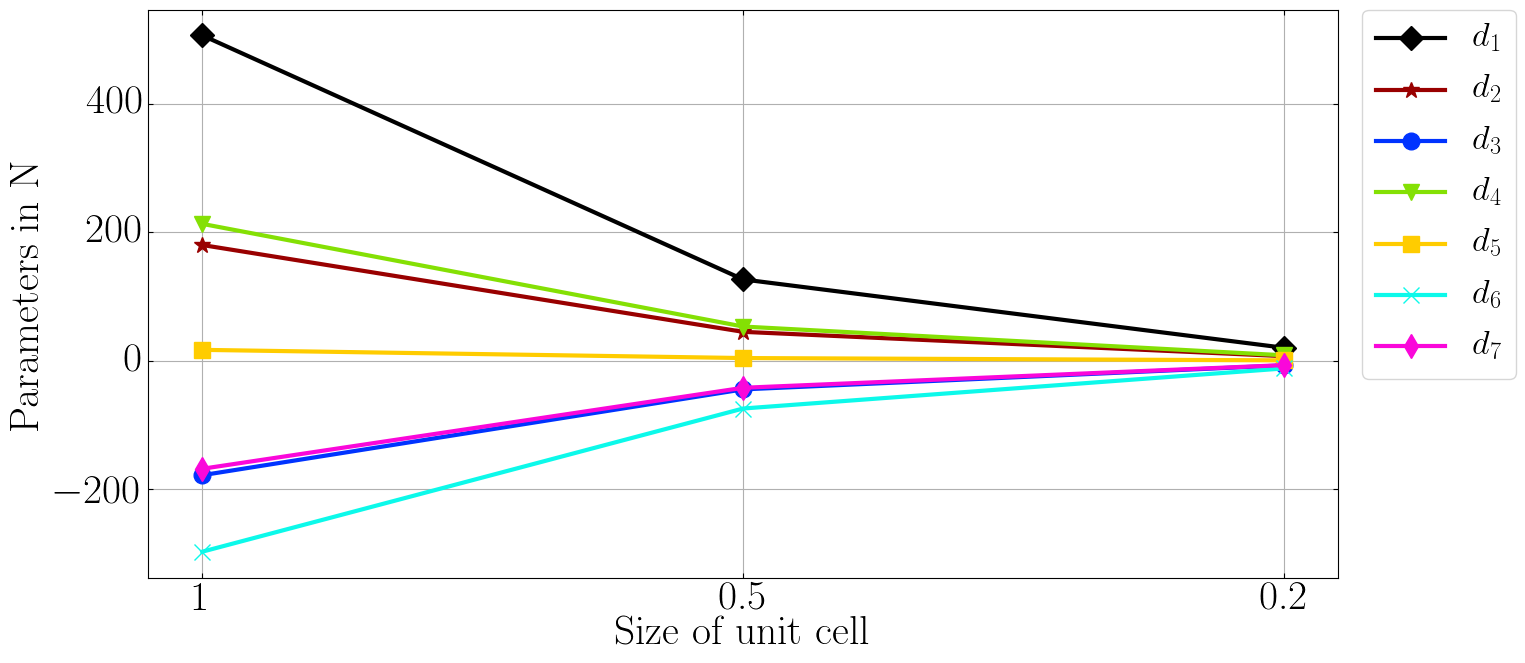}}
	\subfigure[Effective strain gradient stiffness parameters ($d_8$ - $d_{14}$). ]{\includegraphics[width=0.48\textwidth]{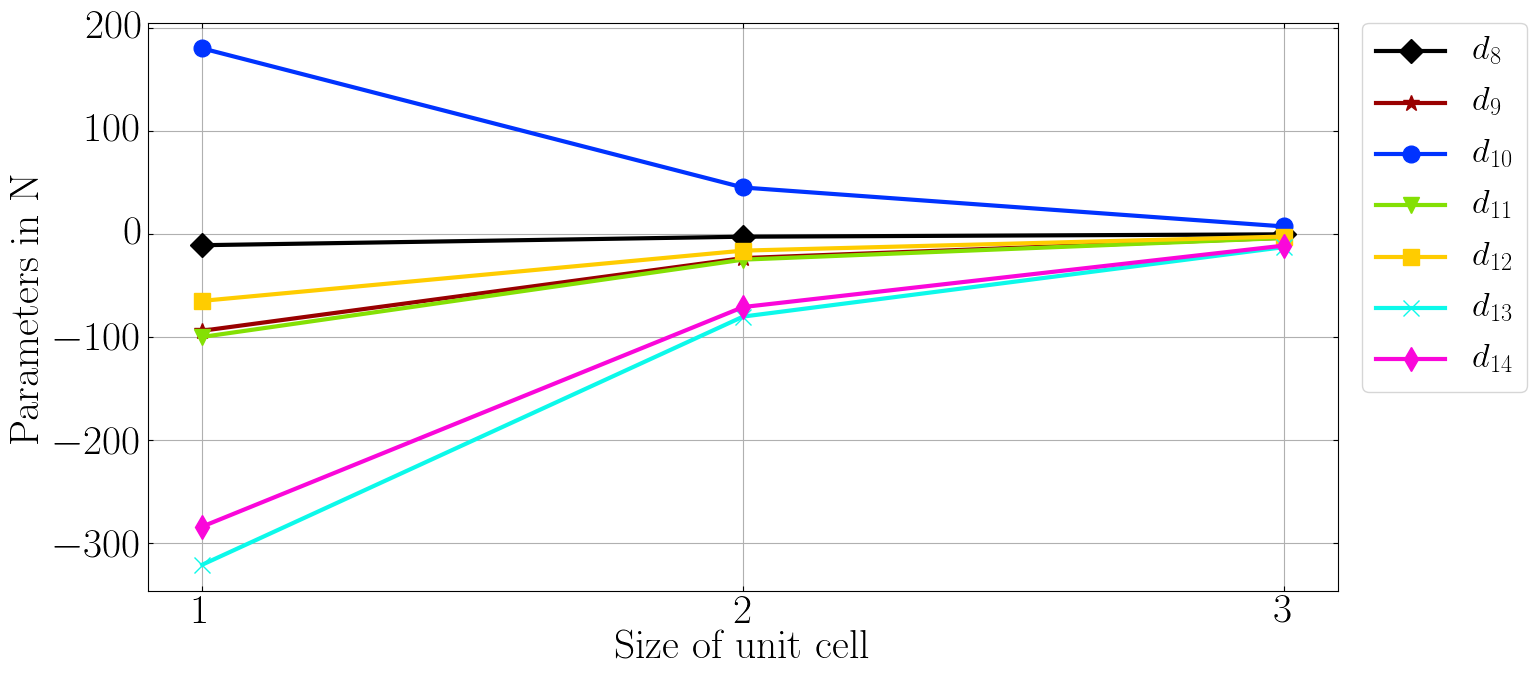}}
	\subfigure[Effective strain gradient stiffness parameters ($d_{15}$ - $d_{21}$). ]{\includegraphics[width=0.48\textwidth]{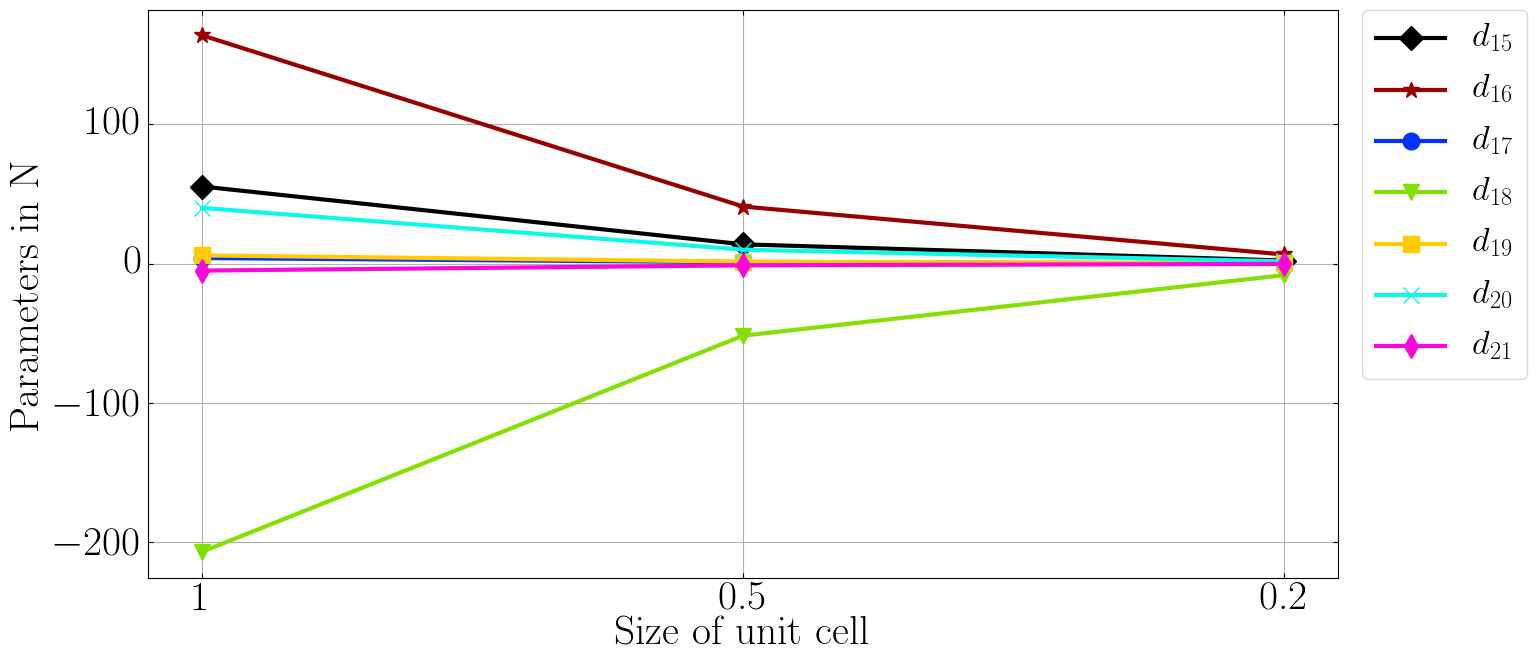}}
	\subfigure[Effective strain gradient stiffness parameters ($d_{22}$ - $d_{28}$). ]{\includegraphics[width=0.48\textwidth]{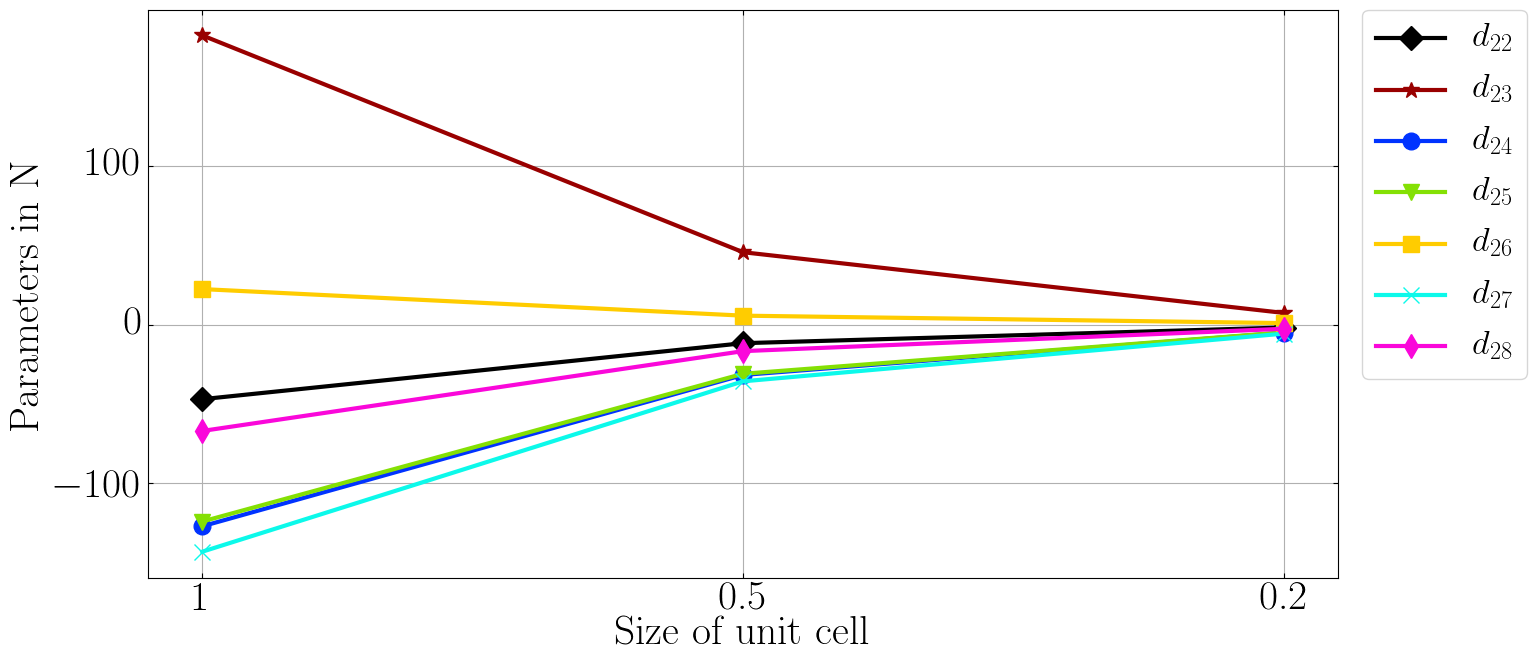}}
	\caption{ Effective material parameters with the changing lengths of unit cells, we emphasize that the substructure remains the same.} 
	\label{Cylinder_UC2}
\end{figure}

Carbon fiber is modeled by using a cylindrical inclusion in 3D. In order to compare and validate the results, the same material properties shown in Table \ref{tab:2D_EC} are used for inclusion and matrix. The radius of the cylinder is of 0.45 mm so that the volume fraction of matrix reads 36.4 \%, which are both equal to the example shown in 2D. The calculated parameters are shown as follows:
\begeq
C\ma_{AB} = \begin{pmatrix}
	38.6 & 17.9 & 18.0 & 0.0 & 0.0 & 0.0 \\ 
	17.9 & 38.6 & 18.0 & 0.0 & 0.0 & 0.0 \\ 
	18.0 & 18.0 & 40.1 & 0.0 & 0.0 & 0.0 \\ 
	0.0 & 0.0 & 0.0 & 10.2 & 0.0 & 0.0 \\ 
	0.0 & 0.0 & 0.0 & 0.0 & 10.2 & 0.0 \\ 
	0.0 & 0.0 & 0.0 & 0.0 & 0.0 & 9.7 \\ 
\end{pmatrix} \text{\,GPa} \ , \notag
\eqend
\begeq
D\ma_{\alpha\beta} =
\setcounter{MaxMatrixCols}{18}
\resizebox{.95\textwidth}{!}{$\displaystyle
	\begin{pmatrix}
		506.2 & 180.1 & -178.8 & 213.5 & 17.3 & 0.0 & 
		0.0 & 0.0 & 0.0 & 0.0 & 0.0 & 0.0 & 
		0.0 & 0.0 & 0.0 & 0.0 & 0.0 & 0.0 \\
		180.1 & -297.1 & -168.8 & -11.4 & -93.9 & 0.0 & 
		0.0 & 0.0 & 0.0 & 0.0 & 0.0 & 0.0 & 
		0.0 & 0.0 & 0.0 & 0.0 & 0.0 & 0.0 \\
		-178.8 & -168.8 & 180.3 & -100.6 & -64.7 & 0.0 & 
		0.0 & 0.0 & 0.0 & 0.0 & 0.0 & 0.0 & 
		0.0 & 0.0 & 0.0 & 0.0 & 0.0 & 0.0 \\
		213.5 & -11.4 & -100.6 & -321.4 & -284.1 & 0.0 & 
		0.0 & 0.0 & 0.0 & 0.0 & 0.0 & 0.0 & 
		0.0 & 0.0 & 0.0 & 0.0 & 0.0 & 0.0 \\
		17.3 & -93.9 & -64.7 & -284.1 & 55.4 & 0.0 & 
		0.0 & 0.0 & 0.0 & 0.0 & 0.0 & 0.0 & 
		0.0 & 0.0 & 0.0 & 0.0 & 0.0 & 0.0 \\
		0.0 & 0.0 & 0.0 & 0.0 & 0.0 & 506.5 & 
		180.2 & -178.8 & 213.6 & 16.9 & 0.0 & 0.0 & 
		0.0 & 0.0 & 0.0 & 0.0 & 0.0 & 0.0
		\\
		0.0 & 0.0 & 0.0 & 0.0 & 0.0 & 180.2 & 
		-297.4 & -169.0 & -11.5 & -94.0 & 0.0 & 0.0 & 
		0.0 & 0.0 & 0.0 & 0.0 & 0.0 & 0.0 \\
		0.0 & 0.0 & 0.0 & 0.0 & 0.0 & -178.8 & 
		-169.0 & 180.2 & -100.6 & -64.7 & 0.0 & 0.0 & 
		0.0 & 0.0 & 0.0 & 0.0 & 0.0 & 0.0 \\
		0.0 & 0.0 & 0.0 & 0.0 & 0.0 & 213.6 & 
		-11.5 & -100.6 & -322.0 & -283.8 & 0.0 & 0.0 & 
		0.0 & 0.0 & 0.0 & 0.0 & 0.0 & 0.0 \\
		0.0 & 0.0 & 0.0 & 0.0 & 0.0 & 16.9 & 
		-94.0 & -64.7 & -283.8 & 55.4 & 0.0 & 0.0 & 
		0.0 & 0.0 & 0.0 & 0.0 & 0.0 & 0.0 \\
		0.0 & 0.0 & 0.0 & 0.0 & 0.0 & 0.0 & 
		0.0 & 0.0 & 0.0 & 0.0 & 164.1 & 4.0 & 
		-207.8 & 4.0 & -207.7 & 0.0 & 0.0 & 0.0 \\
		0.0 & 0.0 & 0.0 & 0.0 & 0.0 & 0.0 & 
		0.0 & 0.0 & 0.0 & 0.0 & 4.0 & 5.9 & 
		39.6 & -4.9 & -47.3 & 0.0 & 0.0 & 0.0 \\
		0.0 & 0.0 & 0.0 & 0.0 & 0.0 & 0.0 & 
		0.0 & 0.0 & 0.0 & 0.0 & -207.8 & 39.6 & 
		181.9 & -47.3 & -126.9 & 0.0 & 0.0 & 0.0 \\
		0.0 & 0.0 & 0.0 & 0.0 & 0.0 & 0.0 & 
		0.0 & 0.0 & 0.0 & 0.0 & 4.0 & -4.9 & 
		-47.3 & 6.2 & 39.3 & 0.0 & 0.0 & 0.0 \\
		0.0 & 0.0 & 0.0 & 0.0 & 0.0 & 0.0 & 
		0.0 & 0.0 & 0.0 & 0.0 & -207.7 & -47.3 & 
		-126.9 & 39.3 & 182.1 & 0.0 & 0.0 & 0.0 \\
		0.0 & 0.0 & 0.0 & 0.0 & 0.0 & 0.0 & 
		0.0 & 0.0 & 0.0 & 0.0 & 0.0 & 0.0 & 
		0.0 & 0.0 & 0.0 & -124.2 & -143.1 & -67.6 \\
		0.0 & 0.0 & 0.0 & 0.0 & 0.0 & 0.0 & 
		0.0 & 0.0 & 0.0 & 0.0 & 0.0 & 0.0 & 
		0.0 & 0.0 & 0.0 & -143.1 & -124.5 & -67.6 \\
		0.0 & 0.0 & 0.0 & 0.0 & 0.0 & 0.0 & 
		0.0 & 0.0 & 0.0 & 0.0 & 0.0 & 0.0 & 
		0.0 & 0.0 & 0.0 & -67.6 & -67.6 & 22.3
	\end{pmatrix} $} \text{\,N} \ . \notag
\eqend
We stress that the algorithm computes $\t G$ as well, but as expected from the centro-symmetry in the substructure, all coefficients of $\t G$ vanish. The unidirectional laminate carbon reinforced epoxy composite is a transverse isotropic material. There are five independent parameters in the classical stiffness tensor, as shown below
\begeq
C\ma_{AB} = \begin{pmatrix}
	c_1 & c_1-2c_5 & c_2 & 0.0 & 0.0 & 0.0 \\ 
	c_1-2c_5 & c_1 &  c_2 & 0.0 & 0.0 & 0.0 \\ 
	c_2 & c_2 & c_3 & 0.0 & 0.0 & 0.0 \\ 
	0.0 & 0.0 & 0.0 & c_4 & 0.0 & 0.0 \\ 
	0.0 & 0.0 & 0.0 & 0.0 & c_4 & 0.0 \\ 
	0.0 & 0.0 & 0.0 & 0.0 & 0.0 & c_5 \\ 
\end{pmatrix}  \ . \notag
\eqend
We stress that the computed parameters are satisfying this condition within a tolerance of $\pm 6.7\% $. After investigating the strain gradient stiffness tensor, we find the relations between higher order parameters as shown in Figure \ref{TV_D}.
Excluding the parameters of the same value, there are 28 parameters in $\t D$ for this transverse isotropic material
\begeq
D\ma_{\alpha\beta} = \\
\setcounter{MaxMatrixCols}{18}
\resizebox{0.75\textwidth}{!}{$\displaystyle
	\begin{pmatrix}
		d_1 & d_2 & d_3 &  d_4  & d_5 & 0 & 0 & 0 & 0 & 0 & 0 &  0 &  0 &  0 &  0 &  0 & 0 & 0 \\
		& d_6 & d_7 &  d_8 & d_9 & 0 & 0 & 0 & 0  & 0 & 0 &  0  & 0 &  0  & 0 & 0 & 0 & 0 \\
		&   &d_{10} &  d_{11} & d_{12} & 0 & 0 & 0 & 0 & 0 & 0 &  0 &  0 &  0  & 0  & 0 & 0 & 0 \\
		&   &  &  d_{13}& d_{14} & 0 & 0 & 0 & 0 & 0 & 0 &  0 &  0 &  0 &  0 & 0 & 0 & 0 \\
		&  &  &   &d_{15} & 0 & 0 & 0 & 0 & 0 & 0 &  0 &  0 & 0 &  0 & 0 & 0 & 0 \\
		&  & &   & & 	d_1 & d_2 & d_3 &  d_4  & d_5 & 0&  0 &  0 & 0 &  0 & 0 & 0 & 0 \\
		&  & &   &  &  &d_6 & d_7 &  d_8 & d_9  & 0 &  0 &  0&  0 & 0 & 0 & 0 & 0 \\
		&  & &   &  &  & & d_{10} &  d_{11} & d_{12} & 0 &  0 &  0 &  0&  0 & 0 &0 &0 \\
		&  &  &   &  &  &  &  &d_{13}& d_{14} & 0 &  0 &  0  &0 &  0 & 0 & 0 & 0 \\
		&  &  &   &  &  &  & & & d_{15} & 0 &  0 &  0 &  0 &  0 & 0 & 0 & 0 \\
		&  &  &   &   & &  &  &   &   & d_{16} & d_{17} & d_{18} &  d_{17}  & d_{18}  & 0 & 0 & 0 \\
		&  &  &   &   & &  &  &   &   &  &  d_{19} & d_{20} &  d_{21} & d_{22}& 0 & 0 & 0 \\
		&  &  &   &   &  &  &  &   &  &  &   & d_{23} &  d_{22} & d_{24} & 0 & 0 & 0 \\
		&  &  &   &   & &  &  &   &   &  &   &  &  d_{19}& d_{20}  & 0 & 0 & 0 \\
		&  &  &   &   & &  &  &   &   &  &   &   &    & d_{23}  & 0 & 0 & 0 \\
		&  &  &   &   & &  &  &   &   &  &  &  &   &  & d_{25} & d_{27} & d_{28} \\
		&  &  &   &   & &  &  &   &  &  &  &  &  &  &  & d_{25} & d_{28} \\
		\text{Sym.} &  &  &   &   & &  &  &   &  &  &  &  &   & &  &  & d_{26} \\
	\end{pmatrix}  $} \notag
\ .
\eqend
We emphasize that some of the 28 parameters could be linearly dependent that leads to a reduction of independent coefficients. 
Moreover, the corresponding parameters in 2D and 3D stiffness tensors are equal within a tolerance of $\pm 4.4\% $, for example, $C_{1111}$ or $D_{111111}$ in the 2D stiffness tensors are equal to those in the 3D tensors. This verifies the calculated results.
In order to further examine the homogenization method, computations for different volume fraction of matrix are conducted as presented in Figure \ref{Vol_Cy}. 
The results are shown in Figure \ref {vol_Cy2}. It is observed that with the increasing of the volume fraction of matrix, absolute values of most of effective parameters decrease. This is due to the fact that matrix (epoxy) is softer than inclusion (carbon). It should be emphasized that when the volume fraction of matrix is 1, namely the material is purely homogeneous, the higher order parameters vanish as expected. 
Further investigations are carried out for RVEs by varying their sizes (1\,mm$\times$1\,mm$\times$1\,mm and 2\,mm$\times$2\,mm$\times$2\,mm as well as 3\,mm$\times$3\,mm$\times$3\,mm ) as shown in Figure \ref{Cylinder_RVE} and Figure \ref{Cylinder_RVE2}.  It is found that all coefficients remain constant, which indicates that the obtained parameters are independent of the repetition of RVEs.
Effective parameters are studied for unit cells with varying sizes as displayed in Figure \ref{Cylinder_UC}. The smaller unit cells are generated by homothetically scaling the larger one. Therefore, the volume fraction of matrix is identical in these cases. It is found in Figure \ref{Cylinder_UC2} that the parameters in the classical stiffness tensor remain the same, but the ones in the strain gradient stiffness tensor vary by changing the unit cell lengths. This fact is because of $C_{ijkl}^{\text M}$ being invariant regarding the microstructural size. However the effective strain gradient ones are sensitive to the homothetic ratio $\epsilon$. These higher order parameters follow a scaling rule. For example, the parameters can be obtained for the unit cell size of $0.5 $ mm $ \times 0.5 $ mm $ \times 0.5 $ mm by multiplying a scaling factor with the effective parameters of the unit cell size of $1 $ mm $ \times 1 $ mm $ \times 1 $ mm. The scaling factor is the square of homothetic ratio $\epsilon^2$, which is numerically equal to the square of ratio of the unit cell lengths herein. 

\subsubsection{SiC/Al Metal Matrix Composite (MMC)}

Aluminum-based MMCs have gained interest in engineering over the past three decades. The insertion of a ceramic material into an aluminum matrix leads to high stiffness and toughness of the composite material. In this section, the effective properties of SiC/Al metal matrix composite are investigated. RVE models have been created for three-dimensional spherical particles embedded into the metal matrix. Filler is used as a reinforcement. Their volume ratios within the MMC vary from 0$\%$ to 38.2 $\%$ by volume. The material parameters taken from \cite{bohm2002multi} are compiled in Table \ref{tab:SiC/Al}.
\begin{table}[H]
	\centering
	\caption{Material properties used for SiC/Al metal matrix composite material. $E$ Young's modulus, $\nu$ Poisson's ratio, and $\rho$ density.}
	\begin{tabular}{lccc}
		\toprule
		Type & $E$ in GPa & $\nu$  & $\rho$ in kg/m$^3$\\
		\midrule
		Matrix (Al2618-T4)  & 70  & 0.3  &  2900 \\
		Inclusion (SiC) & 450 & 0.17  & 3100 \\
		\bottomrule
	\end{tabular}
	\label{tab:SiC/Al}
\end{table}
The identified parameters for 61.8$\%$ volume ratio of matrix are found as follows
\begeq
C\ma_{AB} = \begin{pmatrix}
	163.3 & 50.5 & 50.5 & 0.0 & 0.0 & 0.0 \\ 
	50.5 & 163.5 & 50.5 & 0.0 & 0.0 & 0.0 \\ 
	50.5 & 50.5 & 163.6 & 0.0 & 0.0 & 0.0 \\ 
	0.0 & 0.0 & 0.0 & 46.4 & 0.0 & 0.0 \\ 
	0.0 & 0.0 & 0.0 & 0.0 & 46.3 & 0.0 \\ 
	0.0 & 0.0 & 0.0 & 0.0 & 0.0 & 46.3 \\ 
\end{pmatrix} \text{\,GPa} \ , \notag
\eqend
\begeq
D\ma_{\alpha\beta} =
\resizebox{.95\textwidth}{!}{$\displaystyle
	\begin{pmatrix}
		7120.6 & 1075.0 & -844.4 & 1077.8 & -836.6 & 0.0 & 
		0.0 & 0.0 & 0.0 & 0.0 & 0.0 & 0.0 & 
		0.0 & 0.0 & 0.0 & 0.0 & 0.0 & 0.0 \\
		1075.0 & -2517.5 & -788.5 & -48.3 & -275.8 & 0.0 & 
		0.0 & 0.0 & 0.0 & 0.0 & 0.0 & 0.0 & 
		0.0 & 0.0 & 0.0 & 0.0 & 0.0 & 0.0 \\
		-844.4 & -788.5 & 1914.3 & -276.5 & -347.3 & 0.0 & 
		0.0 & 0.0 & 0.0 & 0.0 & 0.0 & 0.0 & 
		0.0 & 0.0 & 0.0 & 0.0 & 0.0 & 0.0 \\
		1077.9 & -48.3 & -276.5 & -2515.1 & -789.7 & 0.0 & 
		0.0 & 0.0 & 0.0 & 0.0 & 0.0 & 0.0 & 
		0.0 & 0.0 & 0.0 & 0.0 & 0.0 & 0.0 \\
		-836.6 & -275.8 & -347.3 & -789.7 & 1915.1 & 0.0 & 
		0.0 & 0.0 & 0.0 & 0.0 & 0.0 & 0.0 & 
		0.0 & 0.0 & 0.0 & 0.0 & 0.0 & 0.0 \\
		0.0 & 0.0 & 0.0 & 0.0 & 0.0 & 7130.6 & 
		1070.6 & -850.1 & 1076.5 & -840.1 & 0.0 & 0.0 & 
		0.0 & 0.0 & 0.0 & 0.0 & 0.0 & 0.0
		\\
		0.0 & 0.0 & 0.0 & 0.0 & 0.0 & 1070.6 & 
		-2504.4 & -781.2 & -48.1 & -276.2 & 0.0 & 0.0 & 
		0.0 & 0.0 & 0.0 & 0.0 & 0.0 & 0.0 \\
		0.0 & 0.0 & 0.0 & 0.0 & 0.0 & -850.1 & 
		-781.2 & 1915.4 & -275.8 & -347.6 & 0.0 & 0.0 & 
		0.0 & 0.0 & 0.0 & 0.0 & 0.0 & 0.0 \\
		0.0 & 0.0 & 0.0 & 0.0 & 0.0 & 1076.5 & 
		-48.1 & -275.8 & -2513.8 & -790.7 & 0.0 & 0.0 & 
		0.0 & 0.0 & 0.0 & 0.0 & 0.0 & 0.0 \\
		0.0 & 0.0 & 0.0 & 0.0 & 0.0 & -840.1 & 
		-276.2 & -347.6 & -790.8 & 1917.1 & 0.0 & 0.0 & 
		0.0 & 0.0 & 0.0 & 0.0 & 0.0 & 0.0 \\
		0.0 & 0.0 & 0.0 & 0.0 & 0.0 & 0.0 & 
		0.0 & 0.0& 0.0 & 0.0 & 7141.0 & 1072.0 & 
		-848.8 & 1072.0 & -849.8 & 0.0 & 0.0 & 0.0 \\
		0.0 & 0.0 & 0.0 & 0.0 & 0.0 & 0.0 & 
		0.0 & 0.0 & 0.0 & 0.0 & 1072.0 & -2490.1 & 
		-783.5 & -47.5 & -276.5 & 0.0 & 0.0 & 0.0 \\
		0.0 & 0.0 & 0.0 & 0.0 & 0.0 & 0.0 & 
		0.0 & 0.0 & 0.0 & 0.0 & -848.8 & -783.5 & 
		1915.3 & -275.4 & -350.4 & 0.0 & 0.0 & 0.0 \\
		0.0 & 0.0 & 0.0 & 0.0 & 0.0 & 0.0 & 
		0.0 & 0.0 & 0.0 & 0.0 & 1072.0 & -47.5 & 
		-275.4 & -2506.2 & -786.2 & 0.0 & 0.0 & 0.0 \\
		0.0 & 0.0 & 0.0 & 0.0 & 0.0 & 0.0 & 
		0.0 & 0.0 & 0.0 & 0.0 & -849.8 & -276.5 & 
		-350.4 & -786.2 & 1915.4 & 0.0 & 0.0 & 0.0 \\
		0.0 & 0.0 & 0.0 & 0.0 & 0.0 & 0.0 & 
		0.0 & 0.0 & 0.0 & 0.0 & 0.0 & 0.0 & 
		0.0 & 0.0 & 0.0 & -596.3& -711.0 & -709.6 \\
		0.0 & 0.0 & 0.0 & 0.0 & 0.0 & 0.0 & 
		0.0 & 0.0 & 0.0 & 0.0 & 0.0 & 0.0 & 
		0.0 & 0.0 & 0.0 & -711.0 & -593.9 & -708.1 \\
		0.0 & 0.0 & 0.0 & 0.0 & 0.0 & 0.0 & 
		0.0 & 0.0 & 0.0 & 0.0 & 0.0 & 0.0 & 
		0.0 & 0.0 & 0.0 & -709.6 & -708.1 & -589.7
		
	\end{pmatrix} $} \text{\,N} \ . \notag
\eqend
Effected by the cubic material symmetry of the RVE, there are three independent parameters in the classical stiffness tensor,
\begeq
C\ma_{AB} = 
\begin{pmatrix}
	&  c_1 &  c_2  &   c_2  &  0 & 0 & 0 \\
	&  c_2  & c_1  &   c_2  &   0 &   0 &   0\\
	&  c_2  &  c_2  &  c_1  &   0 &   0 &   0\\
	&  0 &  0  &   0  &   c_3 &   0 &   0\\
	&  0 &  0  &   0  &   0 &   c_3 &   0\\
	&  0 &  0  &   0  &   0 &   0 &   c_3\\ 
\end{pmatrix} \notag
\ .
\eqend
\begin{figure}[H]
	\centering
	\includegraphics[width=1.1\textwidth]{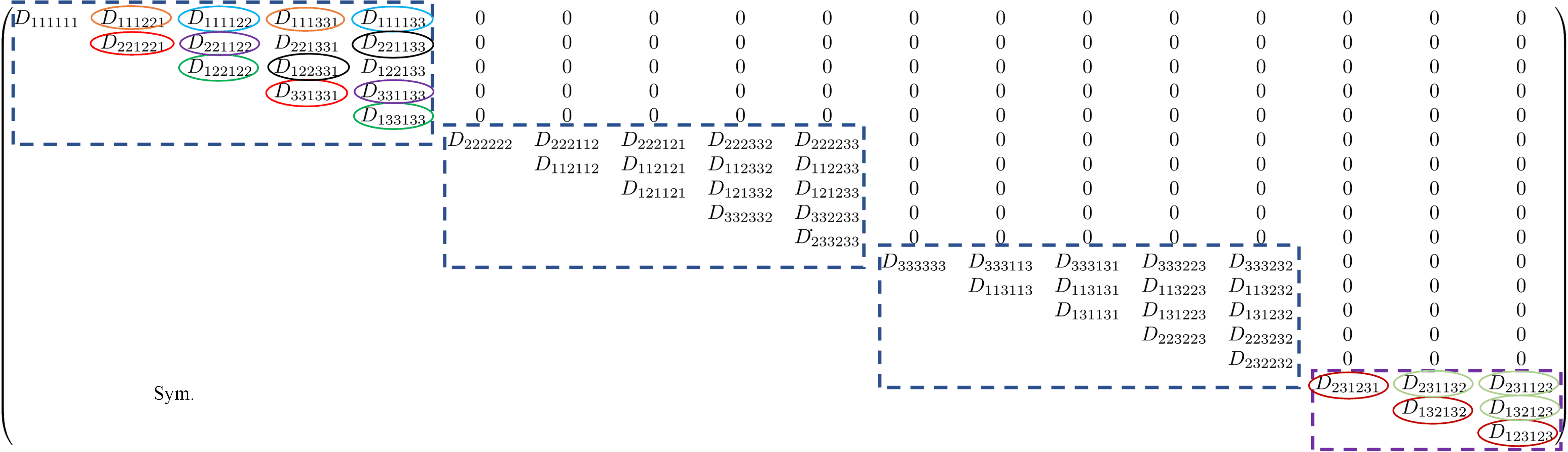}
	\caption{The structure of the strain gradient stiffness tensor for cubic materials. Three 5 $\times$ 5 matrices in the diagonal are equal, For example, $D_{111111} = D_{222222} $. In the first 5 $\times$ 5 matrix, it is also observed that $D_{111221} = D_{111331} $, $ D_{111122} = D_{111133}$, $D_{221221}=D_{331331}$, $D_{122122}=D_{133133} $, $ D_{221122} = D_{331133}$, $ D_{221133} = D_{122331 }$. In the 3 $ \times$ 3 matrix,  $D_{231231} = D_{132132}=D_{123123} $, $D_{231132} = D_{231123} = D_{132123}$.}
	\label{Sp_cubic} 
\end{figure}
As shown in Figure \ref{Sp_cubic}, excluding the parameters of the same value, there are 11 parameters found in the strain gradient stiffness tensor,
\begeq
D\ma_{\alpha\beta} = \\
\setcounter{MaxMatrixCols}{18}
\resizebox{0.95\textwidth}{!}{$\displaystyle
	\begin{pmatrix}
		d_1 & d_2 & d_3 &  d_2  & d_3 & 0 & 0 & 0 & 0 & 0 & 0 &  0 &  0 &  0 &  0 &  0 & 0 & 0 \\
		& d_4 & d_5 &  d_6 & d_7 & 0 & 0 & 0 & 0  & 0 & 0 &  0  & 0 &  0  & 0 & 0 & 0 & 0 \\
		&   &d_8 &  d_7 & d_9 & 0 & 0 & 0 & 0 & 0 & 0 &  0 &  0 &  0  & 0  & 0 & 0 & 0 \\
		&   &  &  d_4& d_5 & 0 & 0 & 0 & 0 & 0 & 0 &  0 &  0 &  0 &  0 & 0 & 0 & 0 \\
		&  &  &   &d_8 & 0 & 0 & 0 & 0 & 0 & 0 &  0 &  0 & 0 &  0 & 0 & 0 & 0 \\
		&  & &   & & d_1 & d_2 & d_3 &  d_2  & d_3 & 0&  0 &  0 & 0 &  0 & 0 & 0 & 0 \\
		&  & &   &  &  &d_4 & d_5 &  d_6 & d_7  & 0 &  0 &  0&  0 & 0 & 0 & 0 & 0 \\
		&  & &   &  &  & & d_8 &  d_7 & d_9 & 0 &  0 &  0 &  0&  0 & 0 &0 &0 \\
		&  &  &   &  &  &  &  & d_4& d_5 & 0 &  0 &  0  &0 &  0 & 0 & 0 & 0 \\
		&  &  &   &  &  &  & & & d_8 & 0 &  0 &  0 &  0 &  0 & 0 & 0 & 0 \\
		&  &  &   &   & &  &  &   &   & d_1 & d_2 & d_3 &  d_2  & d_3  & 0 & 0 & 0 \\
		&  &  &   &   & &  &  &   &   &  & d_4 & d_5 &  d_6 & d_7 & 0 & 0 & 0 \\
		&  &  &   &   &  &  &  &   &  &  &   & d_8 &  d_7 & d_9 & 0 & 0 & 0 \\
		&  &  &   &   & &  &  &   &   &  &   &  &  d_4& d_5  & 0 & 0 & 0 \\
		&  &  &   &   & &  &  &   &   &  &   &   &    & d_8  & 0 & 0 & 0 \\
		&  &  &   &   & &  &  &   &   &  &  &  &   &  & d_{10} & d_{11} & d_{11} \\
		&  &  &   &   & &  &  &   &  &  &  &  &  &  &  & d_{10} & d_{11} \\
		\text{Sym.} &  &  &   &   & &  &  &   &  &  &  &  &   & &  &  & d_{10} \\
	\end{pmatrix}  $} \notag
\ .
\eqend
Please note, among these 11 parameters, some of them might be linearly dependent.
Further investigations are conducted for different volume fraction of matrix, different sizes of selected RVE, and different sizes of unit cells as indicated in Figures \ref{Vol_SP}, \ref{sphere_RVE}, \ref{sphere_UC}. Results are displayed in Figures \ref{Vol_SP2}, \ref{sphere_RVE2}, \ref{sphere_UC2}. It is observed that the higher order parameters are zero when materials are homogeneous; they are independent of the stack of RVEs, and they are sensitive to microstructural sizes as well as following the scaling rule.
\begin{figure}[H]
	\centering
	\includegraphics[width=0.9\textwidth]{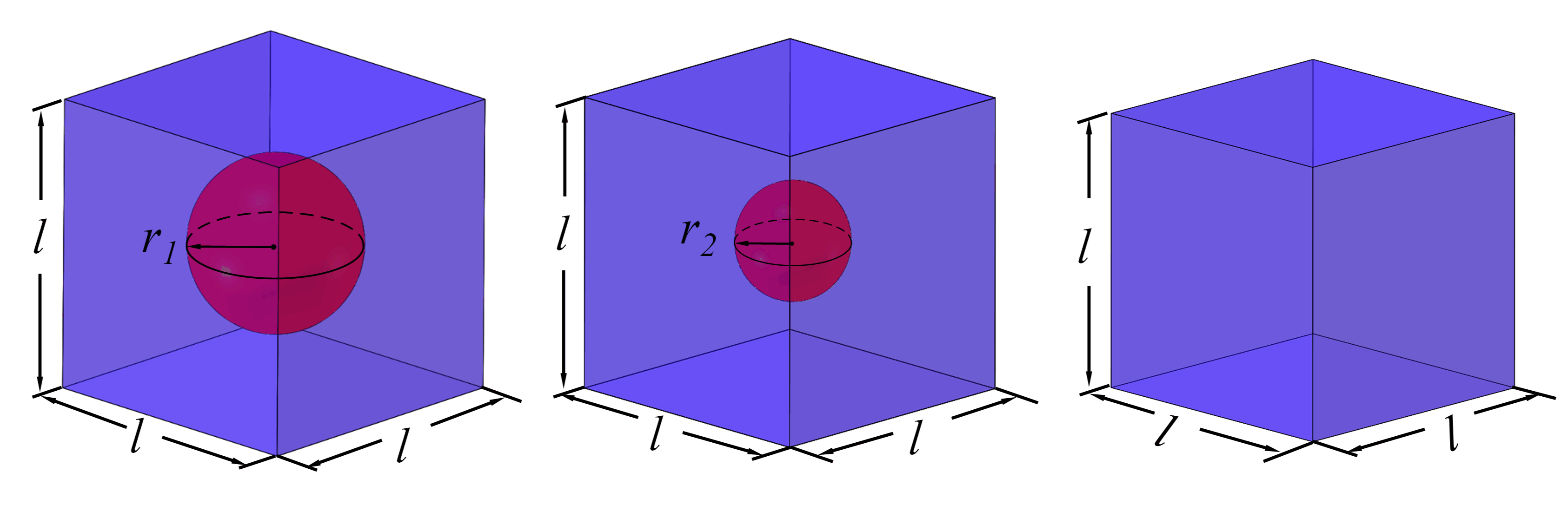}
	\caption{Changing volume fraction of matrix.}
	\label{Vol_SP} 
\end{figure}
\begin{figure}[H]
	\centering
	\subfigure[Effective classical stiffness parameters. ]{\includegraphics[width=0.47\textwidth]{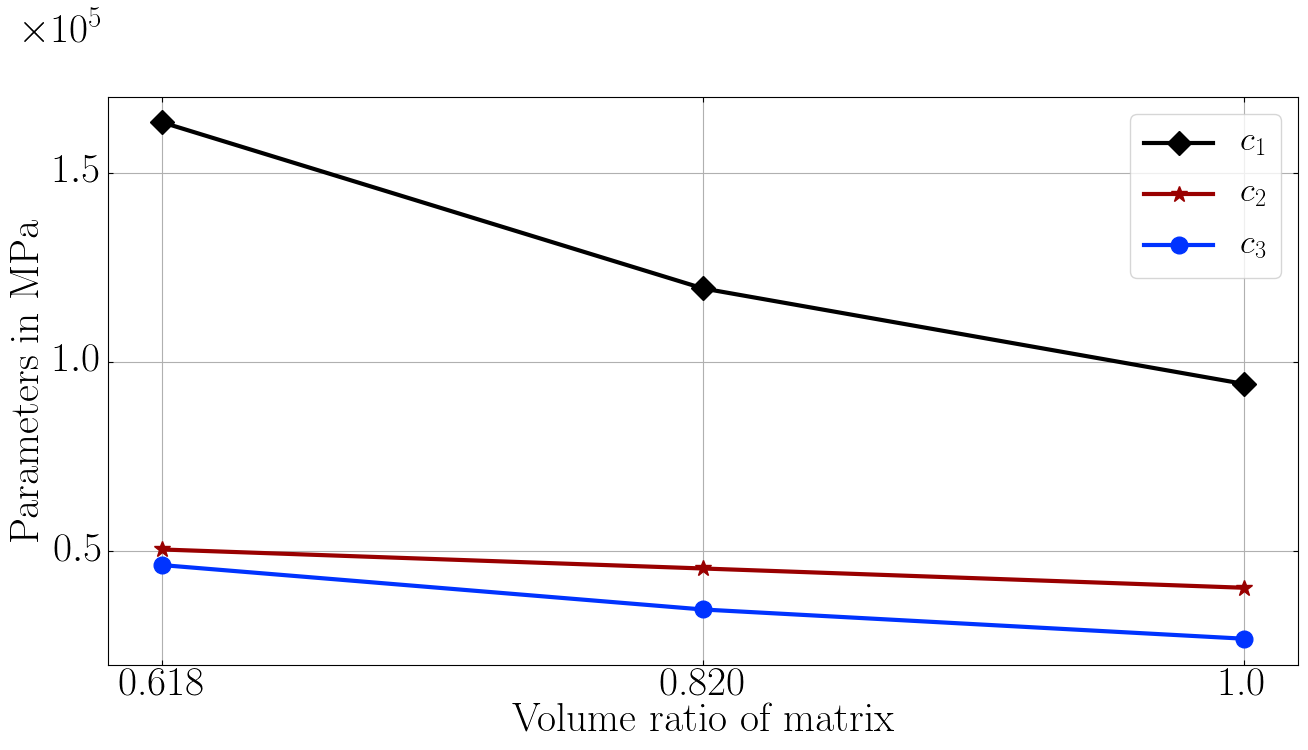}}
	\subfigure[Effective strain gradient stiffness parameters ($d_1$ - $d_6$). ]{\includegraphics[width=0.47\textwidth]{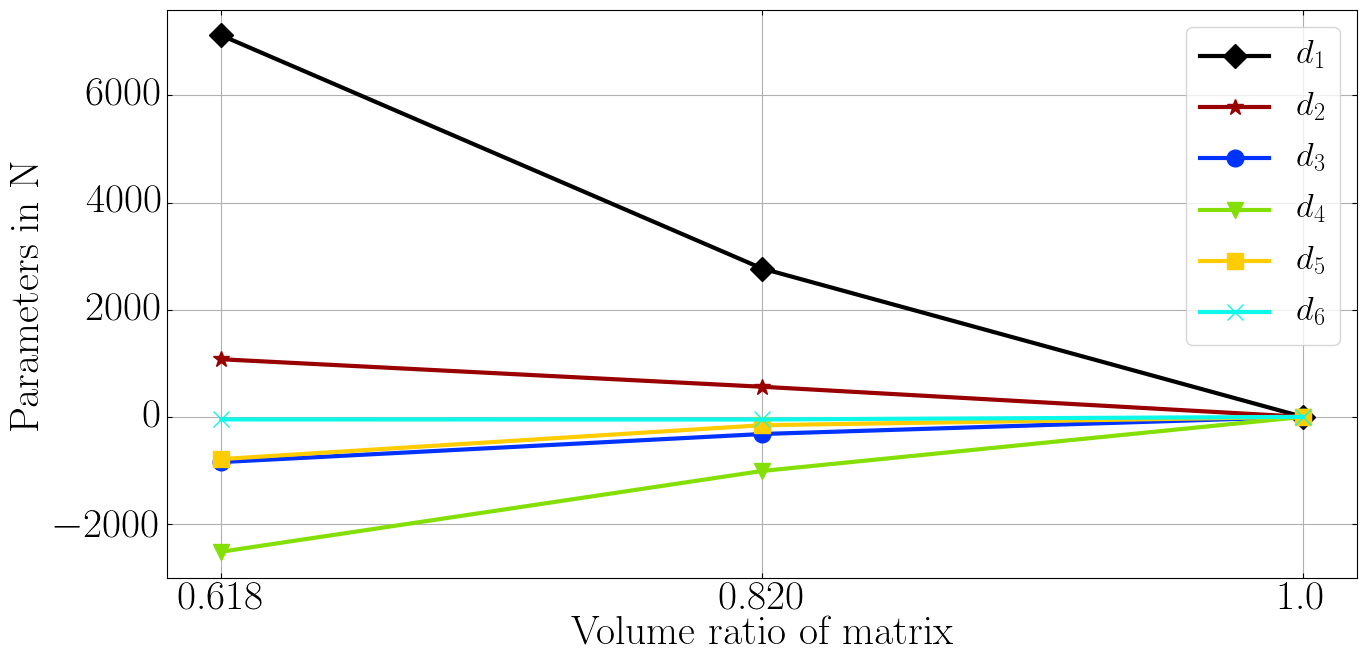}}
	\subfigure[Effective strain gradient stiffness parameters ($d_7$ - $d_{11}$). ]{\includegraphics[width=0.47\textwidth]{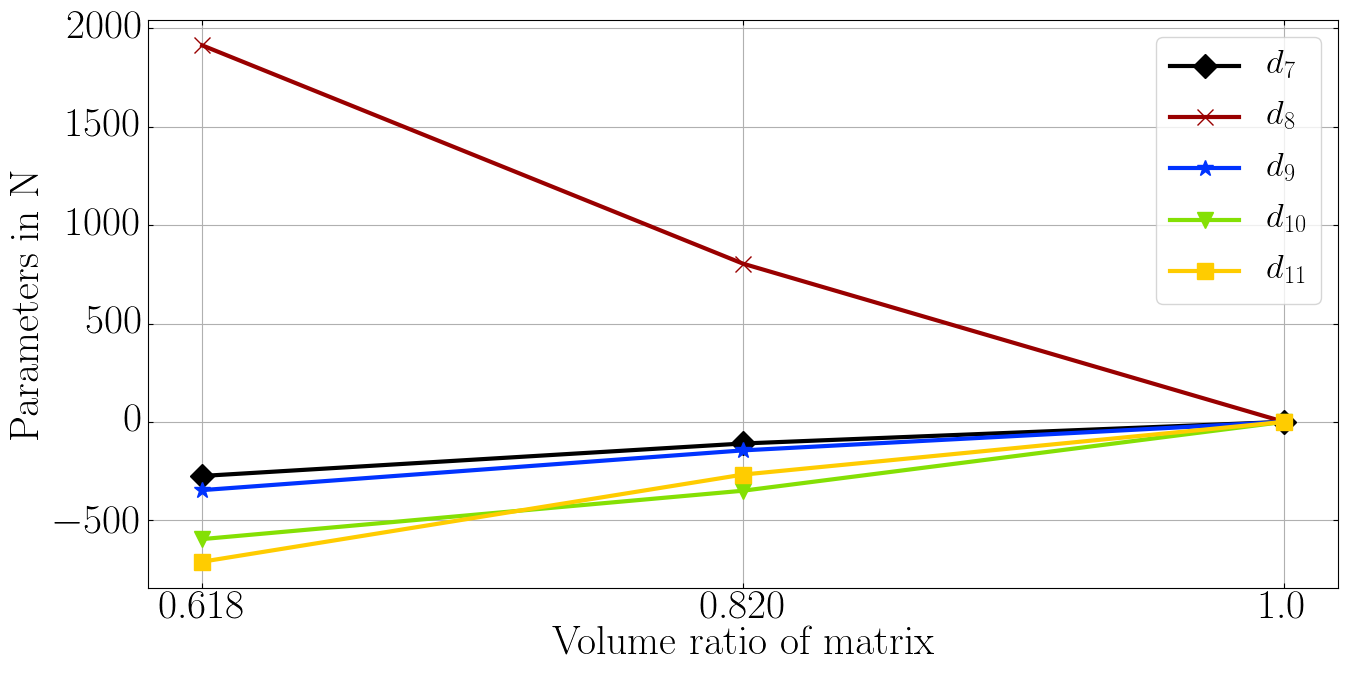}}
	\caption{ Effective material parameters with the changing of volume fraction of matrix.} 
	\label{Vol_SP2}
\end{figure}

\begin{figure}[H]
	\centering
	\includegraphics[width=0.9\textwidth]{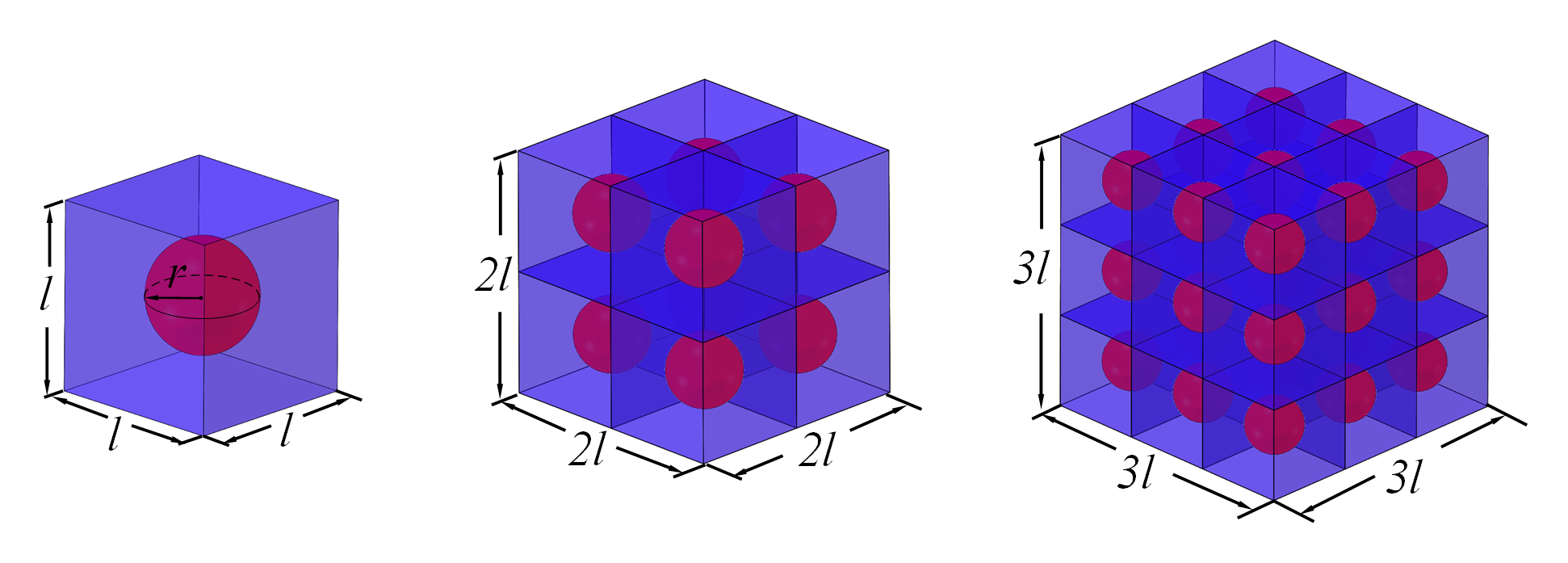}
	\caption{RVEs constructed by 1 unit cell, 8 unit cells, 27 unit cells.}
	\label{sphere_RVE} 
\end{figure}
\begin{figure}[H]
	\centering
	\subfigure[Effective classical stiffness parameters. ]{\includegraphics[width=0.47\textwidth]{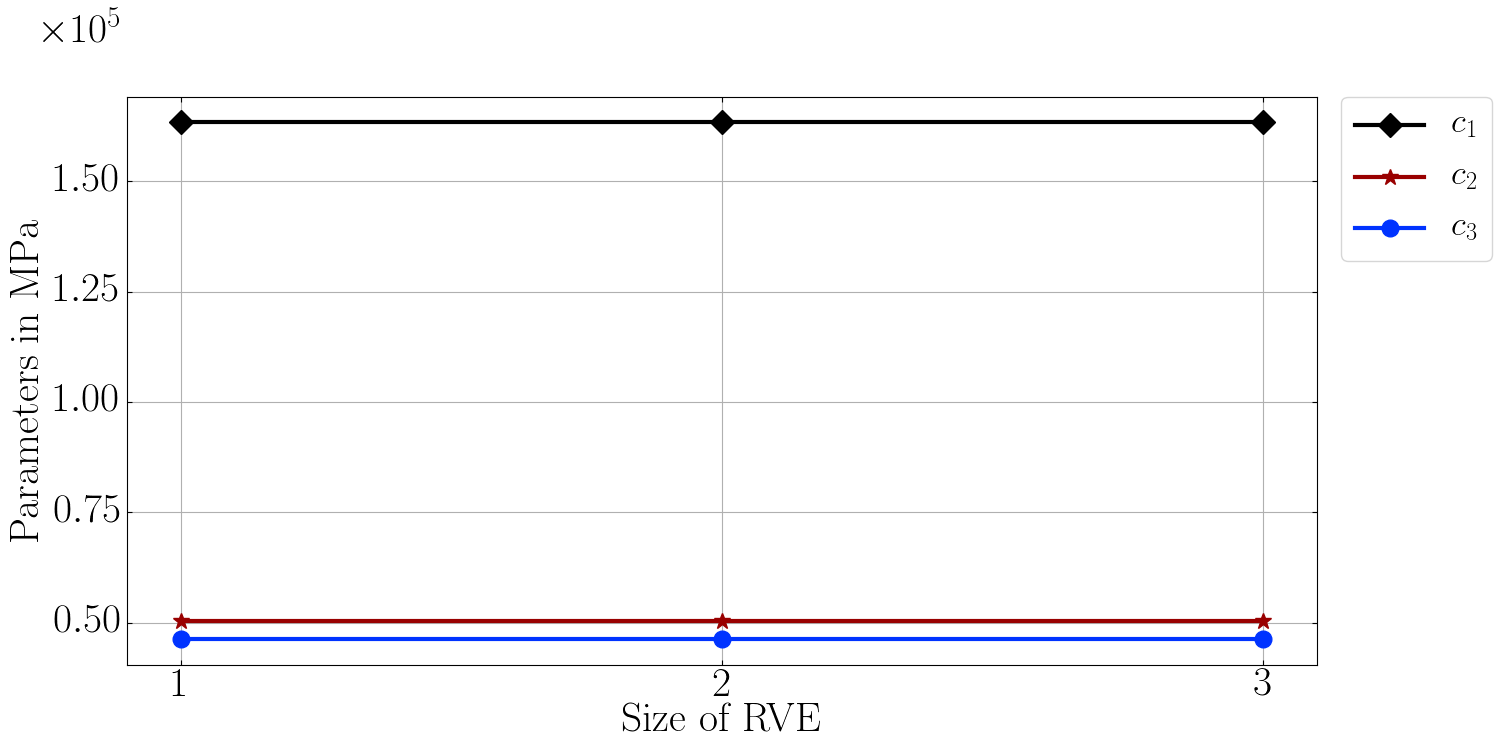}}
	\subfigure[Effective strain gradient stiffness parameters ($d_1$ - $d_6$). ]{\includegraphics[width=0.47\textwidth]{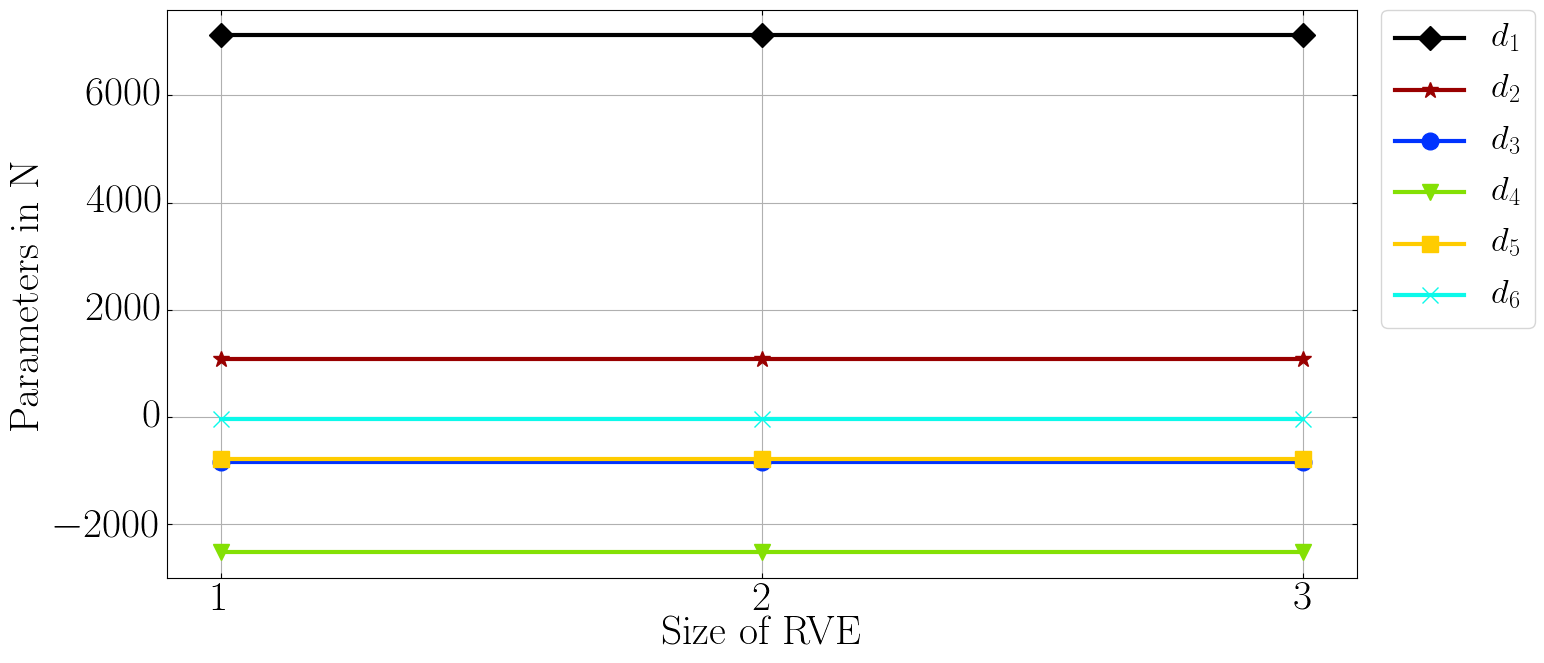}}
	\subfigure[Effective strain gradient stiffness parameters ($d_7$ - $d_{11}$). ]{\includegraphics[width=0.47\textwidth]{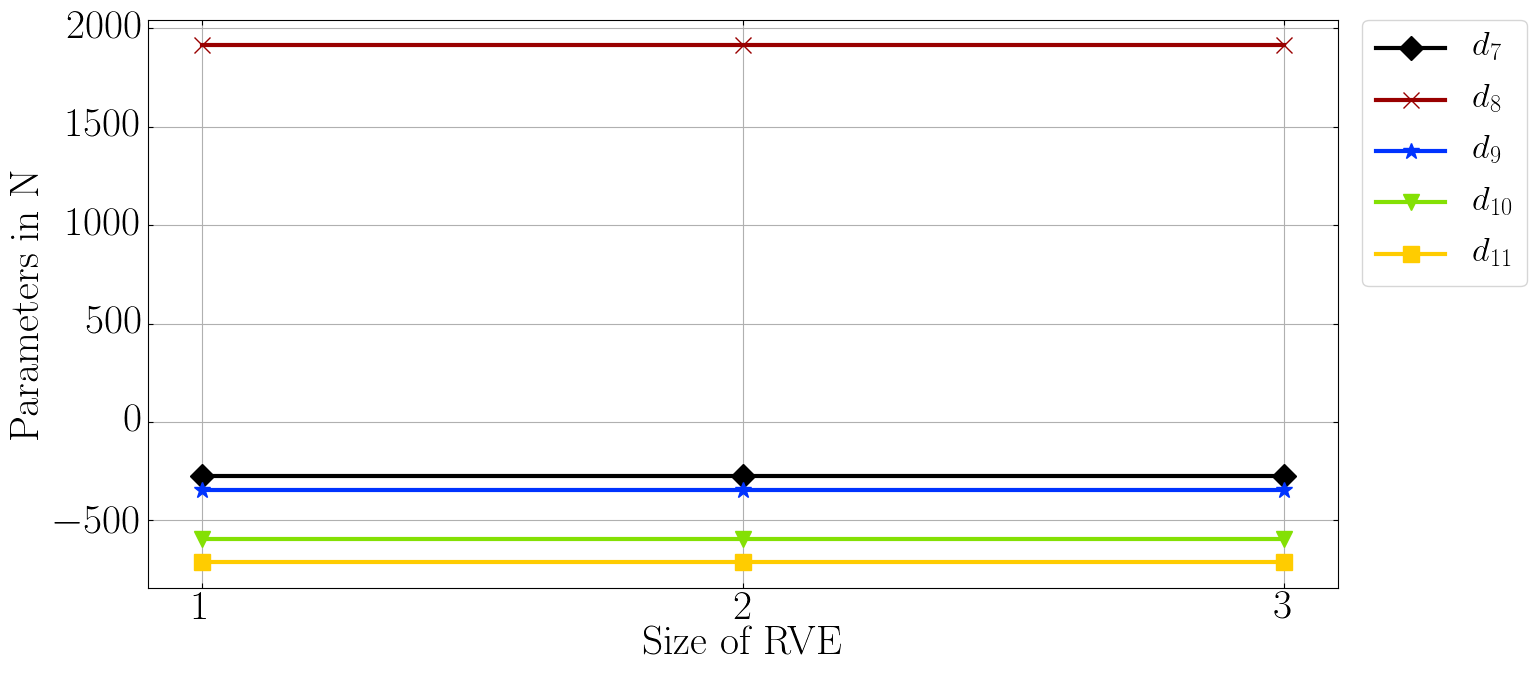}}
	\caption{ Effective material parameters with the changing RVE sizes ($1 \times 1 \times 1$, $2 \times 2 \times 2$, $3 \times 3 \times 3$).} 
	\label{sphere_RVE2}
\end{figure}
\begin{figure}[H]
	\centering
	\includegraphics[width=0.75\textwidth]{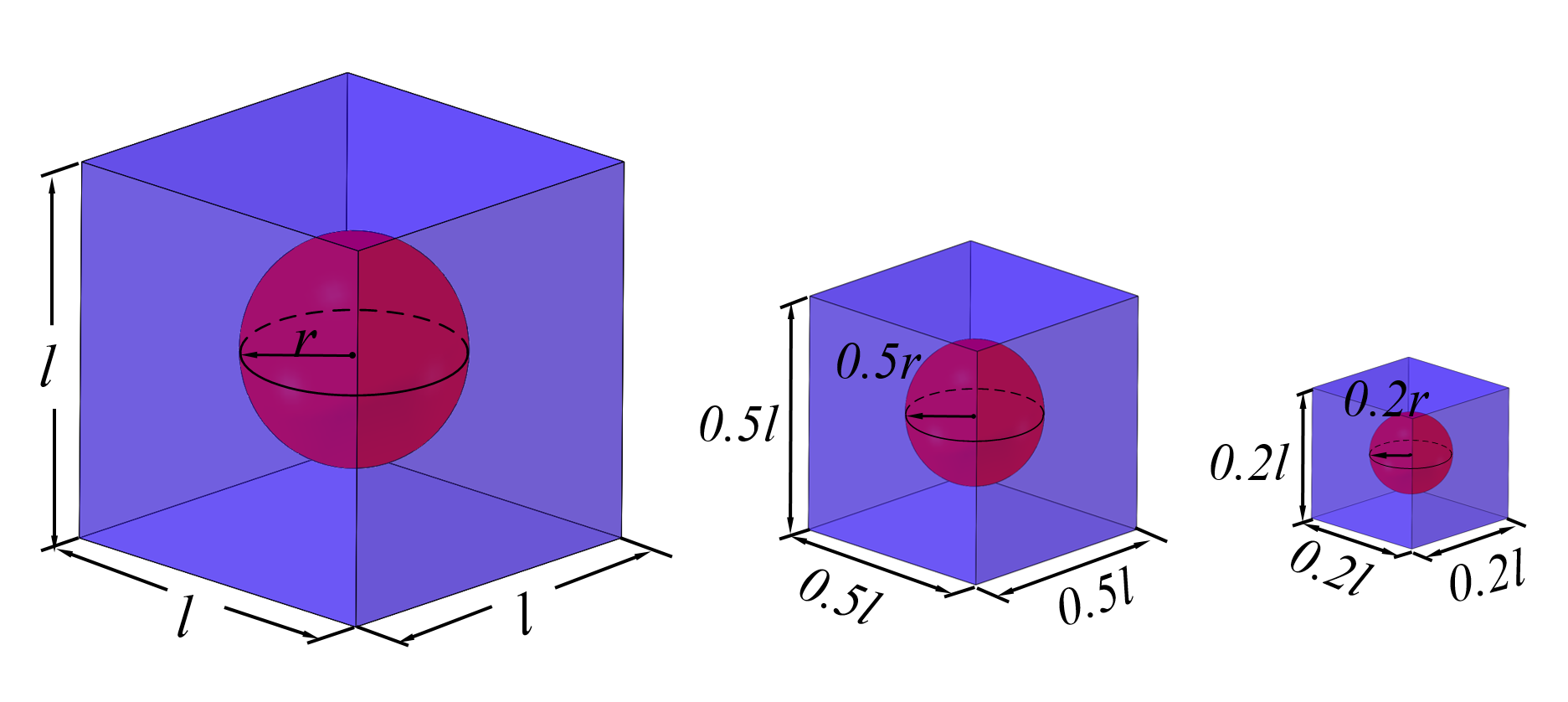}
	\caption{Unit cells with changing lengths.}
	\label{sphere_UC} 
\end{figure}
\begin{figure}[H]
	\centering
	\subfigure[Effective classical stiffness parameters. ]{\includegraphics[width=0.47\textwidth]{Sphere_c1c3}}
	\subfigure[Effective strain gradient stiffness parameters ($d_1$ - $d_6$). ]{\includegraphics[width=0.47\textwidth]{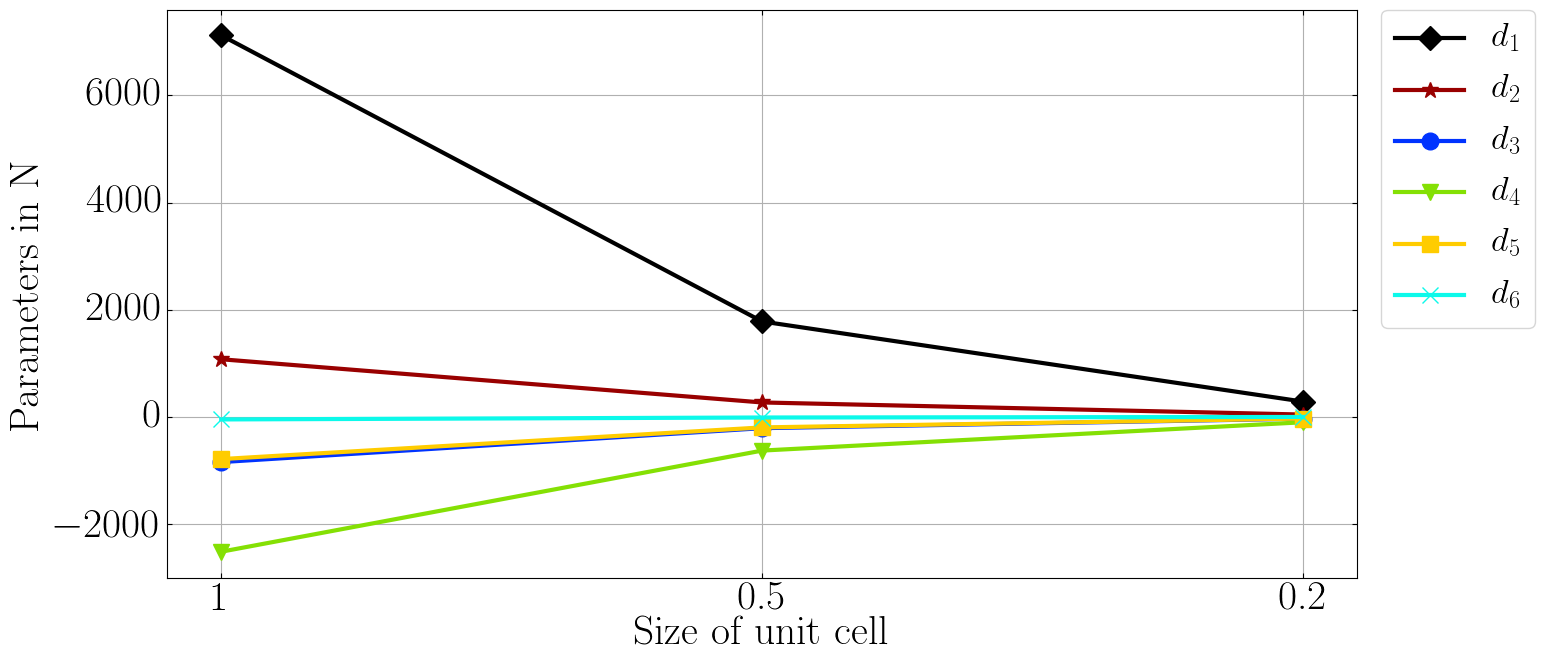}}
	\subfigure[Effective strain gradient stiffness parameters ($d_7$ - $d_{11}$). ]{\includegraphics[width=0.47\textwidth]{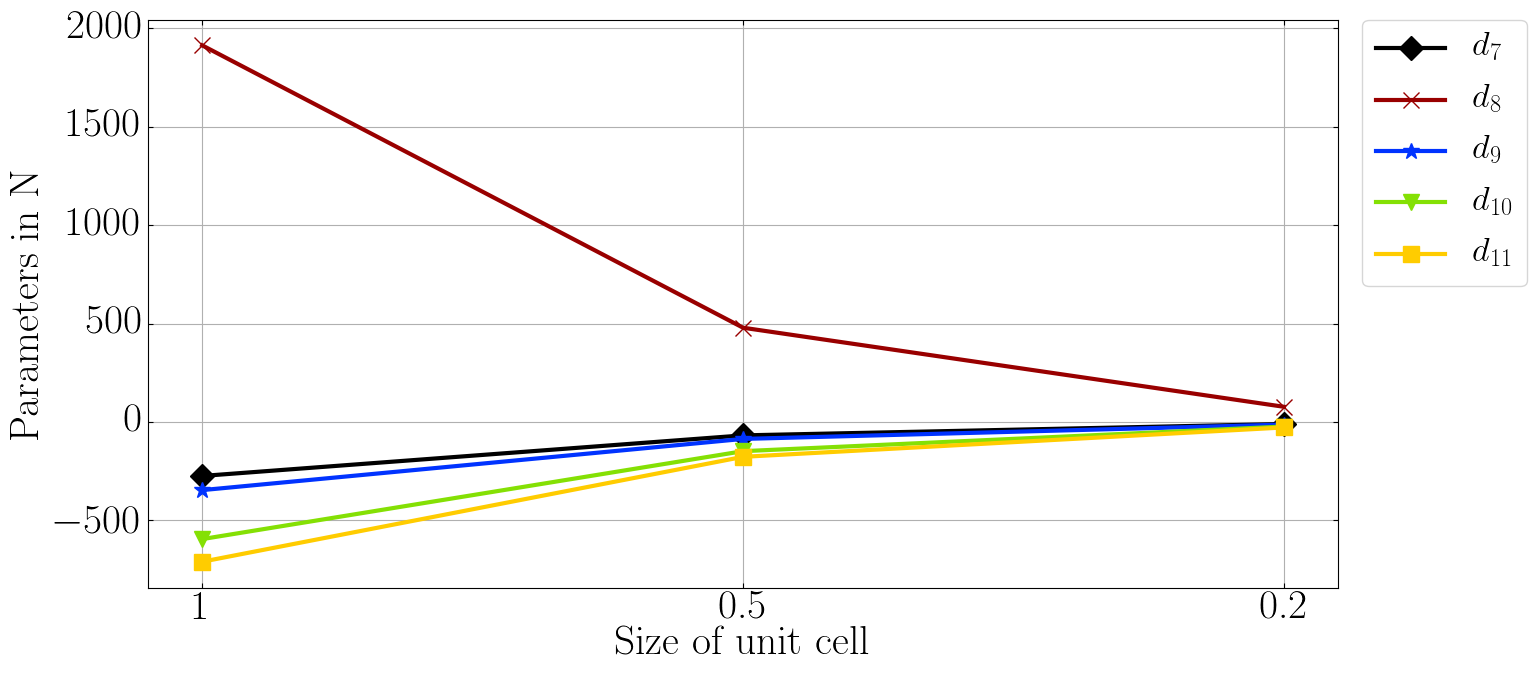}}
	\caption{ Effective material parameters with the changing lengths of unit cells.} 
	\label{sphere_UC2}
\end{figure}

\subsubsection{Aluminum foam}
\begin{figure}[H]
	\centering
	\includegraphics[width=0.85\textwidth]{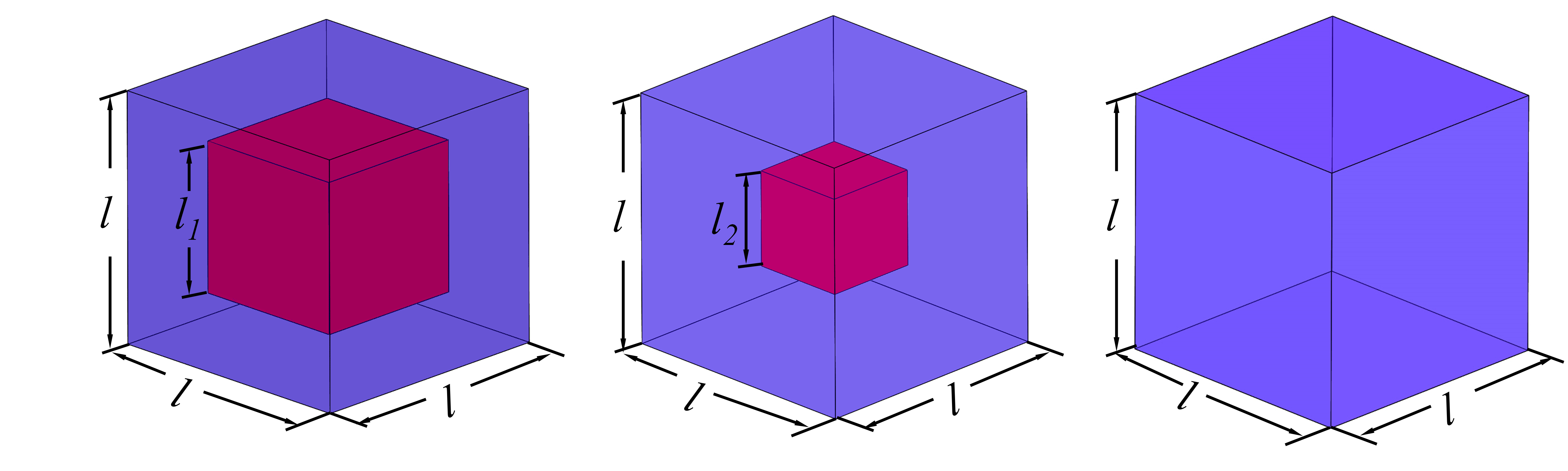}
	\caption{Different volume fraction of matrix for the aluminum foam. }
	\label{Cube_vol} 
\end{figure}

\begin{figure}[H]
	\centering
	\subfigure[Effective classical stiffness parameters. ]{\includegraphics[width=0.48\textwidth]{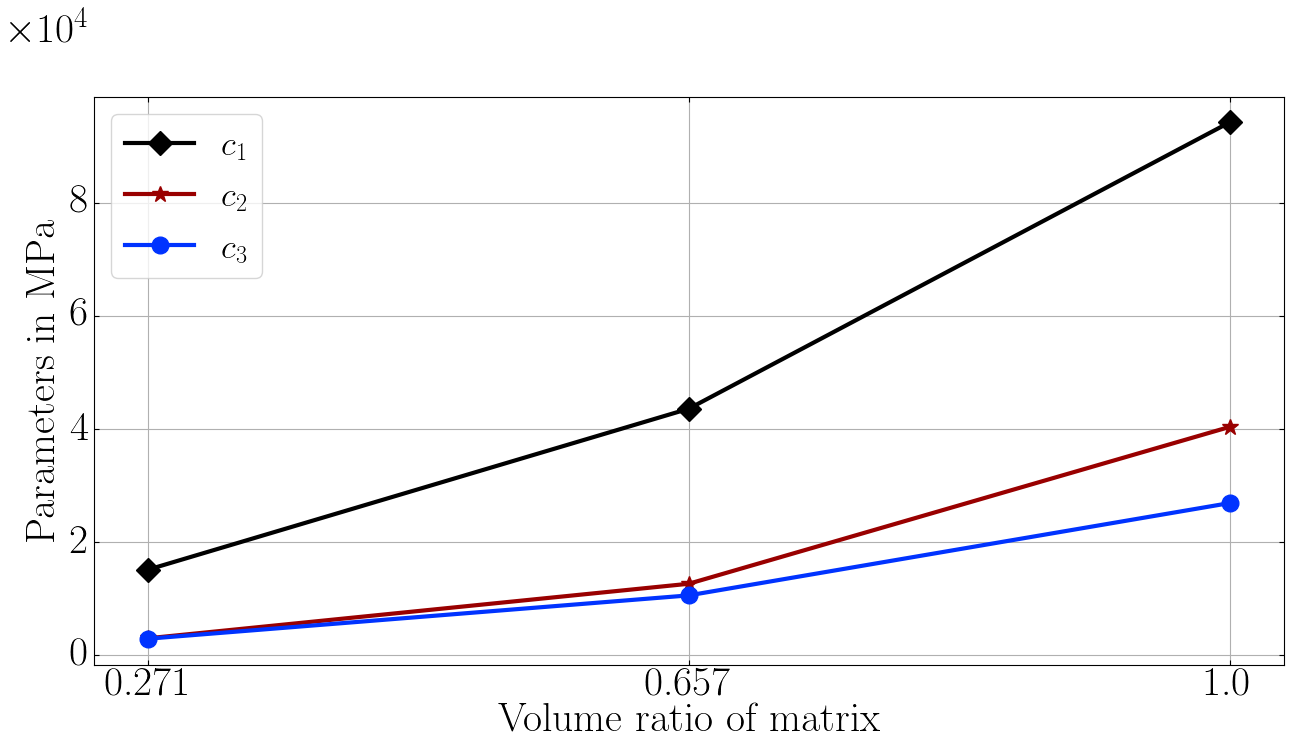}}
	\subfigure[Effective strain gradient stiffness parameters ($d_1$ - $d_6$). ]{\includegraphics[width=0.48\textwidth]{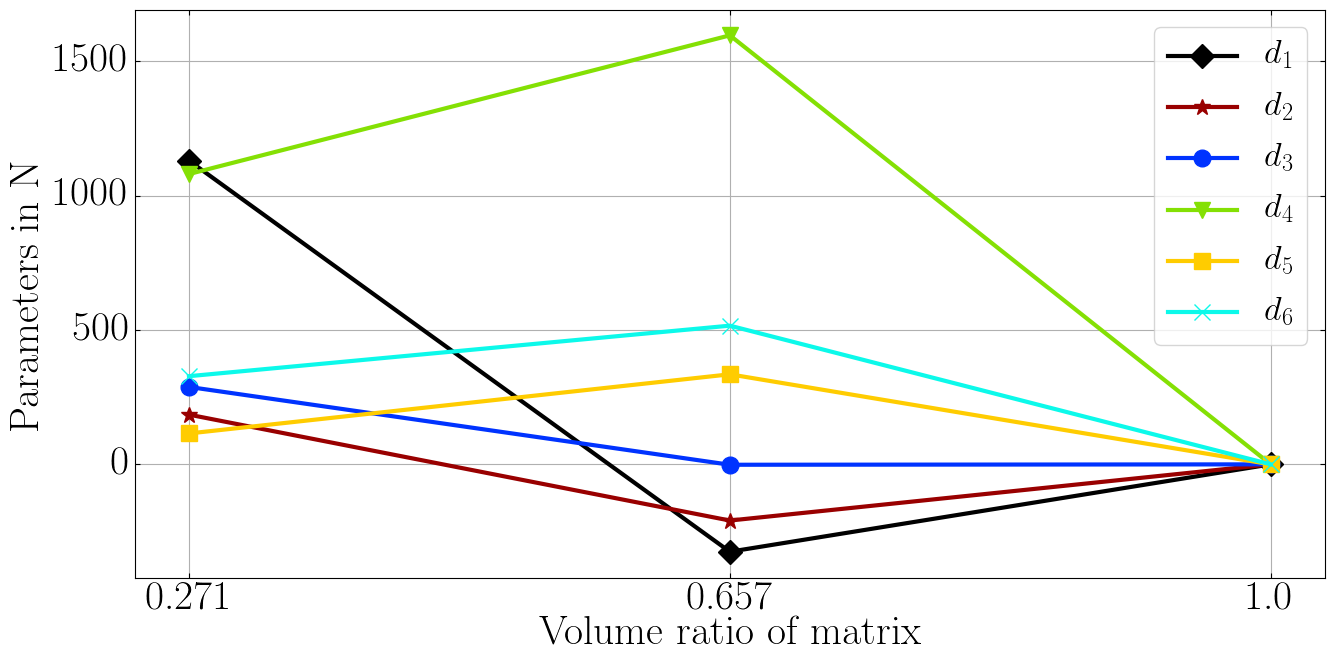}}
	\subfigure[Effective strain gradient stiffness parameters ($d_7$ - $d_{11}$). ]{\includegraphics[width=0.48\textwidth]{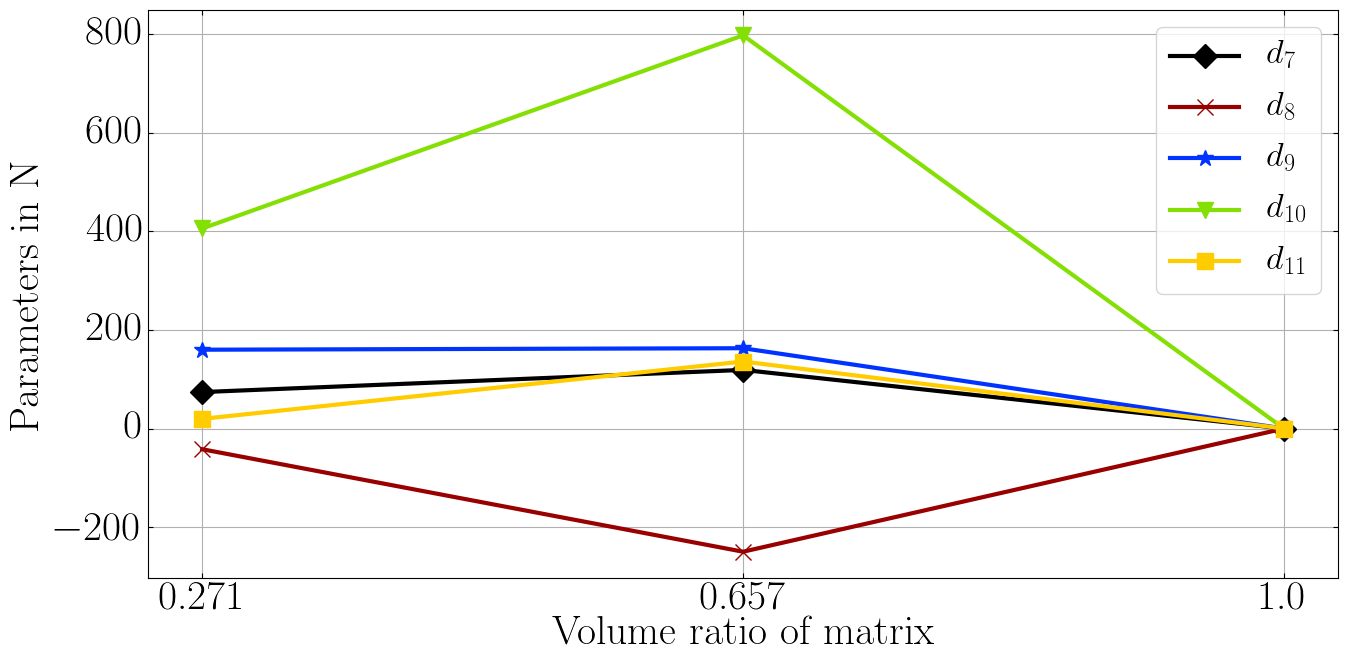}}
	\caption{ Effective material parameters with changing of volume fraction of matrix.} 
	\label{Cube_vol2}
\end{figure}

\begin{figure}[H]
	\centering
	\includegraphics[width=0.75\textwidth]{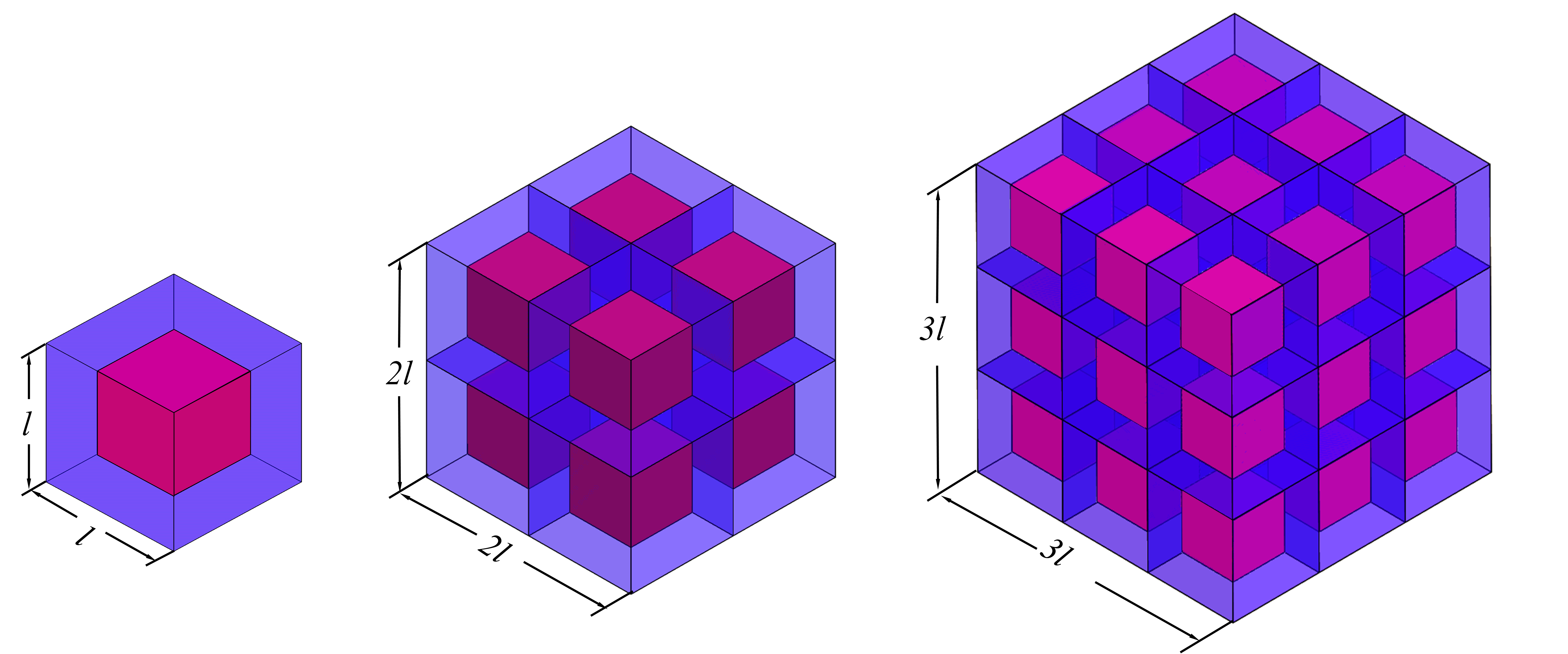}
	\caption{RVEs constructed by 1 unit cell, 8 unit cells, 27 unit cells.}
	\label{Al_RVE} 
\end{figure}

\begin{figure}[H]
	\centering
	\subfigure[Effective classical stiffness parameters. ]{\includegraphics[width=0.48\textwidth]{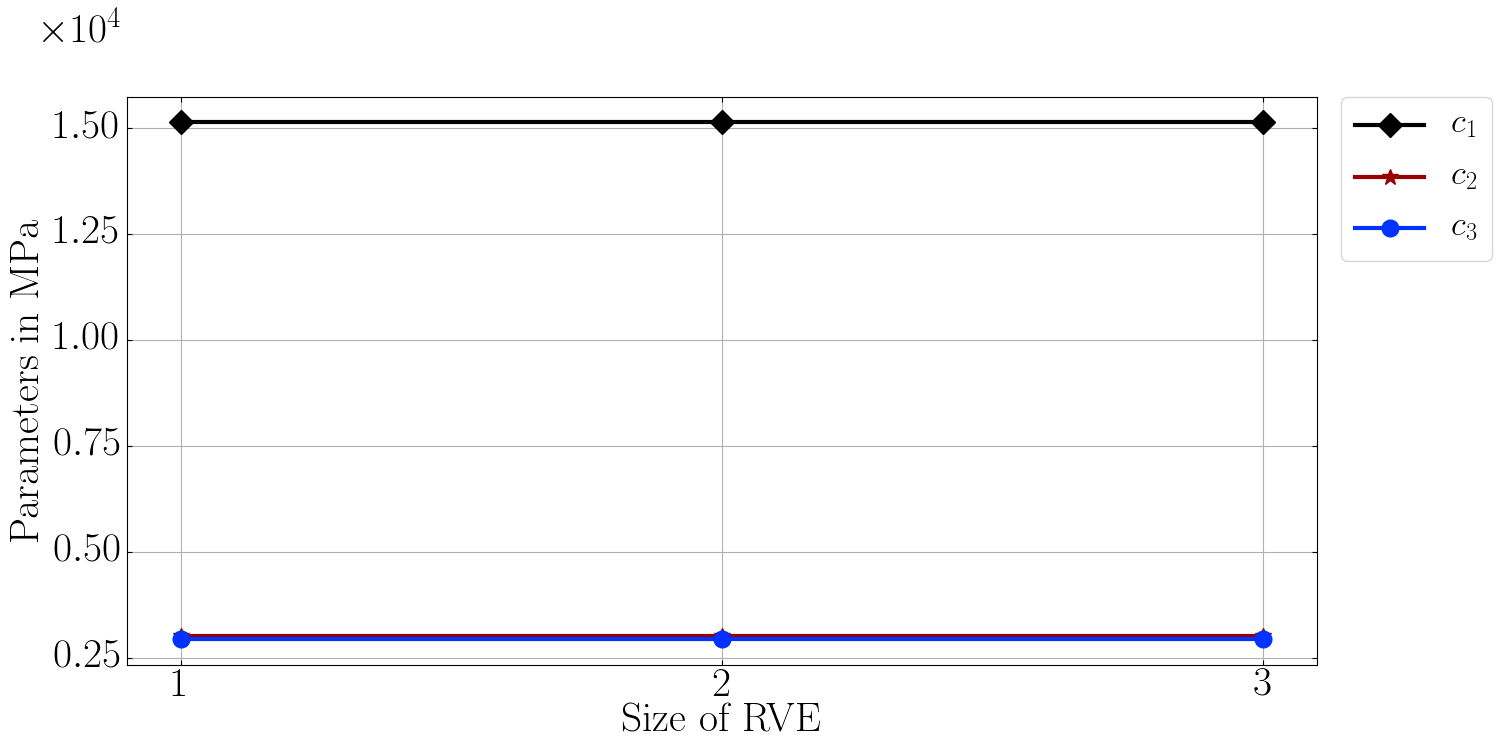}}
	\subfigure[Effective strain gradient stiffness parameters ($d_1$ - $d_6$). ]{\includegraphics[width=0.48\textwidth]{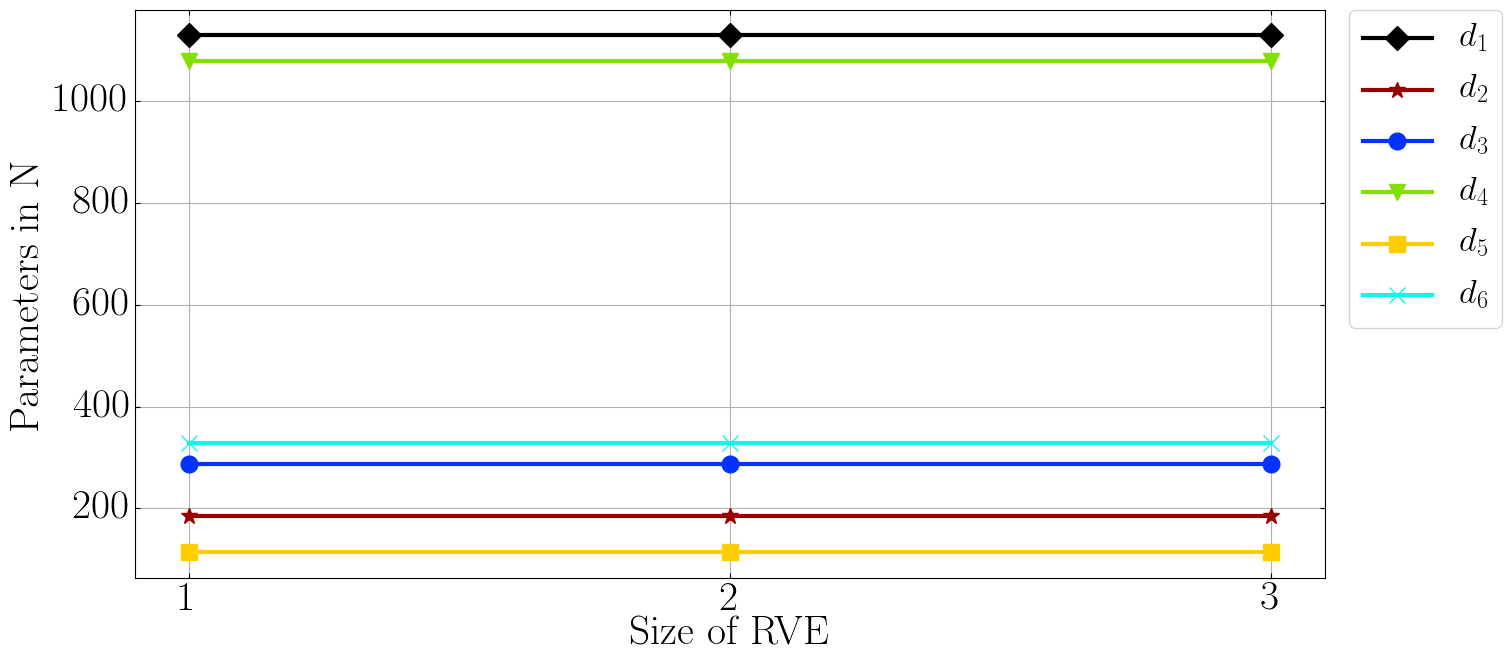}}
	\subfigure[Effective strain gradient stiffness parameters ($d_7$ - $d_{11}$). ]{\includegraphics[width=0.48\textwidth]{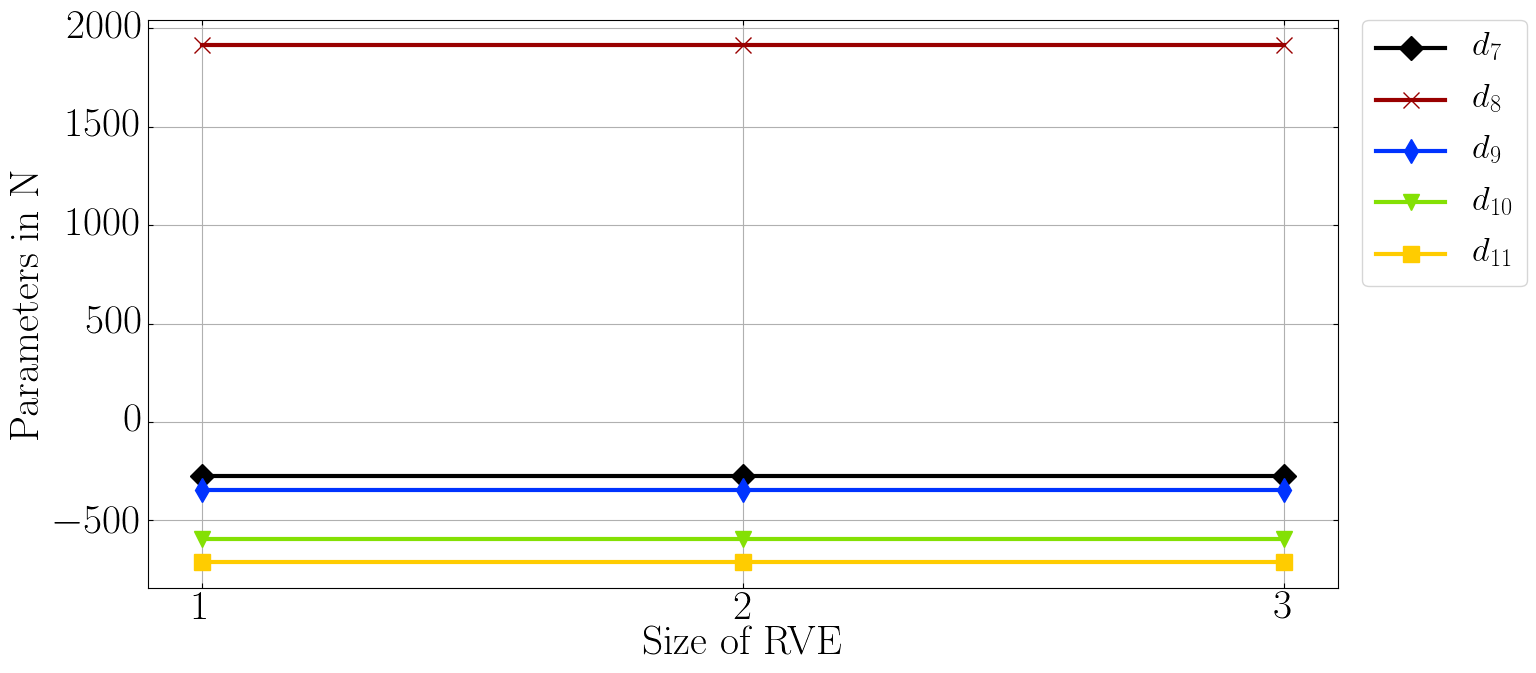}}
	\caption{ Effective material parameters with changing RVE sizes ($1 \times 1 \times 1$, $2 \times 2 \times 2$, $3 \times 3 \times 3$).} 
	\label{Al_RVE2}
\end{figure}

\begin{figure}[H]
	\centering
	\includegraphics[width=0.75\textwidth]{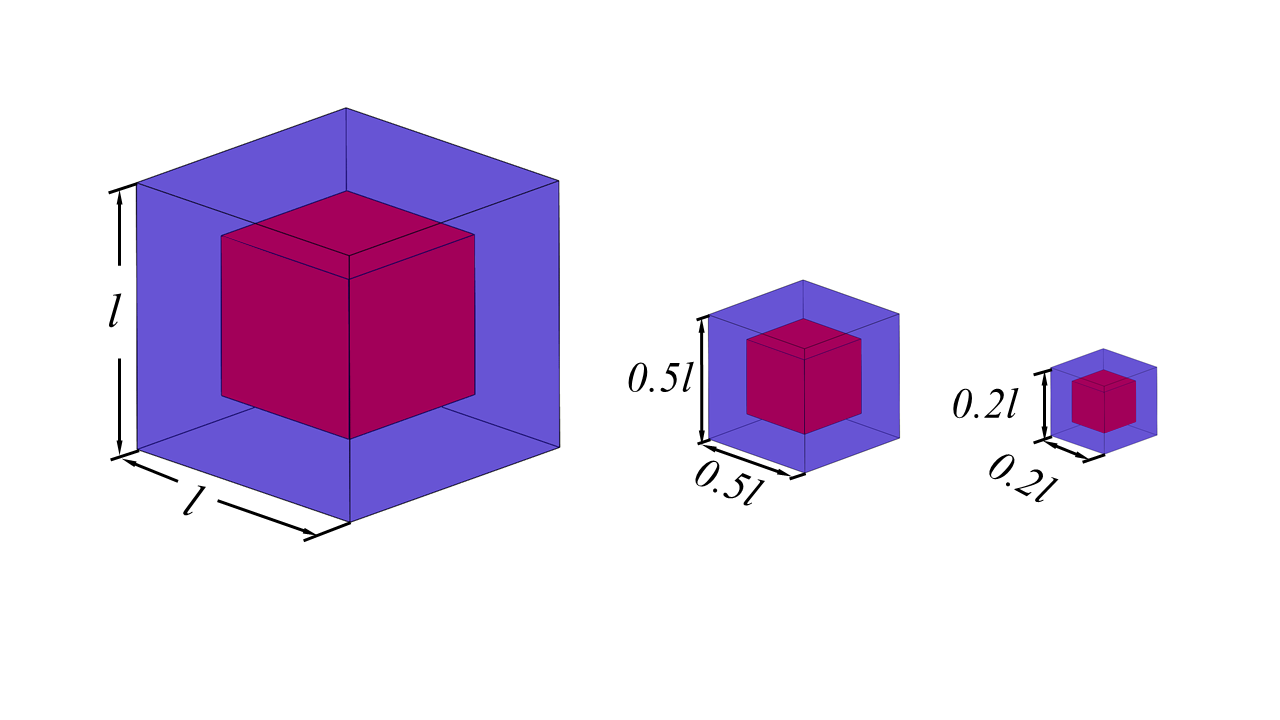}
	\caption{Unit cells with changing lengths.}
	\label{Al_UnitCell} 
\end{figure}

\begin{figure}[H]
	\centering
	\subfigure[Effective classical stiffness parameters. ]{\includegraphics[width=0.48\textwidth]{CubeRVE_c1c3}}
	\subfigure[Effective strain gradient stiffness parameters ($d_1$ - $d_6$). ]{\includegraphics[width=0.48\textwidth]{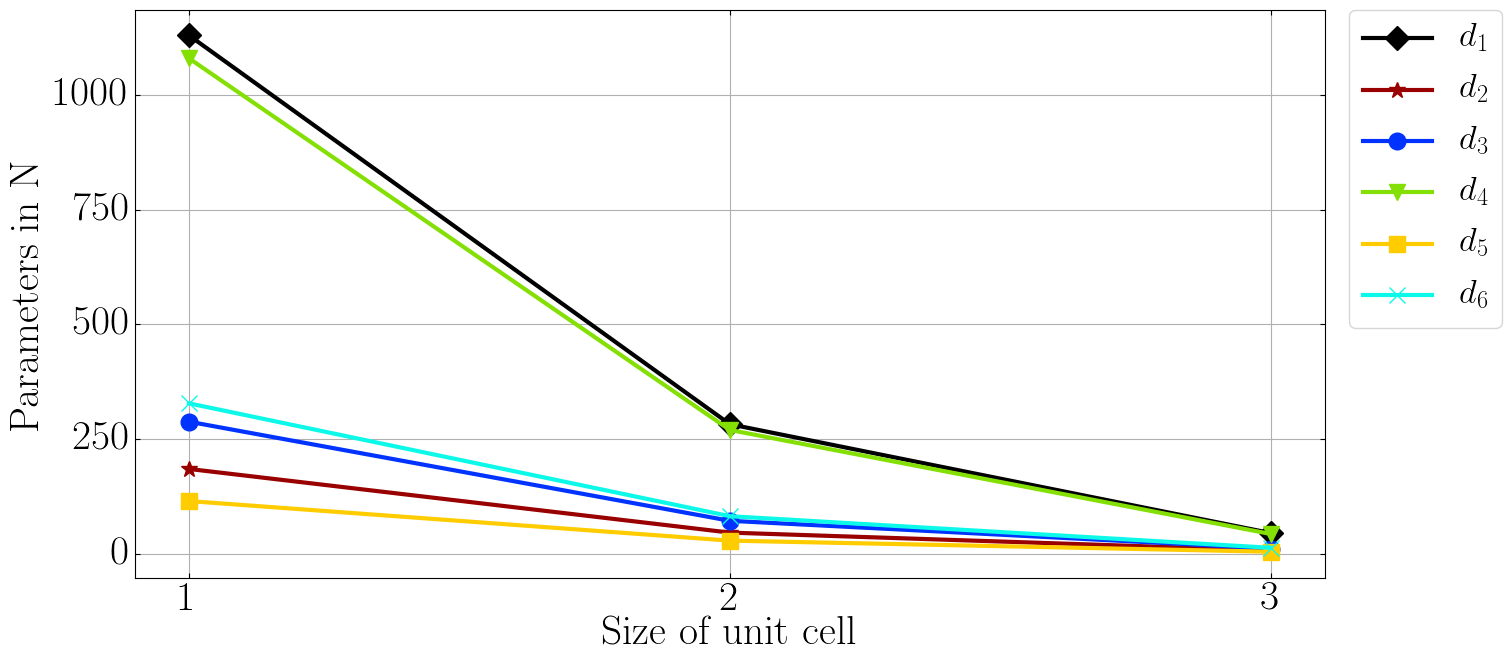}}
	\subfigure[Effective strain gradient stiffness parameters ($d_7$ - $d_{11}$). ]{\includegraphics[width=0.48\textwidth]{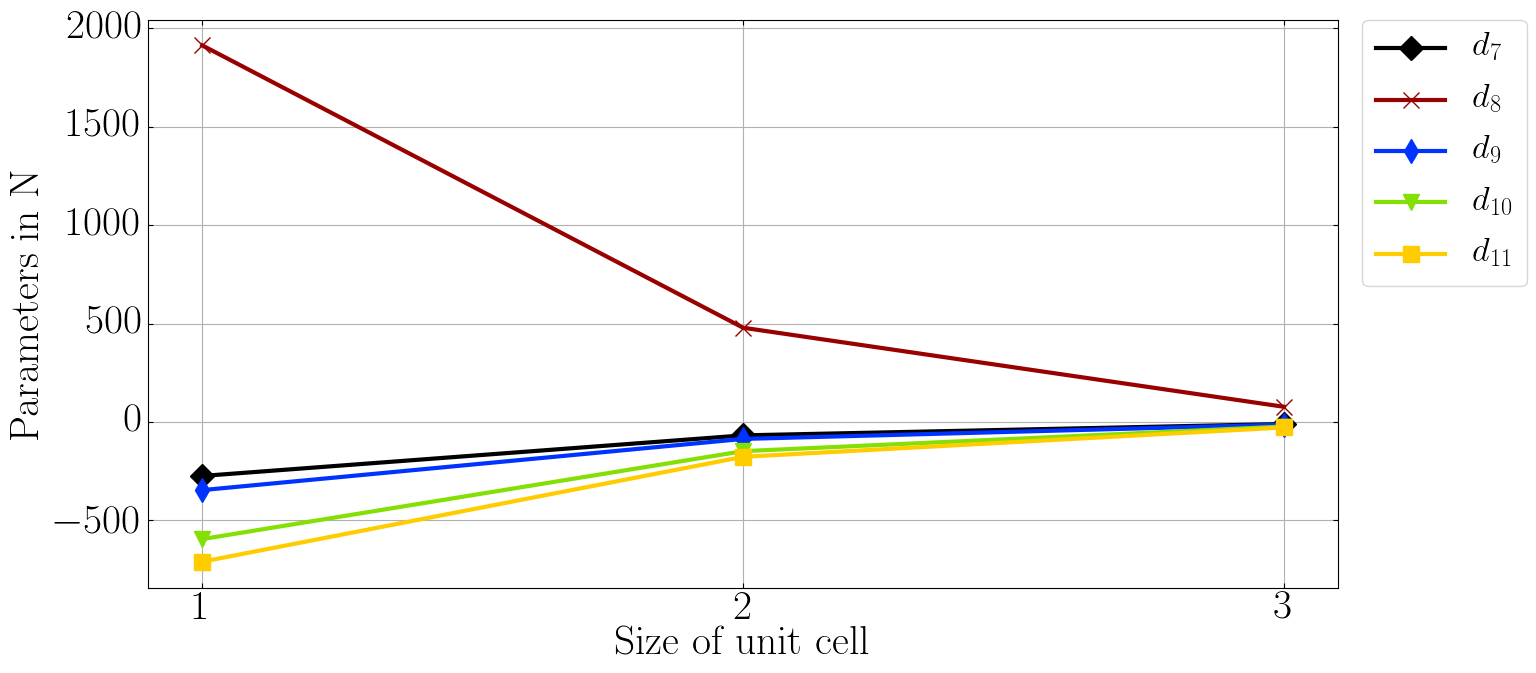}}
	\caption{ Effective material parameters with changing lengths of unit cells.} 
	\label{Al_UnitCell2}
\end{figure}

Aluminum foam is a highly porous metallic material with a cellular substructure. The RVEs of the aluminum foam are modeled by using a cubic inclusion, which are literally voids embedded in a matrix made of aluminum. In order to avoid numerical problems, a small number is assigned to the Young's modulus of voids. Indeed, this benchmark case is challenging to obtain consistently by using other procedures in the literature, where the mass density in the microscale is exchanged with the volume averaged mass density. The parameters do not show a monotonous convergence. Herein, the change in the formulation solves the problem by considering a distinction between mass densities leading to correct results in $\t \psi$ and thus in $\t G$ and $\t D$ parameters. The material properties used for aluminum foam is found in Table \ref{tab:aluminum foam}.
\begin{table}[H]
	\centering
	\caption{Material properties used for aluminum foam. $E$ Young's modulus, $\nu$ Poisson's ratio, and $\rho$ mass density.}
	\begin{tabular}{lccc}
		\toprule
		Type & $E$ in GPa & $\nu$  & $\rho$ in kg/m$^3$\\
		\midrule
		Matrix (Aluminum)  & 70  & 0.3  &  2700 \\
		Inclusion (Voids) & $10^{-10}$ & 0.0  & 0.0 \\
		\bottomrule
	\end{tabular}
	\label{tab:aluminum foam}
\end{table}
The identified parameters are found as follows:
\begeq
C\ma_{AB} = \begin{pmatrix}
	15.1 & 3.0 & 3.0 & 0.0 & 0.0 & 0.0 \\ 
	3.0 & 15.1 & 3.0 & 0.0 & 0.0 & 0.0 \\ 
	3.0 & 3.0 & 15.1 & 0.0 & 0.0 & 0.0 \\ 
	0.0 & 0.0 & 0.0 & 2.9 & 0.0 & 0.0 \\ 
	0.0 & 0.0 & 0.0 & 0.0 & 2.9 & 0.0 \\ 
	0.0 & 0.0 & 0.0 & 0.0 & 0.0 & 2.9 \\ 
\end{pmatrix} \text{\,GPa} \ , \notag
\eqend
\begeq
D\ma_{\alpha\beta} =
\resizebox{.95\textwidth}{!}{$\displaystyle
	\begin{pmatrix}
		1130.3 & 185.4 & 288.8 & 184.8 & 288.6 & 0.0 & 
		0.0 & 0.0 & 0.0 & 0.0 & 0.0 & 0.0 & 
		0.0 & 0.0 & 0.0 & 0.0 & 0.0 & 0.0 \\
		185.4 & 1080.6 & 114.9 & 328.0 & 74.6 & 0.0 & 
		0.0 & 0.0 & 0.0 & 0.0 & 0.0 & 0.0 & 
		0.0 & 0.0 & 0.0 & 0.0 & 0.0 & 0.0 \\
		288.8 & 114.9 & -42.8 & 74.5 & 160.7 & 0.0 & 
		0.0 & 0.0 & 0.0 & 0.0 & 0.0 & 0.0 & 
		0.0 & 0.0 & 0.0 & 0.0 & 0.0 & 0.0 \\
		184.8 & 328.0 & 74.5 & 1080.3 & 114.9 & 0.0 & 
		0.0 & 0.0 & 0.0 & 0.0 & 0.0 & 0.0 & 
		0.0 & 0.0 & 0.0 & 0.0 & 0.0 & 0.0 \\
		288.6 & 74.6 & 160.7 & 114.9 & -42.6 & 0.0 & 
		0.0 & 0.0 & 0.0 & 0.0 & 0.0 & 0.0 & 
		0.0 & 0.0 & 0.0 & 0.0 & 0.0 & 0.0 \\
		0.0 & 0.0 & 0.0 & 0.0 & 0.0 & 1139.1 & 
		187.6 & 290.6 & 186.9 & 290.2 & 0.0 & 0.0 & 
		0.0 & 0.0 & 0.0 & 0.0 & 0.0 & 0.0
		\\
		0.0 & 0.0 & 0.0 & 0.0 & 0.0 & 187.6 & 
		1081.0 & 114.7 & 328.4 & 75.0 & 0.0 & 0.0 & 
		0.0 & 0.0 & 0.0 & 0.0 & 0.0 & 0.0 \\
		0.0 & 0.0 & 0.0 & 0.0 & 0.0 & 290.6 & 
		114.7 & -42.8 & 74.9 & 161.0 & 0.0 & 0.0 & 
		0.0 & 0.0 & 0.0 & 0.0 & 0.0 & 0.0 \\
		0.0 & 0.0 & 0.0 & 0.0 & 0.0 & 186.9 & 
		328.4 & 74.9 & 1080.4 & 115.4 & 0.0 & 0.0 & 
		0.0 & 0.0 & 0.0 & 0.0 & 0.0 & 0.0 \\
		0.0 & 0.0 & 0.0 & 0.0 & 0.0 & 290.2 & 
		75.0 & 161.0 & 115.4 & -42.5 & 0.0 & 0.0 & 
		0.0 & 0.0 & 0.0 & 0.0 & 0.0 & 0.0 \\
		0.0 & 0.0 & 0.0 & 0.0 & 0.0 & 0.0 & 
		0.0 & 0.0 & 0.0 & 0.0 & 1171.8 & 194.3 & 
		296.5 & 194.5 & 296.9 & 0.0 & 0.0 & 0.0 \\
		0.0 & 0.0 & 0.0 & 0.0 & 0.0 & 0.0 & 
		0.0 & 0.0 & 0.0 & 0.0 & 194.3 & 1082.3 & 
		115.6 & 329.8 & 76.3 & 0.0 & 0.0 & 0.0 \\
		0.0 & 0.0 & 0.0 & 0.0 & 0.0 & 0.0 & 
		0.0 & 0.0 & 0.0 & 0.0 & 296.5 & 115.6 & 
		-42.1 & 76.3 & 162.0 & 0.0 & 0.0 & 0.0 \\
		0.0 & 0.0 & 0.0 & 0.0 & 0.0 & 0.0 & 
		0.0 & 0.0 & 0.0 & 0.0 & 194.5 & 329.8 & 
		76.3 & 1082.1 & 115.8 & 0.0 & 0.0 & 0.0 \\
		0.0 & 0.0 & 0.0 & 0.0 & 0.0 & 0.0 & 
		0.0 & 0.0 & 0.0 & 0.0 & 296.9 & 76.3 & 
		162.0 & 115.8 & -42.0 & 0.0 & 0.0 & 0.0 \\
		0.0 & 0.0 & 0.0 & 0.0 & 0.0 & 0.0 & 
		0.0 & 0.0 & 0.0 & 0.0 & 0.0 & 0.0 & 
		0.0 & 0.0 & 0.0 & 406.8 & 19.6 & 19.9 \\
		0.0 & 0.0 & 0.0 & 0.0 & 0.0 & 0.0 & 
		0.0 & 0.0 & 0.0 & 0.0 & 0.0 & 0.0 & 
		0.0 & 0.0 & 0.0 & 19.6 & 406.7 & 19.8 \\
		0.0 & 0.0 & 0.0 & 0.0 & 0.0 & 0.0 & 
		0.0 & 0.0 & 0.0 & 0.0 & 0.0 & 0.0 & 
		0.0 & 0.0 & 0.0 & 19.9 & 19.8 & 406.9
		
	\end{pmatrix} $} \text{\,N} \ . \notag
\eqend
Three independent parameters and eleven parameters are observed in the classical stiffness tensor and the strain gradient stiffness tensor, respectively. This is consistent to the cubic material symmetry as mentioned before.  Investigations on the different volume fraction of matrix, repetition of RVEs, changing sizes of unit cells are conducted as displayed in Figures \ref{Cube_vol}, \ref{Al_RVE}, \ref{Al_UnitCell}. Corresponding outcomes are presented in Figures \ref{Cube_vol2}, \ref{Al_RVE2}, \ref{Al_UnitCell2}.

\section{Remark on positive definiteness} \label{Disc}

As observed from the previous sections, negative values appear in strain gradient stiffness tensors. This fact may raise concerns regarding the positive definiteness of the strain energy function. In \cite{mindlin1968first, dell2009generalized, nazarenko2020positive, eremeyev2020well},  the issue of positive definiteness of the strain energy function for strain gradient materials is addressed and bounds on material parameters are provided. The bounds on strain gradient constants consider the
continuum to be purely local, which means that the strain energy function is convex with respect to every material point \cite{mindlin1968first}. However, when homogenizing the microstructures of composite materials with an equivalent strain gradient continuum, we have a limited non-locality. The non-locality originates from the energy equivalence as shown in Eqn.~\eqref{equivalence of energy1}.  We emphasize that the $\epsilon$ is a finite number, $\epsilon < 1$ but not necessarily $\epsilon \ll 1$, which means that the studied composite material has a finite macroscopic and microscopic sizes. Therefore, the strain energy function averaged over this microstructure size should be positive definite and not the pointwise local strain energy function. Thus coefficients in the strain gradient stiffness tensor could be negative as long as the strain energy density function integrated over the periodic unit cell is positive definite. This interpretation is aligned with in \cite{kumar2004generalized, barboura2018establishment, li2013numerical}. Additionally, this condition is always fulfilled if the microstructure energy density is positive definite, since Eqn.\,\eqref{equivalence of energy1} is enforced.

\section{Verification of the homogenized strain gradient models} \label{Computational}

\begin{figure}[H]
	\centering
	\includegraphics[width=0.75\textwidth]{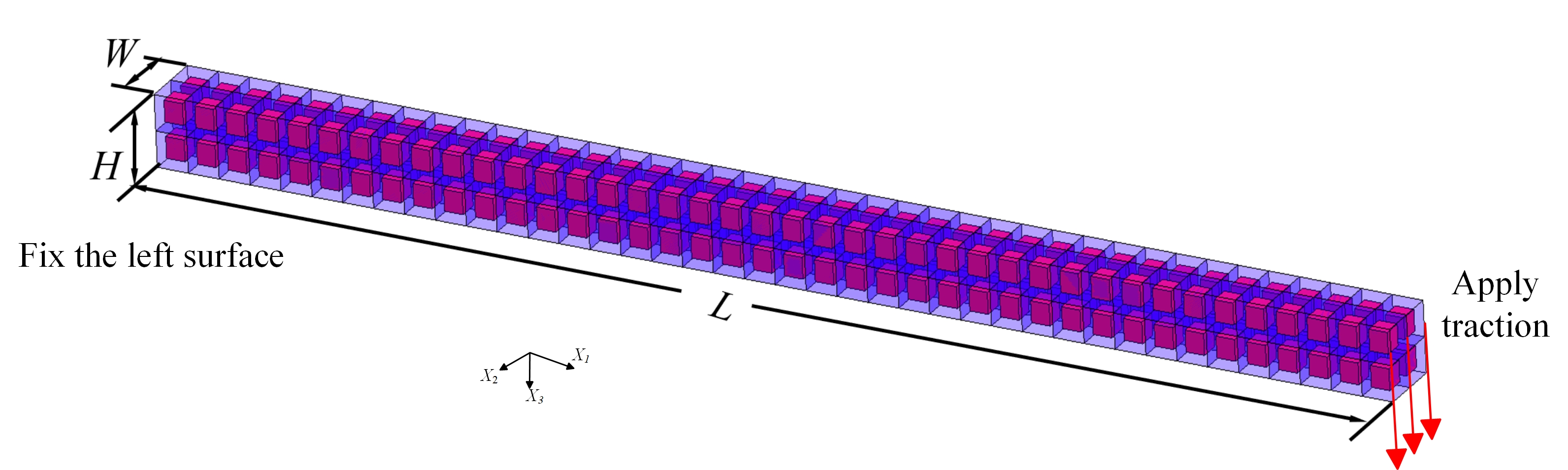}
	\caption{Schematic of a cantilever beam bending problem. The length, width, and height of the beam is $L$, $W$, $H$. The inclusions (voids) are presented as red and matrix is indicated as blue (with less opacity for the sake of visualization).} 
	\label{BeamFig} 
\end{figure}

In order to assess the homogenized strain gradient continuum model developed in this paper, finite element computations are conducted to evaluate the performance of the proposed model. To this end, a cantilever beam bending problem is selected as presented in Figure \ref{BeamFig}. The beam is made out of aluminum foam with periodically aligned microstructures of 1 mm $\times$ 1 mm $\times$ 1 mm. As mentioned above, the inclusions of the microstructures are voids. We use the determined parameters with matrix volume fraction of 0.271 as shown in Figure \ref{Cube_vol2}. The length, width, and height of the beam are assigned to be 50 mm, 2 mm, and 2 mm. The left surface, $X_1=0$, of the beam is clamped ($u_1 = u_2 = u_3 = 0$). A traction is applied at the right surface $X_1=L$.

\begin{figure}[H]
	\centering
	\includegraphics[width=0.75\textwidth]{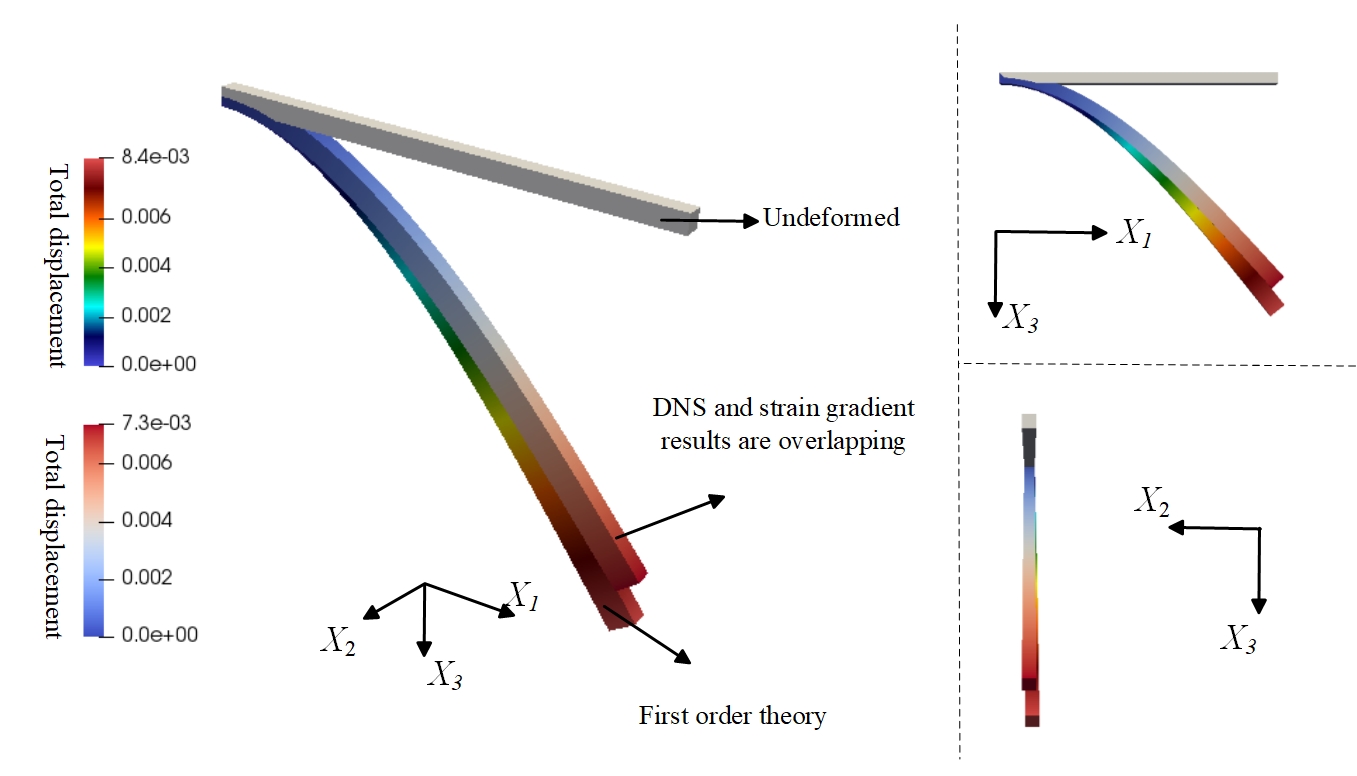}
	\caption{Comparisons of total displacement among the heterogeneous Cauchy continuum, homogenized strain gradient continuum, and the homogenized Cauchy continuum in the case of cantilever beam bending. Scaling factor 5000.  }
	\label{BeamBending} 
\end{figure}

\begin{figure}[H]
	\centering
	\subfigure[Elapsed time for the computations.  ]{\includegraphics[width=0.48\textwidth]{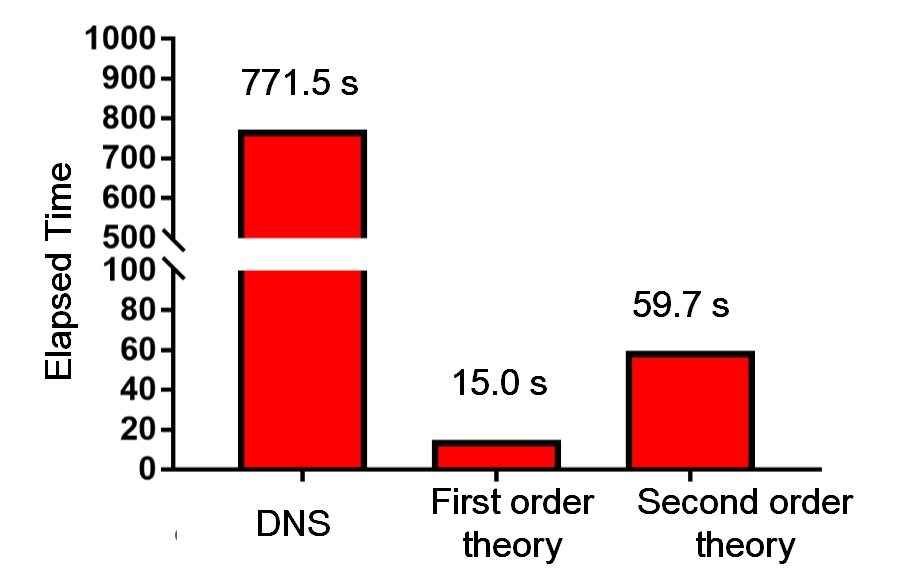}}
	\subfigure[Relative errors. ]{\includegraphics[width=0.48\textwidth]{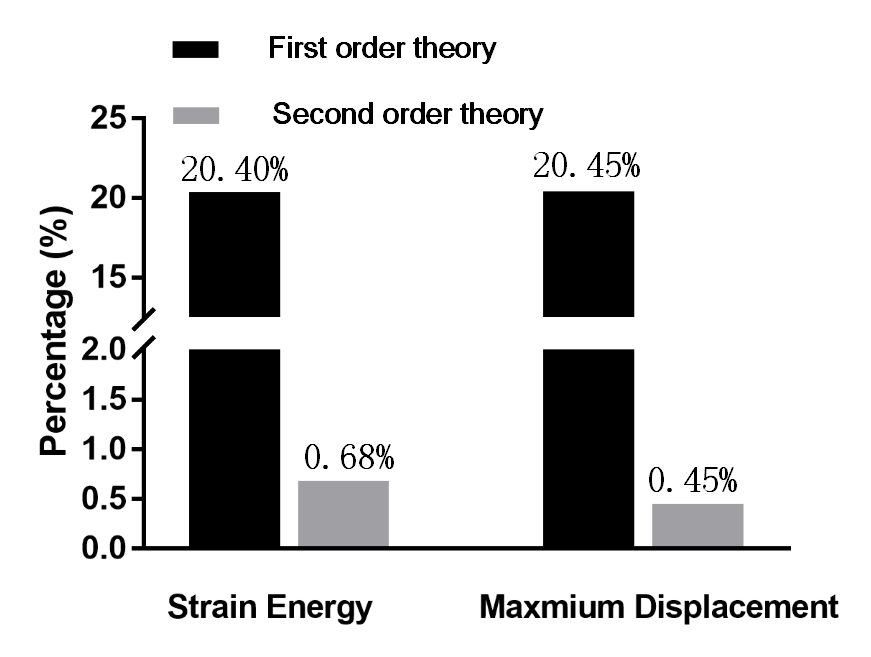}}
	\caption{ Comparisons of the elapsed time, the relative errors of the strain energy and maximum displacement for the first order theory and second order theory results.} 
	\label{bar}
\end{figure}

\begin{figure}[H]
	\centering
	\subfigure[Comparisons of strain energy.  ]{\includegraphics[width=0.48\textwidth]{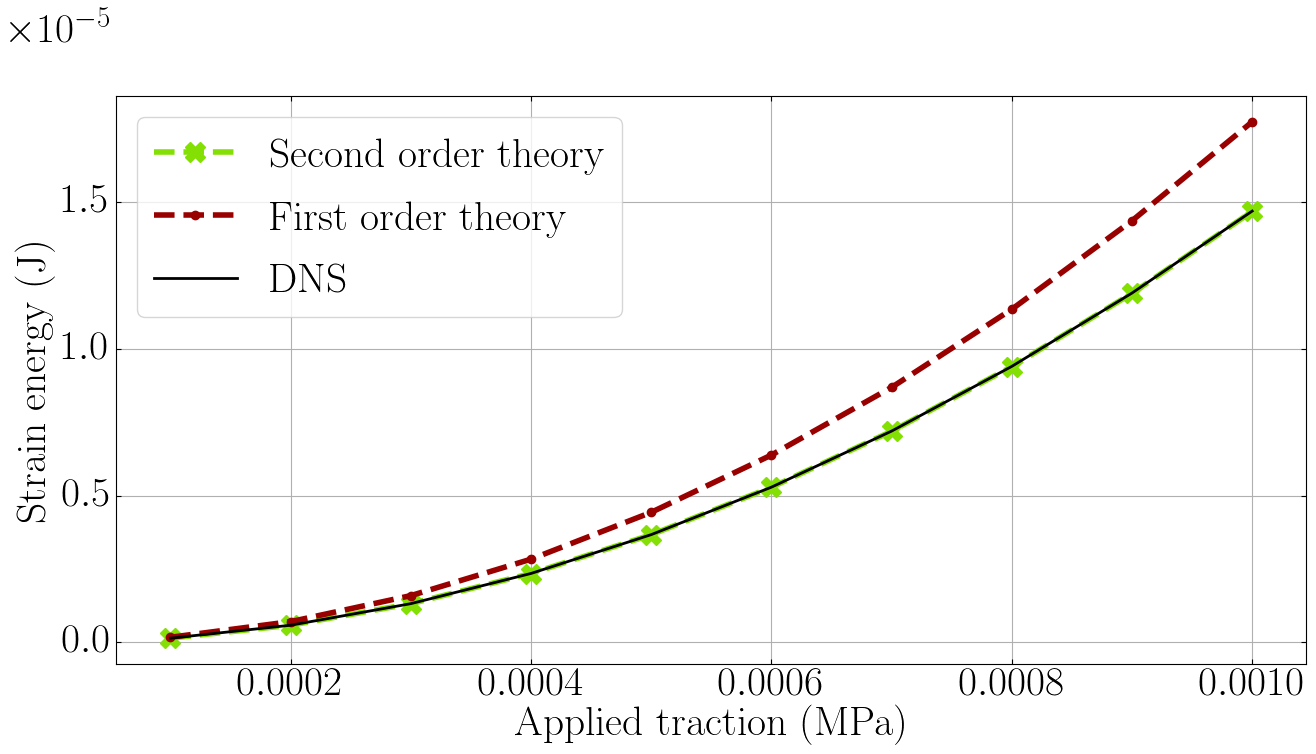}}
	\subfigure[Comparisons of 
	maximum displacement $u_3$. ]{\includegraphics[width=0.48\textwidth]{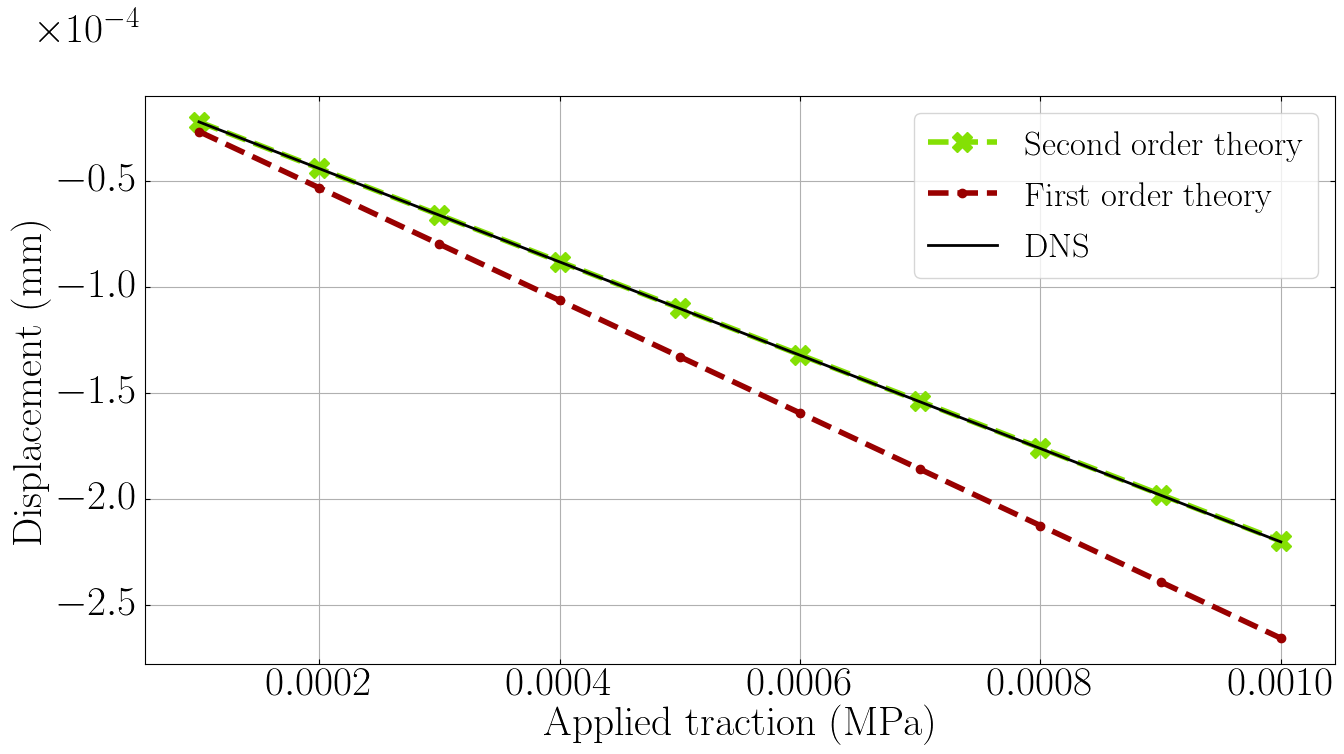}}
	\caption{ Comparisons of the calculated strain energy and 
		maximum displacement among DNS, first order theory, and second order theory results.} 
	\label{curve}
\end{figure}

We conduct three simulations. A Direct Numerical Simulation (DNS), where the microstructure is modeled in detail. This result is accepted as correct. A homogenization simulation with second order (strain gradient) theory, where $\t D$ and $\t G$ tensors are employed. A homogenized simulation with first order theory, in other words, $\t D$ and $\t G$ are set to zero (first order theory is used). As $C^1$ continuity is required for the numerical implementation of the strain gradient computations, the isogeometric analysis is used in the simulations for the homogenized models. The codes developed and verified in \cite{yang2021verification} are used herein. The weak form for linear elastic strain gradient materials is presented as
\begeq \label{weak}
\int_{\Omega} \Big(  \sigma_{ij} \delta u_{i,j} + \tau_{ijk} \delta u_{i,jk}  \Big) \d V = \int_{\p \Omega} t_i  \delta u_i \d A ,
\eqend
where $\sigma_{ij}$ and $\tau_{ijk}$ are the stress tensor and hyperstress tensor defined by 
\begeq \label{stresses}
\sigma_{ij} = \frac{\p w^{\text{M}}}{\p u_{i,j}} \ , \qquad \tau_{ijk} = \frac{\p w^{\text{M}}}{\p u_{i,jk}}  \ .
\eqend
with $w$ the strain energy density. The body forces, double traction, and the so-called wedge forces are all set to be zero, therefore the pertinent terms in the weak form Eqn.~\eqref{weak} are neglected. The traction is applied incrementally from $(0,0,0)$ to $( 0, 0, 0.001 \ \text{MPa})$. The calculated results of total displacement are shown in Figure \ref{BeamBending}. Strain gradient results match accurately DNS results. However, a significant deviation from DNS is observed, if one uses first order theory. DNS shows that the beam "act" stiffer than first order theory suggests, this experimentally well known fact is called size effect. The difference vanishes as homothetic ratio approaches zero, in other words, the same foam in a larger beam shows no size effect. Such phenomenon is also observed in the numerical investigation in \cite{021, 030} In order to assess the models further,  more investigations are made as displayed in Figure \ref{bar} and Figure \ref{curve}. 
The elapsed time for DNS is 771.5 s by using a computer (Intel(R) Core(TM) i7-8565U CPU). The elapsed time for the homogenized Cauchy model and the strain gradient model is 15.0 s and 59.7 s as shown in Figure \ref{bar}(a). It is evident that by using the homogenization techniques, computational efficiency is greatly improved. Additional difficulty is the challenge for the meshing algorithm to construct a high quality mesh for the microstructure. Element quality will not be ensured around sharp edges leading to inconsistencies as well as strain concentrations making the model error-prone. On the other hand, a homogeneous structure faces only macroscopic sharp contours such that a mesh convergence is feasible to minimize the numerical errors. Indeed, the computational efficiency of the first order theory is significantly larger than the second order theory, even in the same type of mesh. But for the chosen homothetic ratio herein, we obtain an inadequate result from the first order theory as indicated in Figure \ref{bar}(b) and Figure \ref{curve}.

\section{Conclusions} \label{Con}
Asymptotic homogenization method has been employed to homogenize composite material into effective homogeneous strain gradient continua. Main conclusions are summarized as follows:
\begin{itemize}
	\item Purely computational analysis determines all the parameters in the strain gradient theory. In particular, the parameters in the rank five tensor and the rank six tensor.	
	\item  Numerical examples for 2D and 3D, stiff and soft inclusions, cubic and transverse material symmetry cases have been conducted.
	\item  In both 2D and 3D numerical examples, the effective strain gradient parameters vanish when materials are purely homogeneous, they are independent of repetitions of RVEs and sensitive to microstructural sizes. 
	\item  Without assuming a specific symmetry group, in the case of cubic symmetry, all expected relations have been captured by the proposed formalism.
	\item  Physical meaning of the homothetic ratio $\epsilon$ is interpreted, a so-called scaling rule for effective strain gradient parameters has been discussed. 	The method is valid when $\epsilon$ is a finite value. $\epsilon < 1$ is required but not necessarily $\epsilon \ll 1$.
	\item An evaluation of the performance of the determined strain gradient parameters is done in 3D for the first time. It is found that including the strain gradient terms in the homogenized model will improve the accuracy of the prediction of the response of aluminum foams compared to classical (first order) homogenization.
\end{itemize}

The homogenization tool is applicable to any composite materials with a periodic substructure at the microscale. Such multiscale are nowadays possible to manufacture by 3D printers. Therefore, effective parameters determination is of interest for a possible topology optimization. Further investigations will focus on the following aspects:
\begin{itemize}
	\item To validate the identified parameters not only in statics but also in vibration responses, buckling critical loads \cite{khakalo2018modelling}, and wave propagation \cite{rosi2016anisotropic, eremeyev2019comparison}. 	
	\item To apply the homogenization method to the analysis of 3D composite materials with finite thickness. This may be achieved by by modeling the full thickness unit cell model and relieving the out-of-plane periodicity of the unit cell \cite{nasution2014novel}.
	\item  To explore the possibility of studying more sophisticated metamaterials such as the so-called pantographic structures \cite{dell2019advances, dell2019pantographic} or the biomimetic spinodoids metamaterials \cite{portela2020extreme} by the homogenization method.  
	\item To extend the homogenization method to non-linear regime \cite{forest2020continuum, elnady2016computation} and multiphysics fields.
\end{itemize}


\bibliographystyle{unsrt}
\bibliography{Yang2020}

\begin{thebibliography}{10}

\bibitem{boutin1996microstructural}
Claude Boutin.
\newblock Microstructural effects in elastic composites.
\newblock {\em International Journal of Solids and Structures},
  33(7):1023--1051, 1996.

\bibitem{dirrenberger2019computational}
Justin Dirrenberger, Samuel Forest, and Dominique Jeulin.
\newblock Computational homogenization of architectured materials.
\newblock In Yuri Estrin, Yves Br{\'e}chet, John Dunlop, and Peter" Fratzl,
  editors, {\em Architectured materials in nature and engineering}, Springer
  Series in Materials Science, pages 89--139. Springer, 2019.

\bibitem{arabnejad2013mechanical}
Sajad Arabnejad and Damiano Pasini.
\newblock Mechanical properties of lattice materials via asymptotic
  homogenization and comparison with alternative homogenization methods.
\newblock {\em International Journal of Mechanical Sciences}, 77:249--262,
  2013.

\bibitem{chen2020extended}
Qiang Chen, George Chatzigeorgiou, and Fodil Meraghni.
\newblock Extended mean-field homogenization of viscoelastic-viscoplastic
  polymer composites undergoing hybrid progressive degradation induced by
  interface debonding and matrix ductile damage.
\newblock {\em International Journal of Solids and Structures}, 210:1--17,
  2020.

\bibitem{yvonnet2020computational}
J~Yvonnet, Nicolas Auffray, and V~Monchiet.
\newblock Computational second-order homogenization of materials with effective
  anisotropic strain-gradient behavior.
\newblock {\em International Journal of Solids and Structures}, 2020.

\bibitem{jakabvcin2020periodic}
Luk{\'a}{\v{s}} Jakab{\v{c}}in and Pierre Seppecher.
\newblock On periodic homogenization of highly contrasted elastic structures.
\newblock {\em Journal of the Mechanics and Physics of Solids}, 144:104104,
  2020.

\bibitem{hollister1992comparison}
Scott~J Hollister and Noboru Kikuchi.
\newblock A comparison of homogenization and standard mechanics analyses for
  periodic porous composites.
\newblock {\em Computational mechanics}, 10(2):73--95, 1992.

\bibitem{muller2020experimental}
Wolfgang~H M{\"u}ller.
\newblock The experimental evidence for higher gradient theories.
\newblock In A~Bertram and S~Forest, editors, {\em Mechanics of Strain Gradient
  Materials}, volume 600 of {\em CISM International Centre for Mechanical
  Sciences}, pages 1--18. Springer, 2020.

\bibitem{mindlin1968first}
Raymond~David Mindlin and NN~Eshel.
\newblock On first strain-gradient theories in linear elasticity.
\newblock {\em International Journal of Solids and Structures}, 4(1):109--124,
  1968.

\bibitem{eringen1999theory}
A~Cemal Eringen.
\newblock Theory of micropolar elasticity.
\newblock In {\em Microcontinuum field theories}, pages 101--248. Springer,
  1999.

\bibitem{altenbach2016generalized}
Holm Altenbach and Samuel Forest.
\newblock {\em Generalized continua as models for classical and advanced
  materials}.
\newblock Springer, 2016.

\bibitem{kumar2004generalized}
Rajesh~S Kumar and David~L McDowell.
\newblock Generalized continuum modeling of 2-{D} periodic cellular solids.
\newblock {\em International Journal of solids and structures},
  41(26):7399--7422, 2004.

\bibitem{dos2012construction}
F~Dos~Reis and JF~Ganghoffer.
\newblock Construction of micropolar continua from the asymptotic
  homogenization of beam lattices.
\newblock {\em Computers \& Structures}, 112:354--363, 2012.

\bibitem{skrzat2020effective}
Andrzej Skrzat and Victor~A Eremeyev.
\newblock On the effective properties of foams in the framework of the couple
  stress theory.
\newblock {\em Continuum Mechanics and Thermodynamics}, pages 1--23, 2020.

\bibitem{kouznetsova2002multi}
Varvara Kouznetsova, Marc~GD Geers, and WA~Marcel Brekelmans.
\newblock Multi-scale constitutive modelling of heterogeneous materials with a
  gradient-enhanced computational homogenization scheme.
\newblock {\em International journal for numerical methods in engineering},
  54(8):1235--1260, 2002.

\bibitem{goda2016construction}
Ibrahim Goda and Jean-Fran{\c{c}}ois Ganghoffer.
\newblock Construction of first and second order grade anisotropic continuum
  media for 3d porous and textile composite structures.
\newblock {\em Composite Structures}, 141:292--327, 2016.

\bibitem{abdoul2019homogenization}
Houssam Abdoul-Anziz, Pierre Seppecher, and C{\'e}dric Bellis.
\newblock Homogenization of frame lattices leading to second gradient models
  coupling classical strain and strain-gradient terms.
\newblock {\em Mathematics and Mechanics of Solids}, 24(12):3976--3999, 2019.

\bibitem{weeger2021numerical}
Oliver Weeger.
\newblock Numerical homogenization of second gradient, linear elastic
  constitutive models for cubic 3d beam-lattice metamaterials.
\newblock {\em International Journal of Solids and Structures}, 2021.

\bibitem{forest2011generalized}
Samuel Forest and Duy~Khanh Trinh.
\newblock Generalized continua and non-homogeneous boundary conditions in
  homogenisation methods.
\newblock {\em ZAMM-Journal of Applied Mathematics and Mechanics/Zeitschrift
  f{\"u}r Angewandte Mathematik und Mechanik}, 91(2):90--109, 2011.

\bibitem{rokovs2019micromorphic}
Ond{\v{r}}ej Roko{\v{s}}, Maqsood~M Ameen, Ron~HJ Peerlings, and Mark~GD Geers.
\newblock Micromorphic computational homogenization for mechanical
  metamaterials with patterning fluctuation fields.
\newblock {\em Journal of the Mechanics and Physics of Solids}, 123:119--137,
  2019.

\bibitem{liu2009effective}
Shutian Liu and Wenzheng Su.
\newblock Effective couple-stress continuum model of cellular solids and size
  effects analysis.
\newblock {\em International Journal of Solids and Structures},
  46(14-15):2787--2799, 2009.

\bibitem{eremeyev2016effective}
Victor~A Eremeyev.
\newblock On effective properties of materials at the nano-and microscales
  considering surface effects.
\newblock {\em Acta Mechanica}, 227(1):29--42, 2016.

\bibitem{ganghoffer2021variational}
JF~Ganghoffer and H~Reda.
\newblock A variational approach of homogenization of heterogeneous materials
  towards second gradient continua.
\newblock {\em Mechanics of Materials}, page 103743, 2021.

\bibitem{bacigalupo2018identification}
Andrea Bacigalupo, Marco Paggi, F~Dal~Corso, and D~Bigoni.
\newblock Identification of higher-order continua equivalent to a {Cauchy}
  elastic composite.
\newblock {\em Mechanics Research Communications}, 93:11--22, 2018.

\bibitem{boutin2017linear}
Claude Boutin, Francesco dell’Isola, Ivan Giorgio, and Luca Placidi.
\newblock Linear pantographic sheets: {A}symptotic micro-macro models
  identification.
\newblock {\em Mathematics and Mechanics of Complex Systems}, 5(2):127--162,
  2017.

\bibitem{kouznetsova2004multi}
VG~Kouznetsova, Marc~GD Geers, and WAM1112 Brekelmans.
\newblock Multi-scale second-order computational homogenization of multi-phase
  materials: a nested finite element solution strategy.
\newblock {\em Computer methods in applied Mechanics and Engineering},
  193(48-51):5525--5550, 2004.

\bibitem{rosi2018validity}
Giuseppe Rosi, Luca Placidi, and Nicolas Auffray.
\newblock On the validity range of strain-gradient elasticity: a mixed
  static-dynamic identification procedure.
\newblock {\em European Journal of Mechanics-A/Solids}, 69:179--191, 2018.

\bibitem{rosi2020waves}
Giuseppe Rosi.
\newblock Waves and generalized continua.
\newblock In H~Altenbach and A~\"{O}chsner, editors, {\em Encyclopedia of
  Continuum Mechanics}, pages 2756--2765. Springer, 2020.

\bibitem{dell2019advances}
Francesco dell’Isola, Pierre Seppecher, Mario Spagnuolo, Emilio Barchiesi,
  François Hild, Tomasz Lekszycki, Ivan Giorgio, Luca Placidi, Ugo Andreaus,
  Massimo Cuomo, Simon~R. Eugster, Aron Pfaff, Klaus Hoschke, Ralph Langkemper,
  Emilio Turco, Rizacan Sarikaya, Aviral Misra, Michele~De Angelo, Francesco
  D’Annibale, Amine Bouterf, Xavier Pinelli, Anil Misra, Boris Desmorat,
  Marek Pawlikowski, Corinne Dupuy, Daria Scerrato, Patrice Peyre, Marco
  Laudato, Luca Manzari, Peter Göransson, Christian Hesch, Sofia Hesch,
  Patrick Franciosi, Justin Dirrenberger, Florian Maurin, Zacharias Vangelatos,
  Costas Grigoropoulos, Vasileia Melissinaki, Maria Farsari, Wolfgang Müller,
  Bilen~Emek Abali, Christian Liebold, Gregor Ganzosch, Philip Harrison, Rafał
  Drobnicki, Leonid Igumnov, Faris Alzahrani, and Tasawar Hayat.
\newblock Advances in pantographic structures: design, manufacturing, models,
  experiments and image analyses.
\newblock {\em Continuum Mechanics and Thermodynamics}, 31(4):1231--1282, 2019.

\bibitem{dell2019pantographic}
Francesco dell’Isola, Pierre Seppecher, Jean~Jacques Alibert, Tomasz
  Lekszycki, Roman Grygoruk, Marek Pawlikowski, David Steigmann, Ivan Giorgio,
  Ugo Andreaus, Emilio Turco, Maciej Gołaszewski, Nicola Rizzi, Claude Boutin,
  Victor~A. Eremeyev, Anil Misra, Luca Placidi, Emilio Barchiesi, Leopoldo
  Greco, Massimo Cuomo, Antonio Cazzani, Alessandro~Della Corte, Antonio
  Battista, Daria Scerrato, Inna~Zurba Eremeeva, Yosra Rahali, Jean-François
  Ganghoffer, Wolfgang Müller, Gregor Ganzosch, Mario Spagnuolo, Aron Pfaff,
  Katarzyna Barcz, Klaus Hoschke, Jan Neggers, and François Hild.
\newblock Pantographic metamaterials: an example of mathematically driven
  design and of its technological challenges.
\newblock {\em Continuum Mechanics and Thermodynamics}, 31(4):851--884, 2019.

\bibitem{misra2015identification}
Anil Misra and Payam Poorsolhjouy.
\newblock Identification of higher-order elastic constants for grain assemblies
  based upon granular micromechanics.
\newblock {\em Mathematics and Mechanics of Complex Systems}, 3(3):285--308,
  2015.

\bibitem{alibert2019homogenization}
Jean-Jacques Alibert and Alessandro Della~Corte.
\newblock Homogenization of nonlinear inextensible pantographic structures by
  $\gamma$-convergence.
\newblock {\em Mathematics and Mechanics of Complex Systems}, 7(1):1--24, 2019.

\bibitem{rahali2015homogenization}
Y~Rahali, I~Giorgio, JF~Ganghoffer, and F~dell'Isola.
\newblock Homogenization {\`a} la {Piola} produces second gradient continuum
  models for linear pantographic lattices.
\newblock {\em International Journal of Engineering Science}, 97:148--172,
  2015.

\bibitem{li2011micromechanics}
Jia Li.
\newblock A micromechanics-based strain gradient damage model for fracture
  prediction of brittle materials--part i: Homogenization methodology and
  constitutive relations.
\newblock {\em International journal of solids and structures},
  48(24):3336--3345, 2011.

\bibitem{li2013numerical}
Jia Li and Xiao-Bing Zhang.
\newblock A numerical approach for the establishment of strain gradient
  constitutive relations in periodic heterogeneous materials.
\newblock {\em European Journal of Mechanics-A/Solids}, 41:70--85, 2013.

\bibitem{barboura2018establishment}
Salma Barboura and Jia Li.
\newblock Establishment of strain gradient constitutive relations by using
  asymptotic analysis and the finite element method for complex periodic
  microstructures.
\newblock {\em International Journal of Solids and Structures}, 136:60--76,
  2018.

\bibitem{yang2019determination}
Hua Yang, Bilen~Emek Abali, Dmitry Timofeev, and Wolfgang~H M{\"u}ller.
\newblock Determination of metamaterial parameters by means of a homogenization
  approach based on asymptotic analysis.
\newblock {\em Continuum Mechanics and Thermodynamics}, pages 1--20, 2019.

\bibitem{abali2020additive}
Bilen~Emek Abali and Emilio Barchiesi.
\newblock Additive manufacturing introduced substructure and computational
  determination of metamaterials parameters by means of the asymptotic
  homogenization.
\newblock {\em Continuum Mechanics and Thermodynamics}, pages 1--17, 2020.

\bibitem{compreal}
B.~E. Abali.
\newblock Supply code for computations.
\newblock http://bilenemek.abali.org/, 2020.

\bibitem{075}
K.~K. Mandadapu, B.~E. Abali, and P.~Papadopoulos.
\newblock On the polar nature and invariance properties of a thermomechanical
  theory for continuum-on-continuum homogenization.
\newblock {\em Mathematics and Mechanics of Solids}, pages 1--18, 2021.

\bibitem{bleyer2018numericaltours}
Jeremy Bleyer.
\newblock {\em Numerical Tours of Computational Mechanics with {FE}ni{CS}},
  2018.

\bibitem{027}
B.~E. Abali.
\newblock {\em Computational {R}eality}, volume~55 of {\em Advanced Structured
  Materials}.
\newblock Springer Nature, Singapore, 2017.

\bibitem{auffray2015complete}
Nicolas Auffray, Justin Dirrenberger, and Giuseppe Rosi.
\newblock A complete description of bi-dimensional anisotropic strain-gradient
  elasticity.
\newblock {\em International Journal of Solids and Structures}, 69:195--206,
  2015.

\bibitem{auffray2009derivation}
Nicolas Auffray, Regis Bouchet, and Yves Brechet.
\newblock Derivation of anisotropic matrix for bi-dimensional strain-gradient
  elasticity behavior.
\newblock {\em International Journal of Solids and Structures}, 46(2):440--454,
  2009.

\bibitem{auffray2013matrix}
Nicolas Auffray, Hung Le~Quang, and Qi-Chang He.
\newblock Matrix representations for 3{D} strain-gradient elasticity.
\newblock {\em Journal of the Mechanics and Physics of Solids},
  61(5):1202--1223, 2013.

\bibitem{bohm2002multi}
Helmut~J B{\"o}hm, Anton Eckschlager, and W~Han.
\newblock Multi-inclusion unit cell models for metal matrix composites with
  randomly oriented discontinuous reinforcements.
\newblock {\em Computational materials science}, 25(1-2):42--53, 2002.

\bibitem{dell2009generalized}
Francesco dell'Isola, Giulio Sciarra, and Stefano Vidoli.
\newblock Generalized {H}ooke's law for isotropic second gradient materials.
\newblock {\em Proceedings of the Royal Society A: Mathematical, Physical and
  Engineering Sciences}, 465(2107):2177--2196, 2009.

\bibitem{nazarenko2020positive}
Lidiia Nazarenko, Rainer Gl{\"u}ge, and Holm Altenbach.
\newblock Positive definiteness in coupled strain gradient elasticity.
\newblock {\em Continuum Mechanics and Thermodynamics}, pages 1--13, 2020.

\bibitem{eremeyev2020well}
Victor~A Eremeyev, Sergey~A Lurie, Yury~O Solyaev, and Francesco dell’Isola.
\newblock On the well posedness of static boundary value problem within the
  linear dilatational strain gradient elasticity.
\newblock {\em Zeitschrift f{\"u}r angewandte Mathematik und Physik},
  71(6):1--16, 2020.

\bibitem{yang2021verification}
Hua Yang, Dmitry Timofeev, B~Emek Abali, Baotong Li, and Wolfgang~H M{\"u}ller.
\newblock Verification of strain gradient elasticity computation by analytical
  solutions.
\newblock {\em ZAMM-Journal of Applied Mathematics and Mechanics/Zeitschrift
  f{\"u}r Angewandte Mathematik und Mechanik}, page e202100023, 2021.

\bibitem{021}
B.~E. Abali, W.~H. M\"uller, and V.~A. Eremeyev.
\newblock Strain gradient elasticity with geometric nonlinearities and its
  computational evaluation.
\newblock {\em Mechanics of Advanced Materials and Modern Processes},
  1(1):1--11, 2015.

\bibitem{030}
B.~E. Abali, W.~H. M\"uller, and F.~dell'Isola.
\newblock Theory and computation of higher gradient elasticity theories based
  on action principles.
\newblock {\em Archive of Applied Mechanics}, 87(9):1495--1510, 2017.

\bibitem{khakalo2018modelling}
Sergei Khakalo, Viacheslav Balobanov, and Jarkko Niiranen.
\newblock Modelling size-dependent bending, buckling and vibrations of 2d
  triangular lattices by strain gradient elasticity models: applications to
  sandwich beams and auxetics.
\newblock {\em International Journal of Engineering Science}, 127:33--52, 2018.

\bibitem{rosi2016anisotropic}
Giuseppe Rosi and Nicolas Auffray.
\newblock Anisotropic and dispersive wave propagation within strain-gradient
  framework.
\newblock {\em Wave Motion}, 63:120--134, 2016.

\bibitem{eremeyev2019comparison}
Victor~A Eremeyev, Giuseppe Rosi, and Salah Naili.
\newblock Comparison of anti-plane surface waves in strain-gradient materials
  and materials with surface stresses.
\newblock {\em Mathematics and mechanics of solids}, 24(8):2526--2535, 2019.

\bibitem{nasution2014novel}
Muhammad Ridlo~Erdata Nasution, Naoyuki Watanabe, Atsushi Kondo, and Arief
  Yudhanto.
\newblock A novel asymptotic expansion homogenization analysis for {3-D}
  composite with relieved periodicity in the thickness direction.
\newblock {\em Composites science and technology}, 97:63--73, 2014.

\bibitem{portela2020extreme}
Carlos~M Portela, A~Vidyasagar, Sebastian Kr{\"o}del, Tamara Weissenbach,
  Daryl~W Yee, Julia~R Greer, and Dennis~M Kochmann.
\newblock Extreme mechanical resilience of self-assembled nanolabyrinthine
  materials.
\newblock {\em Proceedings of the National Academy of Sciences},
  117(11):5686--5693, 2020.

\bibitem{forest2020continuum}
Samuel Forest.
\newblock Continuum thermomechanics of nonlinear micromorphic, strain and
  stress gradient media.
\newblock {\em Philosophical Transactions of the Royal Society A},
  378(2170):20190169, 2020.

\bibitem{elnady2016computation}
Khaled ElNady, Ibrahim Goda, and Jean-Fran{\c{c}}ois Ganghoffer.
\newblock Computation of the effective nonlinear mechanical response of lattice
  materials considering geometrical nonlinearities.
\newblock {\em Computational Mechanics}, 58(6):957--979, 2016.

\end{thebibliography}
\end{document}